\documentclass[11pt]{article}              

\usepackage[margin=25mm]{geometry}
\usepackage{graphicx,adjustbox}
\usepackage{colortbl}
\usepackage{fancyhdr}
\usepackage{amssymb,amsfonts,amsmath,amsthm}
\usepackage{natbib}
\usepackage{bstnotations}

\usepackage{authblk}
	
\graphicspath{{./},{./Figures/}}

\usepackage{algorithm}
\usepackage{algorithmic}

\usepackage{multirow}
\usepackage{color}
\usepackage{epsfig}
\usepackage{mathrsfs}
\usepackage{epstopdf}
\usepackage{tabularx}
\usepackage{pdfpages}
\usepackage{float}
\usepackage{subcaption}
\usepackage{caption}
\usepackage{tikz}
\usetikzlibrary{calc,patterns,decorations.pathmorphing,decorations.markings}
\newtheorem{theorem}{Theorem}
\newtheorem{corollary}{Corollary}

\begin{document}
\title{Sparse polynomial chaos expansions of frequency response functions using stochastic frequency transformation} 

\author[1]{V. Yaghoubi} \author[1]{S. Marelli} \author[2]{B. Sudret}  \author[1]{T. Abrahamsson}

\affil[1]{Department of Applied Mechanics, Chalmers University of Technology, Horsalsvagen 7A, 412 96 Gothenburg, Sweden}

\affil[2]{Chair of Risk, Safety and Uncertainty Quantification,
	
	ETH Zurich, Stefano-Franscini-Platz 5, 8093 Zurich, Switzerland}

\date{}
\maketitle

\abstract{Frequency response functions (FRFs) are important for assessing the behavior of stochastic linear dynamic systems. For large systems, their evaluations are time-consuming even for a single simulation. In such cases, uncertainty quantification by crude Monte-Carlo simulation is not feasible. In this paper, we propose the use of sparse adaptive polynomial chaos expansions (PCE) as a surrogate of the full model. 
	To overcome known limitations of PCE when applied to FRF simulation, we propose a frequency transformation strategy that maximizes the similarity between FRFs prior to the calculation of the PCE surrogate. This strategy results in lower-order PCEs for each frequency. Principal component analysis is then employed to reduce the number of random outputs. The proposed approach is applied to two case studies: a simple 2-DOF system and a 6-DOF system with 16 random inputs. The accuracy assessment of the results indicates that the proposed approach can predict single FRFs accurately. Besides, it is shown that the first two moments of the FRFs obtained by the PCE converge to the reference results faster than with the Monte-Carlo (MC) methods. \\[1em] 

}

\maketitle

\section{Introduction}
\label{intro}

Interest towards working with large engineering systems is increasing recently, but long simulation time is one of the main limiting factors. Although the development of the computational power of modern computers has been very fast in recent years, increasing model complexity, more precise description of model properties and more detailed representation of the system geometry still result in considerable execution time and memory usage. Model reduction \citep{khorsand2012improved, rahrovani2014modal}, efficient simulation \citep{yaghoubi2016efficient, avitabile2009efficient, liu2012efficient}  and parallel simulation methods \citep{yaghoubiparallel, tak2013high} are different strategies to address this issue. 

Consequently, uncertainty propagation in these systems cannot be carried out by classical approaches such as crude Monte-Carlo (MC) simulation. More advanced methods such as stochastic model reduction \citep{amsallem2011online} or surrogate modeling \citep{frangos2010surrogate} are required to replace the computationally expensive model with an approximation that can reproduce the essential features faster. Of interest here are surrogate models. They can be created intrusively or non-intrusively. In intrusive approaches, the equations of the system are modified such that one explicit function relates the stochastic properties of the system responses to the random inputs. The perturbation method \citep{schueller2009uncertain} is a classical tool used for this purpose but it is only accurate when the random inputs have small coefficients of variation (COV). An alternative method is intrusive polynomial chaos expansion \citep{ghanem2003stochastic}. It was first introduced for Gaussian input random variables \citep{wiener1938homogeneous} and then extended to the other types of random variables leading to generalized polynomial chaos \citep{xiu2002wiener,soize2004physical}. 

In non-intrusive approaches, already existing deterministic codes are evaluated at several sample points selected over the parameter space. This selection depends on the methods employed to build the surrogate model, namely regression \citep{blatman2010adaptive, berveiller2006stochastic} or projection methods \citep{gilli2013uncertainty, knio2001stochastic}.  Kriging \citep{fricker2011probabilistic, jones1998efficient} and non-intrusive PCE \citep{Blatman2011a} or combination thereof \citep{kersaudy2015new, SchoebiIJUQ2015} are examples of the non-intrusive approaches. 
The major drawback of PCE methods, both intrusive and non-intrusive, is the large number of unknown coefficients in problems with large parameter spaces, which is referred to as the curse of dimensionality \citep{Sudret2007}. Sparse \citep{blatman2008sparse} and adaptive sparse \citep{blatman2011adaptive} polynomial chaos expansions have been developed to dramatically reduce the computational cost in this scenario. 

To propagate and quantify the uncertainty in a Quantity of Interest (QoI) of a system, its response should be monitored all over the parameter space. This response could be calculated in time, frequency or modal domain. For dynamic systems, the frequency response is important because it provides information over a frequency range with a clear physical interpretation. This is the main reason of the recent focus on frequency response functions (FRF) for uncertainty quantification of dynamic systems and their surrogates \citep{fricker2011probabilistic, goller2011interpolation, kundu2014hybrid,adhikari2011doubly,Chatterjee2015}.

Several attempts have been made to find a surrogate model for the FRF by using modal properties or random eigenvalue problems. 
\citet{pichler2009mode} proposed a mode-based meta-model for the frequency response functions of stochastic structural systems.  \citet{yuhermite} used Hermite polynomials to solve the random eigenvalue problem and then employed modal assurance criteria (MAC) to detect the phenomenon of modal intermixing.
\citet{manan2010prediction} used non-intrusive polynomial expansions to find the modal properties of a system and predict the bounds for stochastic FRFs. They implemented the method on models with one or two parameters and COV $\leq$ 2\%.

Very few and recent papers addressed the direct implementation of PCE on the frequency responses of systems. \citet{kundu2015dynamic} proposed to obtain the frequency response of a stochastic system by projecting the response on a reduced subspace of eigenvectors of a set of complex, frequency-adaptive, rational stochastic weighting functions.

\citet{pagnacco2013polynomial} investigated the use of polynomial chaos expansions for modeling multimodal dynamic systems using the intrusive approach by studying a single degree of freedom (DOF) system. 
They showed that the direct use of the polynomial chaos results in some spurious peaks and proposed to use multi-element PCE to model the stochastic frequency response but, to the knowledge of the authors, they did not publish anything on more complex systems yet.
\citet{jacquelinpolynomial2015} studied a 2-DOF system to investigate the possibility of direct implementation of PCE for the moments of the FRFs and they also reported the problem of spurious peaks. They showed that the PCE converges slowly on the resonance parts. They accelerate the convergence of the first two statistical moments by using Aitken's method and its generalizations \citep{Jacquelin2015144}.  

In general, there are two main difficulties to make the PCE surrogate model directly for the FRFs: $(i)$ their non-smooth behavior over the frequency axis due to abrupt changes of the amplitude that occur close to the resonance frequencies. At such frequencies, the amplitudes are driven by damping \citep{craig2006fundamentals}. In \citet{adhikari2016damping}, Adhikari and Pascal investigated the effect of damping in the dynamic response of stochastic systems and explain why making surrogate models in the areas close to the resonance frequencies is very challenging. $(ii)$ the frequency shift of the eigenfrequencies due to uncertainties in the parameters.
This results in very high-order PCEs even for the FRFs obtained from cases with 1 or 2 DOFs. The main contribution of this work is to propose a method that can solve both problems.

The proposed approach consists of two steps. First, the FRFs are transformed via a stochastic frequency transformation such that their associated eigenfrequencies are aligned in the transformed frequency axis, called \textit{scaled frequency}. 
Then, PCE is performed on the \textit{scaled frequency} axis.

The advantage of this procedure is the fact that after the transformation, the behavior of the FRFs at each \textit{scaled frequency} is smooth enough to be surrogated with low-order PCEs. However, since PCE is made for each \textit{scaled frequency}, this approach results in a very large number of random outputs. To solve this issue, an efficient version of principal component analysis is employed. Moreover, the problem of the curse of dimensionality is resolved here by means of adaptive sparse PCEs.

The outline of the paper is as follows. 
In Section \ref{frf}, the required equations for deriving the FRFs of a system are presented. 
In Section \ref{method}, all appropriate mathematics for approximating a model by polynomial chaos expansion are presented. The main challenges for building PCEs for FRFs are elaborated and the proposed solutions are presented.
In Section \ref{example}, the method is applied to two case studies, a simple case and a case with a relatively large number of input parameters.

\section{Frequency response function (FRF)}
\label{frf}

Consider the spatially-discretized governing second-order equation of motion of a structure as

\begin{equation}
\label{eq1}
\ve{M}\ddot{\ve{q}} +\ve{V}\dot{\ve{q}}+\ve{K}\ve{q}=\ve{f}(t)
\end{equation}

where for an $n$-DOF system with $n_u$ system inputs and $n_y$ system outputs, $\ve{q}(t)\in \mathbb{R}^n $ is the displacement vector, $\ve{f}(t)$ is the external load vector which is governed by a Boolean transformation of stimuli vector $\ve{f}(t)=\ve{P_uu}(t)$; with $\ve{u}(t)\in \mathbb{R}^{n_u}$. Real positive-definite symmetric matrices $\ve{M},\ve{V},\ve{K} \in \mathbb{R}^{n\times n}$ are mass, damping and stiffness matrices, respectively. 
The state-space realization of the equation of motion in Eq. (\ref{eq1}) can be written as 

\begin{equation}
\label{eq2}
\dot{\ve{x}}(t)=\ve{Ax}(t)+\ve{Bu}(t), \hspace{1cm} \ve{y}(t)=\ve{Cx}(t)+\ve{Du}(t)
\end{equation}
where $\ve{A}\in \mathbb{C}^{2n\times 2n}$, $\ve{B}\in \mathbb{C}^{2n\times n_u}$, $\ve{C}\in \mathbb{C}^{n_y\times 2n}$, and $\ve{D}\in \mathbb{C}^{n_y\times n_u}$. $ \ve{x}^T(t)=[\ve{q}(t)^T, \dot{\ve{q}}^T(t)]\in \mathbb{R}^{2n} $ is the state vector, and $ \ve{y}(t)\in \mathbb{R}^{n_y}$ is the system output. $\ve{A}$ and $\ve{B}$ are related to mass, damping and stiffness as follows
\begin{equation}
\label{eq3}
\ve{A}= \left[\begin{array}{cc}
\ve{0} & \ve{I}\\
-\ve{M}^{-1}\ve{V} & -\ve{M}^{-1}\ve{K}  
\end{array}\right],     
\ve{B}= \left[\begin{array}{c}
\ve{0} \\ \ve{M}^{-1}\ve{P}_u
\end{array}\right].
\end{equation}

The output matrix $\ve{C}$, which has application dependent elements, linearly maps the states to the output $\ve{y}$ and $\ve{D}$ is the associated direct throughput matrix. The frequency response of the model (\ref{eq2}) can be written as
\begin{equation}
\label{eq4}
\ve{\mathcal{H}}(j\omega)=\ve{C}(j\omega \ve{I}-\ve{A})^{-1}\ve{B}+\ve{D},
\end{equation}
where $\ve{\mathcal{H}}=[\mathcal{H}_1, \mathcal{H}_2,\cdots,\mathcal{H}_{n_u \times n_y}]^\text{T} \in \mathbb{C}^{(n_y \times n_u)\times 1}, \forall \omega$ and $j=\sqrt{-1}$. $(\bullet)^\text{T}$ stands for the transpose of the matrix. It should be mentioned that the eigenvalues of $\ve{A}$ are the poles of the system. They are complex and their imaginary parts can be approximated as the frequencies, in rad/s, at which the maximum amplitude occurs.

\section{Methodology}
\label{method}
This section first, briefly reviews polynomial chaos expansion for real-valued responses. Then, the method of stochastic frequency transformation is explained in conjunction with the proposed method as well as its application to the complex-valued FRF responses.

\subsection{Polynomial chaos expansions}
\label{PCE}
Let $\cm$ be a computational model with \emph{M}-dimensional random inputs $\ve{X}$=$\{X_1, X_2, ...,X_M\}^\text{T}$ and a scalar output $Y$. Further, let us denote the joint probability distribution function (PDF) of the random inputs by $f_{\ve{X}}(\ve{x}) $ defined in the probability space ($\Omega$,$\mathscr{F}$, $\mathbb{P}$).

Assume that the system response $Y=\cm(\Ve{X})$ is a second-order random variable, \ie $\Esp{Y^2}<+\infty$ and therefore it belongs to the Hilbert space $\mathscr{H}=\mathscr{L}_{f_{\Ve{X}}}^2(\mathbb{R}^M, \mathbb{R}) $ of $f_{\Ve{X}} $-square integrable functions of $\Ve{X}$ with respect to the inner product: 

\begin{equation}
\Esp{\psi(\Ve{X})\phi(\Ve{X})}=\int_{\cd_{\Ve{X}}} \psi(\Ve{x})\phi(\Ve{x})f_{\Ve{X}}(\Ve{x}) \di \Ve{x}
\label{eq:PCE:Hilbert}
\end{equation} 
where $\cd_{\Ve{X}}$ is the support of $\ve{X}$. Further assume that the input variables are independent, \ie $f_{\ve{X}}(\ve{x})=\prod_{i=1}^{\emph{M}}\emph{f}_{X_i}(x_i)$. Then the generalized polynomial chaos representation of $Y$ reads \citep{xiu2002wiener}:

\begin{equation}
Y=\sum_{\boldsymbol{\alpha\in \mathbb{N}^{\emph{M}}}}\tilde{u}_{\ua} \psi_{\ve{\alpha}}(\ve{X})
\label{eq:PCE:infinite}
\end{equation}
in which $\tilde{u}_{\ua}$ is a set of unknown deterministic coefficients, $\boldsymbol{\alpha} = (\alpha_1, \alpha_2, ..., \alpha_M)$  is a multi-index set which indicates the polynomial degree of $\psi_{\ve{\alpha}}(\ve{X})$ in each of the \emph{M} input variables. $\psi_{\ve{\alpha}}$s are multivariate orthonormal polynomials with respect to the joint PDF $f_{\Ve{X}}(\Ve{x})$, \ie:

\begin{equation}
\Esp{\psi_{\ve{\alpha}}(\Ve{X})\psi_{\ve{\beta}}(\Ve{X})}=\int_{\cd_{\Ve{X}}} \psi_{\ve{\alpha}}(\Ve{x})\psi_{\ve{\beta}}(\Ve{x}) f_{\Ve{X}}(\Ve{x}) \di \Ve{x}=\delta_{\ve{\alpha}\ve{\beta}}
\label{eq:PCE:orthonormal}
\end{equation}
where $\delta_{\ve{\alpha}\ve{\beta}}$ is the Kronecker delta. Since the input variables are assumed to be independent, these multivariate polynomials can be constructed by a tensorization of univariate orthonormal polynomials with respect to the marginal PDFs, \ie $ {\psi}_{\ve{\alpha}}(\ve{X})=\prod_{i=1}^M \psi_{\alpha_i}^{(i)}({X}_i) $. For instance, if the inputs are standard normal or uniform variables, the corresponding univariate polynomials are Hermite or Legendre polynomials, respectively.

In practice, the infinite series in Eq. (\ref{eq:PCE:infinite}) has to be truncated. 
Given a maximum polynomial degree $p$, the standard truncation scheme includes all polynomials corresponding to the set $\ca^{M,p} = \{ \ua \in \Nn^M\; \colon \; |\ua| \le p \}, $ where $|\ua|= \sum_{i = 1}^M \alpha_i$ is the total degree of polynomial  $\psi_{\ve{\alpha}}$. The cardinality of the set $\ca^{M,p}=\binom{M+p}{p}=P$ increases rapidly by increasing the number of parameters \emph{M} and the order of polynomials $p$. However, it can be controlled with suitable truncation strategies such as $q$-norm hyperbolic truncation \citep{blatman2010adaptive}, that drastically reduce the number of unknowns when $M$ is large.

The estimation of the vector of coefficient $\tilde{u}_{\ua}$ can be done non-intrusively by projection  \citep{Ghiocel98,Ghiocel2002} or least square regression methods \citep{blatman2010adaptive, berveiller2006stochastic}. The latter is based on minimizing the truncation error $\epsilon$ via least square as follows: 
\begin{equation}
\label{eqn:PCE:coeff:LS:1}
{Y}=\cm(\Ve{X} )=\sum_{\ua \in \ca^{M,p}} {\tilde{u}_{\ua}} \, {{\psi}_{\ve{\alpha}}({\ve{X}})} +
{{\epsilon}} \equiv \ve{\tilde{U}}^\text{T} {\ve{\Psi}}(\Ve{X}) +
{\ve{\epsilon}}
\end{equation}
This can be formulated as

\begin{equation}
\label{eqn:PCE:coeff:LS:2}
\ve{\hat{\tilde{U}}} = \arg \min \Esp{ \left(\ve{\tilde{U}}^\text{T}
	\ve{\Psi}(\Ve{X}) - \cm(\Ve{X}) \right)^2}. 
\end{equation} 
Let $ \ve{\cx} = \{\ve{x}^{(1)}, \ve{x}^{(2)}, ..., \ve{x}^{(N_{ED})}\} $ and $\ve{\cy}=\{{y}^{(1)}=\cm(\ve{x}^{(1)}), {y}^{(2)}=\cm(\ve{x}^{(2)}), ..., {y}^{(N_{ED})}=\cm(\ve{x}^{(N_{ED})})\}$ be an experimental design with $N_{ED}$ space-filling samples of $\Ve{X}$ and the corresponding system responses, respectively. Then, the minimization problem (\ref{eqn:PCE:coeff:LS:2}) admits a closed form solution
\begin{equation}
\label{eqn:PCE:coeff:LS:3}
\ve{\hat{\tilde{U}}} = (\ve{\Psi}^\text{T}\ve{\Psi})^{-1} \ve{\Psi}^\text{T} \ve{\cy},
\end{equation}
in which $\ve{\Psi}$ is the matrix containing the evaluations of the Hilbertian bases, that is $\ve{\Psi}_{ij}={\psi}_{\ua_j}(\ve{x}^{(i)}), i=1,2,...,N_{ED}, j=1,2,...,P$.

The accuracy of PCE will be improved by reducing the effect of over-fitting in least square regression. This can be done by using sparse adaptive regression algorithms proposed in \citet{hastie2007forward, efron2004least}. In particular, the Least Angle Regression (LAR) algorithm has been demonstrated to be effective in the context of PCE by \citet{Blatman2011a}. 

\subsection{Vector-valued response}
\label{PCE:PCA}

In the case of vector-valued response, \ie $\ve{Y} \in \mathbb{R}^N, N>1$, the presented approach may be applied componentwise. This can make the algorithm computationally cumbersome for models with large number of random outputs. To decrease the computational cost, one can extract the main statistical features of the vector random response by principal component analysis (PCA). The concept has been adapted to the context of PCE by \citet{Blatman2013}. 

To perform sample-based PCA, let us rewrite the $\ve{\cy}$ as the combination of its mean $\bar{\ve{\cy}}$ and covariance matrix as follows:
\begin{equation}
\label{eqn:PCE:PCA:1}
\ve{\cy}=\bar{\ve{\cy}}+\sum_{i=1}^{N}\ve{u}_i\ve{v}^\text{T}_i 
\end{equation}
where the $\ve{v}_i$'s are the eigenvectors of the covariance matrix:
\begin{equation}
\label{eqn:PCE:PCA:2}
COV(\ve{\cy}) = \Esp{(\ve{\cy}-\bar{\ve{\cy}})^\text{T} (\ve{\cy}-\bar{\ve{\cy})}}
=[\ve{v}_1, ..., \ve{v}_{N}]
\left[\begin{array}{ccc}
l_1 & \dots & 0 \\
& \ddots & \\
0 & \dots & l_{N} \\
\end{array} \right]
\left[\begin{array}{c}
\ve{v}_1^\text{T}\\
\vdots \\
\ve{v}_{N}^\text{T} 
\end{array}\right]^\text{T}
\end{equation}
and the $\ve{u}_i$'s are vectors such that 
\begin{equation}
\label{eqn:PCE:PCA:3}
\ve{u}_i=(\ve{\cy}-\bar{\ve{\cy}})\ve{v}_i.
\end{equation}

One can approximate $\ve{\cy}$ by the $\hat{N}$-term truncation:

\begin{equation}
\label{eqn:PCE:PCA:4}
\ve{\cy}=\bar{\ve{\cy}}+\sum_{i=1}^{\hat{N}}\ve{u}_i\ve{v}^\text{T}_i,  \quad \hat{N} \ll {N}.
\end{equation}

Since $\bar{\ve{\cy}}$ and $\ve{v}^\text{T}_i$ are the mean and the eigenvectors of the system responses, respectively, they are independent of the realization. Therefore, PC expansion can be applied directly on the $\hat{N} \ll N$ auxiliary variables $\ve{u}_i$. Besides, acknowledging the fact that the PCA is an invertible transform, the original output can be retrieved directly from Eq. (\ref{eqn:PCE:PCA:4}) for every new prediction of $\ve{u}$. 

\subsubsection{Vector-valued data with extremely large output size}
\label{method:PCA}
Assume $\ve{\cy} \in \mathbb{R}^{{N}_{ED} \times N}$, has an extremely large $N$ and $N \gg N_{ED}$. Then $ COV(\ve{\cy}) \in \mathbb{R}^{N \times N}$ is exceptionally large and solving the eigenvalue problem numerically may be unfeasible. To address this issue, the following well-known theorem and the associated corollary is presented. 

\begin{theorem} (Singular value decomposition)
	\label{thm:PCE:PCA}
	Let $A \in \mathbb{R}^{n \times N}, n<N$ and $rank(A)=n$ then there exist two orthogonal matrices, $U$ and $V$ and two diagonal matrices $S$ and $\Sigma$ such that 
	$A=USV^\text{T}=U
	\left[\begin{array}{ll}
	\Sigma & 0\\
	0 & 0
	\end{array}\right]
	V^\text{T}$ in which $ U \in \mathbb{R}^{n \times n}$, $ S \in \mathbb{R}^{n \times N}$, $ \Sigma \in \mathbb{R}^{n \times n}$ and $ V \in \mathbb{R}^{N \times N}$. Furthermore, this decomposition can be written as eigenvalue decomposition as $AA^\text{T} U=U\Sigma$ and $A^\text{T} AV=VS$. 
\end{theorem}
\begin{corollary}
	\label{cor:PCE:PCA}
	The nonzero eigenvalues of $A^\text{T} A$ and $AA^\text{T}$ are equal. Furthermore, U and V are related to each other by 
	\begin{equation}
	\label{eq:cor:PCE:PCA}
	U = AVS^{-1}
	\end{equation}
\end{corollary}
The proof of the theorem can be found in any matrix analysis book, \eg \citet{laub2005matrix} and the corollary directly follows from the theorem.

Therefore, instead of the eigenvalue calculation of $(\ve{\cy}-\bar{\ve{\cy}})^\text{T} (\ve{\cy}-\bar{\ve{\cy}}) \in \mathbb{R}^{N \times N}$, which may be an extremely large matrix, one can consider $(\ve{\cy}-\bar{\ve{\cy}}) (\ve{\cy}-\bar{\ve{\cy}})^\text{T} \in \mathbb{R}^{N_{ED} \times N_{ED}}$, which is much smaller. The associated eigenvectors can be transformed to the ones in Eq. (\ref{eqn:PCE:PCA:4}) through Eq. (\ref{eq:cor:PCE:PCA}).

\subsection{Stochastic frequency transformation}
\label{transformation}

In this section, the method of stochastic frequency transformation is developed to address the challenge of frequency shift at eigenfrequencies due to uncertainty in the parameters. The idea is basically to apply a transformation to the system responses to maximize their similarity before building PCE, as first proposed by \citet{mai2015polynomial}. 
Here, the technique is extended and adopted into the frequency domain to obtain PCEs of the FRFs.   

\begin{figure}[H]
	\centering
	\begin{subfigure}[b]{1\columnwidth}
		\centering
		\label{fig:FRF:2DOF:indirect}
		\includegraphics[width=.75\linewidth]{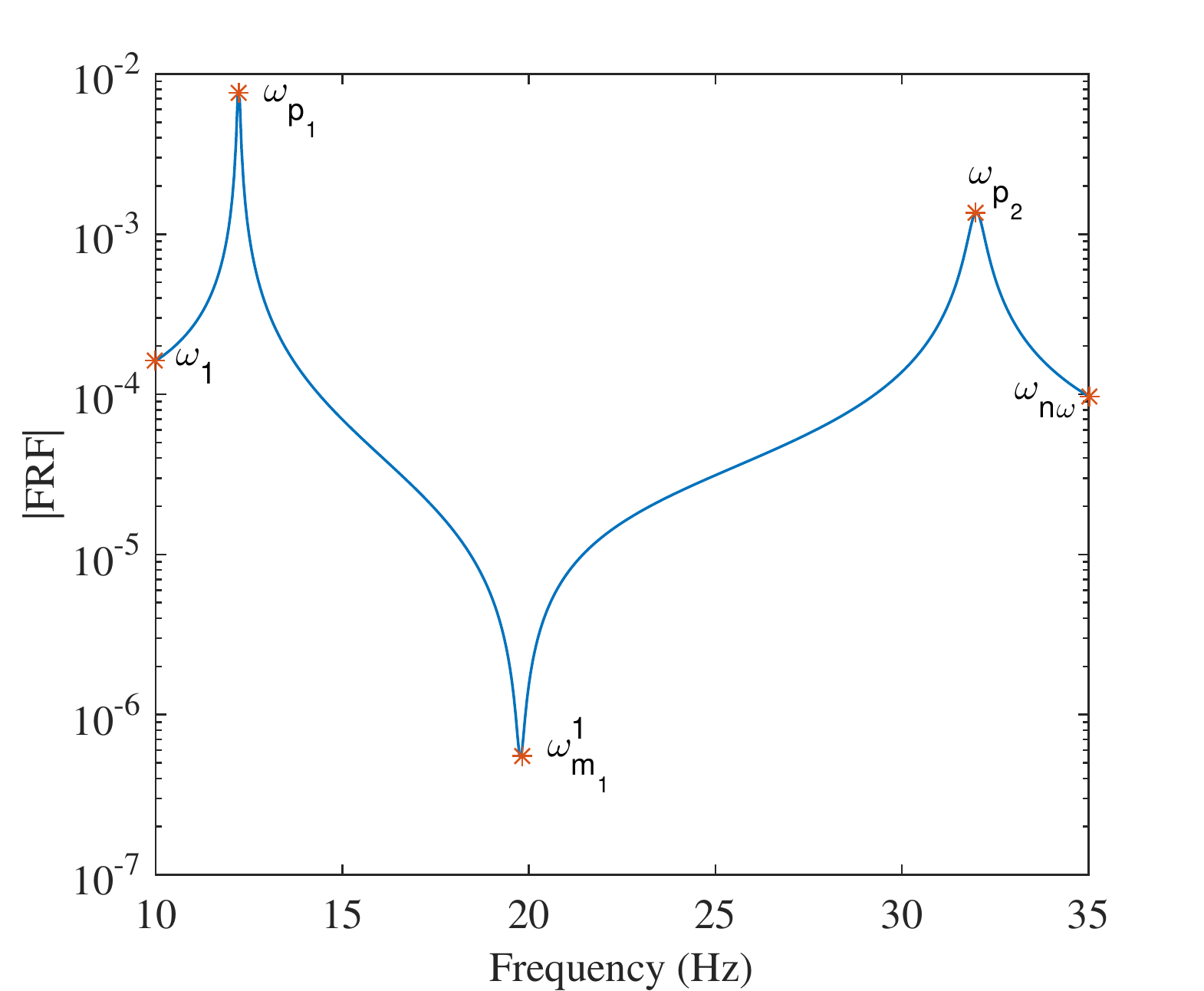}
		\caption{FRFs calculated at first system output}
	\end{subfigure}
	\\
	\begin{subfigure}[b]{1\columnwidth}
		\centering
		\label{fig:FRF:2DOF:direct}
		\includegraphics[width=.75\linewidth]{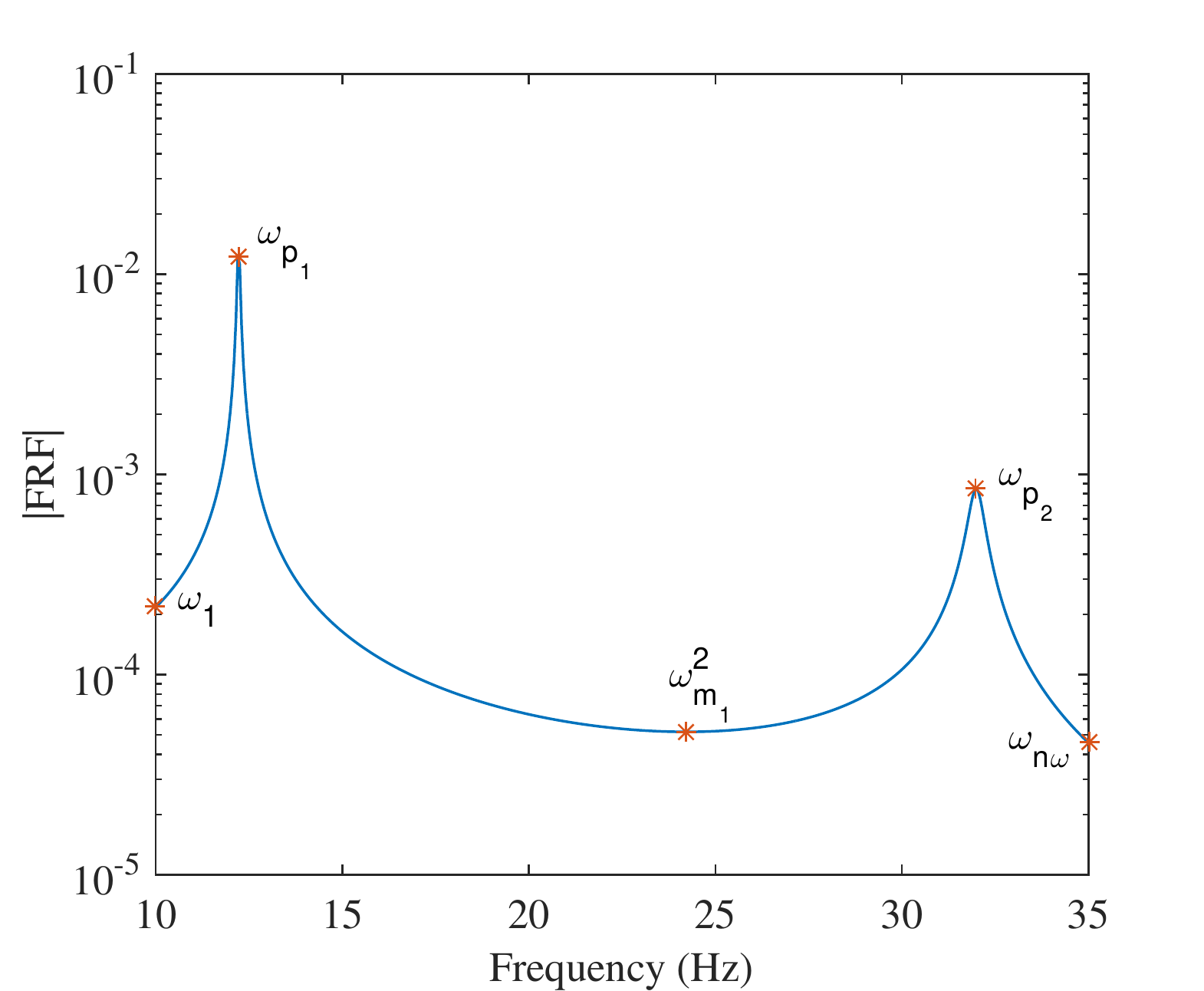}
		\caption{FRF calculated at second system output}
	\end{subfigure}
	\caption{FRFs of the 2-DOF system presented in Figure \ref{fig:2DOF:simple}. The \textit{selected frequencies} $\ve{\mathcal{F}}$ and the associated notations are illustrated with asterisks (\textasteriskcentered).}
	\label{fig:FRF:2DOF}
\end{figure}

To this end, the following algorithm is proposed. First, an experimental design $\ve{\mathcal{X}}$ and the corresponding model responses $\ve{\cy}$ are evaluated. Each system response will be called a trajectory in the remainder of this paper. Let the frequency range of interest be discretized to $n_{\omega}$ equidistant frequencies $\Omega_d=\left[\omega_1, \omega_2,..., \omega_{n_{\omega}}\right]$. Then, the required system responses are matrices $\ve{\mathcal{H}}(\Omega_d) \in \mathbb{C}^{n_u n_y \times n_\omega} $  and $\ve{\ve{\mathcal{F}}}\in \mathbb{R}^{n_u n_y \times n_{sf}}$. The matrix $\ve{\mathcal{H}}
(\Omega_d)$ is obtained by evaluating Eq. (\ref{eq4}) at frequencies $\Omega_d$. The matrix $\ve{\ve{\mathcal{F}}}$ consists of all the resonance and antiresonance frequencies of the system's input-output relations for one system realization, as follows:

\begin{equation}
\label{eq16}
\ve{\mathcal{F}}=
\left[\begin{array}{ccccccccc}
\omega_1 & \omega_{p_1} & \omega^1_{m_1} & \omega_{p_2} & \ldots & \omega_{p_{n_p-1}} & \omega^1_{m_{n_p-1}} & \omega_{p_{n_p}} & \omega_{n_\omega} \\ 
\omega_1 & \omega_{p_1} & \omega^2_{m_1} & \omega_{p_2} & \ldots & \omega_{p_{n_p-1}} & \omega^2_{m_{n_p-1}} & \omega_{p_{n_p}} & \omega_{n_\omega} \\
\vdots & \vdots & \vdots & \vdots & \ddots & \vdots & \vdots & \vdots & \vdots \\
\omega_1 & \omega_{p_1} & \omega^{n_u \times n_y}_{m_1} & \omega_{p_2} & \ldots & \omega_{p_{n_p-1}} & \omega^{n_u \times n_y}_{m_{n_p-1}} & \omega_{p_{n_p}} & \omega_{n_\omega}
\end{array}\right]=
\left[\begin{array}{c}
{\mathcal{F}}_1\\
{\mathcal{F}}_2\\
\vdots\\
{\mathcal{F}}_{n_u \times n_y}\\
\end{array}\right]
\end{equation}
in which $n_p$ is the number of eigenvalues of the system. Furthermore, $\{\omega_{p_i}    i=1,2,...,n_p\}$ are the resonant frequencies and $\{\omega^l_{m_i}  i=1,2,...,n_p-1,l=1,2,...,n_u \times n_y\}$ are frequencies between each two consecutive resonant frequencies at which the minimum amplitude occurs. Throughout the paper, these important frequencies, shown by red asterisks in Figure \ref{fig:FRF:2DOF} for a typical frequency response, will be referred to as \emph{selected frequencies}. Their number $n_{sf}$ is assumed to be constant across different realizations of the system inputs.
$\{\mathcal{F}_i, i=1,2,\cdots,n_u\times n_y\}$ includes all the \textit{selected frequencies} for the $i^{th}$ input-output relation.

For the next step of the algorithm, let $ \ve{x}^{(ref)} $ be selected randomly among the sample points in the ED to have its associated trajectory as the reference, \ie: $$\ve{\mathcal{H}}^{ref}=\ve{\mathcal{H}}(\ve{x}^{(ref)},\Omega_d), \quad  \ve{\mathcal{F}}^{ref}=\ve{\mathcal{F}}(\ve{x}^{(ref)};\omega).$$ Then, the other trajectories are transformed in the frequency axis so as to have the peaks and valleys as close to the corresponding locations in the reference trajectory as possible \ie: 

\begin{equation}
\label{eq:sTrans:PLT}
\mathcal{T}^k_i=\mathcal{T}^{(k)}_i(\omega, \nu^{(k)}_i)=\{\nu^{(k)}_i = f(\omega)|\mathcal{F}_i(\ve{x}^{(k)};\nu^{(k)})=\mathcal{F}^{ref}_i\}
\end{equation}
where $i=1,2,\cdots,n_u\times n_y$, $k=1,2,\cdots,N_{ED}$ and $\nu$ is the transformed frequency axis called \textit{scaled frequency}.
The transform $\mathcal{T}^k_i$ consists of a continuous piecewise linear transform of the intervals between the identified \textit{selected frequencies} that align them to the corresponding ones of the reference trajectory as follows,

\begin{equation}
\label{eq:trans:scaling}
\mathcal{T}^k_i : \nu^{(k)}_{i,l}=a^{(k)}\omega_l+b^{(k)} \quad  \mathcal{F}^{(k)}_i(j)\leq \omega_l \leq \mathcal{F}^{(k)}_i(j+1)
\end{equation}
where 
$$ a^{(k)}=\frac{\mathcal{F}^{ref}_i(j)-\mathcal{F}^{ref}_i(j+1)}{\mathcal{F}^{(k)}_i(j)-\mathcal{F}^{(k)}_i(j+1)}, $$
$$ b^{(k)}=\frac{\mathcal{F}^{ref}_i(j)\mathcal{F}^{(k)}_i(j+1)-\mathcal{F}^{ref}_i(j+1)\mathcal{F}^{(k)}_i(j+1)}{\mathcal{F}^{(k)}_i(j)-\mathcal{F}^{(k)}_i(j+1)},$$
$j=1,2,\cdots,n_{sf}-1$ and $l=1,2,\cdots,n_\omega$. 
This transformation results in the FRFs which are similar to the reference one in the \textit{scaled frequency} domain:

\begin{equation}
\label{eq:sTrans:sFRF}
\tilde{\mathcal{H}}_i(\ve{x}^{(k)},\cn^{(k)}_i)=\mathcal{H}_i(\ve{x}^{(k)},\Omega_d) \circ \mathcal{T}^k_i
\end{equation}
where the set $\cn^{(k)}_i=\{\nu_{i,1}^{(k)}, \nu_{i,2}^{(k)}, \cdots, \nu_{i,n_\omega}^{(k)}\}$ consists of the discretized \textit{scaled frequencies} which are non-equidistantly spread over the frequency range of interest.

\begin{figure}[H]
	\centering
	\begin{subfigure}[b]{1\columnwidth}
		\centering
		\includegraphics[width=.85\linewidth]{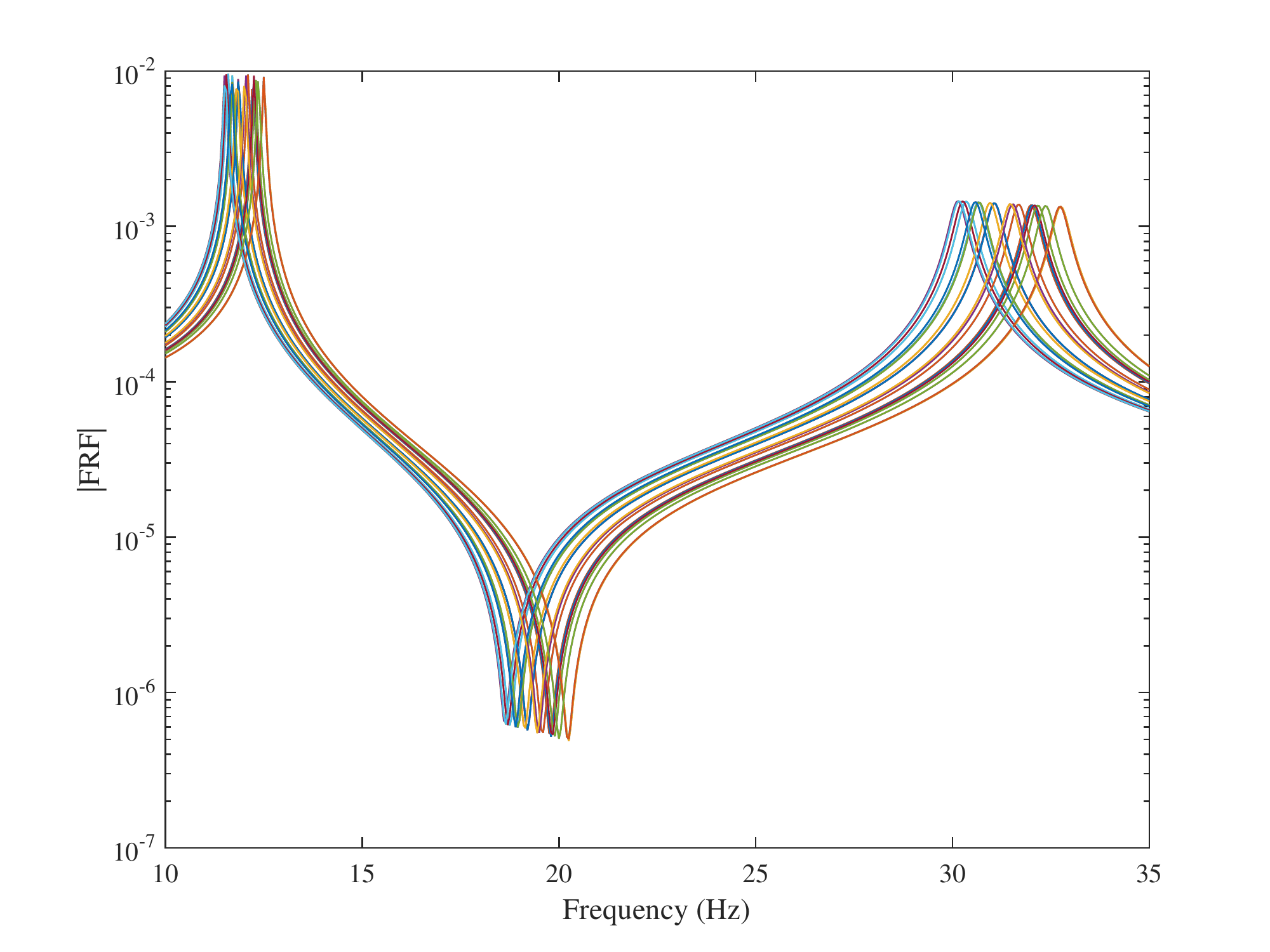}
		\caption{FRFs before frequency transformation.}
		\label{fig:FRF:2DOF:freq}
	\end{subfigure}
	\\
	\begin{subfigure}[b]{1\columnwidth}
		\centering
		\includegraphics[width=.85\linewidth]{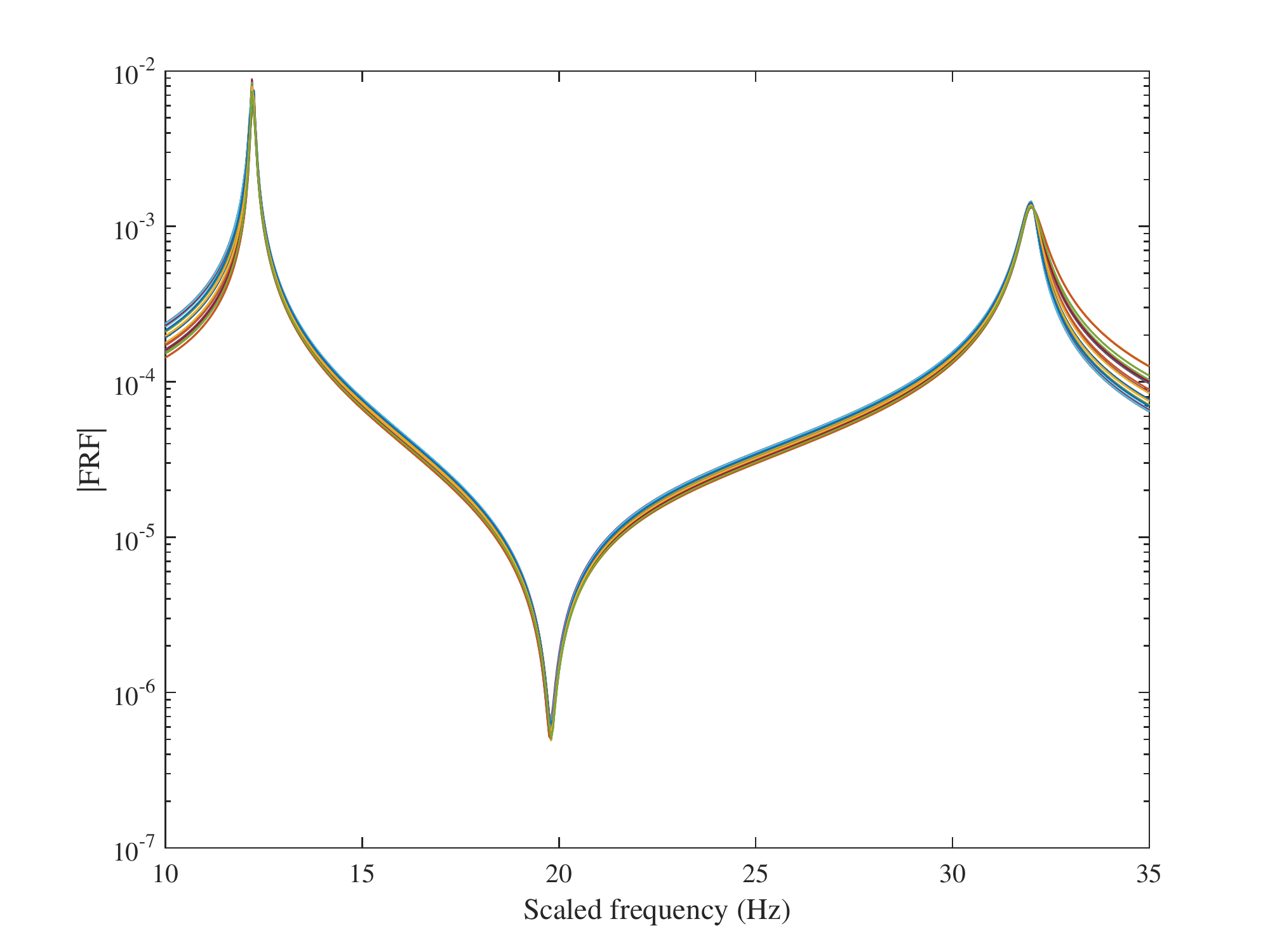}
		\caption{FRFs after frequency transformation.}
		\label{fig:FRF:2DOF:Gfreq}
	\end{subfigure}
	\caption{Several realizations of the FRFs of the 2-DOF system at first system output.}
\end{figure}

\begin{figure}[H]
	\centering
	\includegraphics[width=0.8\columnwidth]{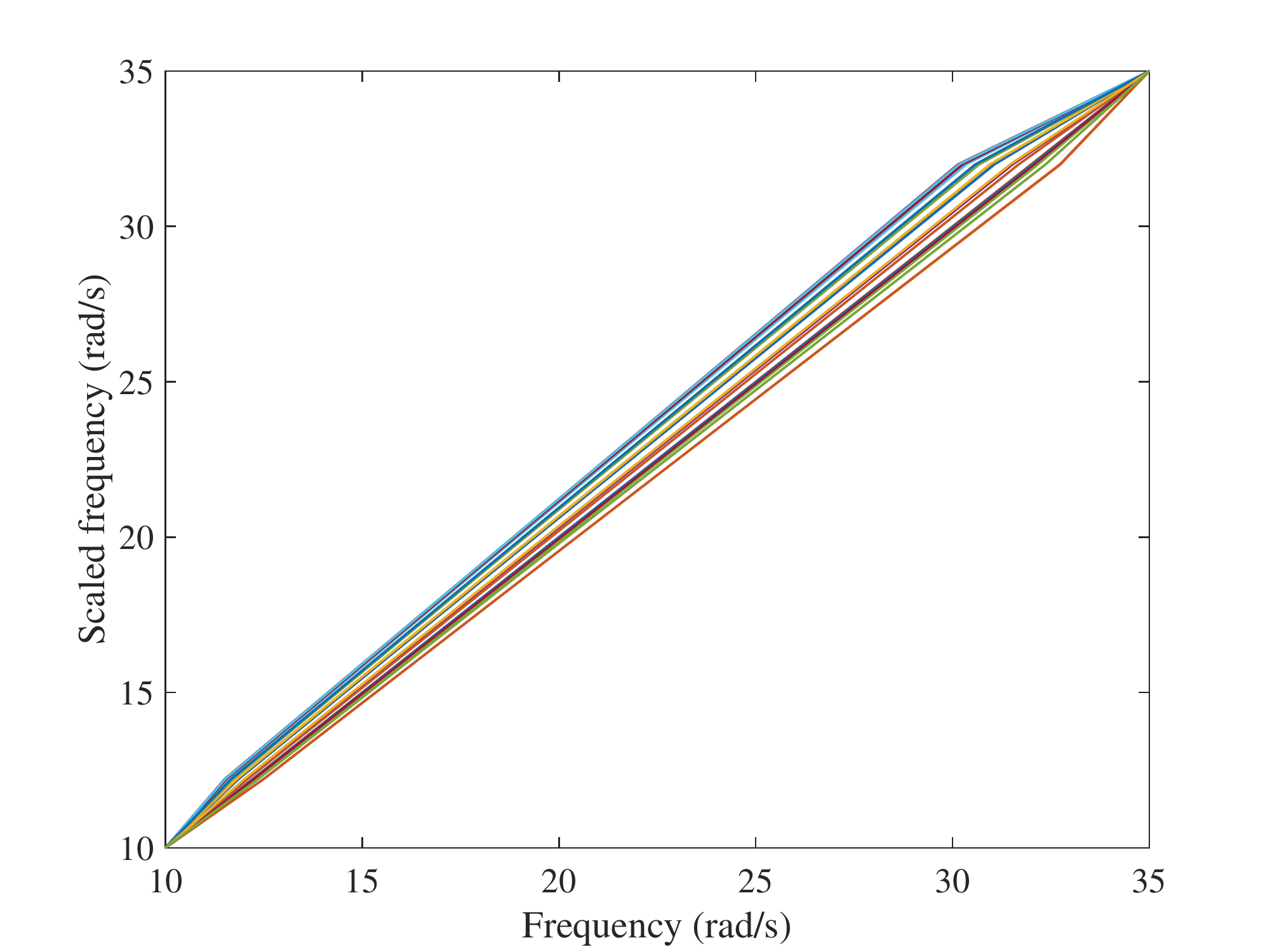}
	\caption{Continuous piecewise-linear function used to transform the FRFs in Figure \ref{fig:FRF:2DOF:freq} into Figure \ref{fig:FRF:2DOF:Gfreq}}
	\label{fig:FRF:2DOF:CPL}
\end{figure}

Figures \ref{fig:FRF:2DOF:freq} and \ref{fig:FRF:2DOF:Gfreq} illustrate the FRFs of a 2-DOF system versus frequency and \textit{scaled frequency}, respectively. 
An example of such a transform used for transforming the FRFs of a 2-DOF is presented in Figure \ref{fig:FRF:2DOF:CPL}. 

One should notice that since $\cn^{(k)}_i$ contains the non-equidistant \textit{scaled frequencies}, a final interpolation is required to obtain a common discretized \textit{scaled frequency} $\cn^{ref}=\{\nu_1^{ref}, \nu_2^{ref}, \cdots, \nu_{n_\omega}^{ref}\}$ between the reference and all other trajectories.
To reduce interpolation error in the system response, small frequency steps should be selected.
The proposed approach for preprocessing the FRFs is summarized in Algorithm \ref{alg:method:scale}.
\clearpage
\begin{algorithm}
	\begin{algorithmic}[1]
		\STATE {\bf Input}: ${\ve{\cx}}=\{{\ve{x}^{(1)}, \ve{x}^{(2)},...,\ve{x}^{(N_{ED})}}$\}
		\STATE $\ve{\mathcal{H}}^{ref}$=$\ve{\mathcal{H}}(\ve{x}^{(r)},\Omega_d)$,
		$\ve{\mathcal{F}}^{ref}=\ve{\mathcal{F}}(\ve{x}^{(r)};\omega) $,   for a random $r \in [1,...,N_{ED}]$   
		\FOR{$k=1$ \TO $N_{ED}$}
		\STATE $\ve{\mathcal{F}}(\ve{x}^{(k)};\omega)$=$[\mathcal{F}_1(\ve{x}^{(k)};\omega) \mathcal{F}_2(\ve{x}^{(k)};\omega), \cdots, \mathcal{F}_{n_u\times n_y}(\ve{x}^{(k)};\omega)]^\text{T}$ using Eq. (\ref{eq16})
		\STATE $\ve{\mathcal{H}}(\ve{x}^{(k)};\Omega_d)$=$[\mathcal{H}_1(\ve{x}^{(k)};\Omega_d), \mathcal{H}_2(\ve{x}^{(k)};\Omega_d), \cdots, \mathcal{H}_{n_u\times n_y}(\ve{x}^{(k)};\Omega_d) ]^\text{T}$ using Eq. (\ref{eq4})
		\FOR{$i=1$ \TO $n_u \times n_y$}
		\STATE Evaluate $\mathcal{T}^k_i$ using Eq. (\ref{eq:trans:scaling})
		\STATE $\tilde{\mathcal{H}}_i(\ve{x}^{(k)},\cn^{(k)}_i)$=$\mathcal{H}_i(\ve{x}^{(k)},\Omega_d) \circ \mathcal{T}^k_i$
		\STATE $\tilde{\mathcal{H}}_i(\ve{x}^{(k)},\cn^{ref})$=interpolate$(\tilde{\mathcal{H}}_i(\ve{x}^{(k)},\cn^{(k)}_i), \cn^{(k)}_i, \cn^{ref})$
		\ENDFOR
		\STATE $\tilde{\ve{\mathcal{H}}}(\ve{x}^{(k)};\cn^{ref})$=$[\tilde{\mathcal{H}}_1(\ve{x}^{(k)};\cn^{ref}), \tilde{\mathcal{H}}_2(\ve{x}^{(k)};\cn^{ref}), \cdots, \tilde{\mathcal{H}}_{n_u\times n_y}(\ve{x}^{(k)};\cn^{ref}) ]^\text{T}$
		\ENDFOR
		\STATE {\bf F}=\{vect$(\ve{\mathcal{F}}(\ve{x}^{(1)};\omega))$, vect$(\ve{\mathcal{F}}(\ve{x}^{(2)};\omega)),...,\text{vect}(\ve{\mathcal{F}}(\ve{x}^{(N_{ED})};\omega))\}^\text{T}$
		\STATE $\tilde{\bf H}$($\cn^{ref}$)= \{vec$(\tilde{\ve{\mathcal{H}}}(\ve{x}^{(1)},\cn^{ref})), \text{vec}( \tilde{\ve{\mathcal{H}}}(\ve{x}^{(2)},\cn^{ref})),...,\text{vec}(\tilde{\ve{\mathcal{H}}}(\ve{x}^{(N_{ED})},\cn^{ref}))\}^\text{T} $,  	
		\STATE {\bf Output}: {\bf F}, {$\ve{\mathcal{G}}^\mathfrak{R}$}=real($\tilde{\bf H}(\cn^{ref})$), {$\ve{\mathcal{G}}^\mathfrak{I}$}=imag($\tilde{\bf H}(\cn^{ref})$)
	\end{algorithmic}	
	\caption{Data preprocessing: continuous piecewise-linear transformation}
	\label{alg:method:scale}
\end{algorithm}


\subsection{Polynomial chaos representation}
\label{PCE:transformation}

The non-smooth behavior of the FRFs makes their surrogation by polynomials a problematic task. To solve this issue, one PCE could be calculated for each \textit{scaled frequency}. This means that to compute PCEs for the FRFs, two sets of PCEs are required.
The first set is to predict the \textit{selected frequencies}, collected in the matrix $\ve{\mathcal{F}}$ (\ref{eq16}), which are required for performing the stochastic transformation as explained in Section \ref{transformation}. This matrix includes eigenfrequencies of the system, therefore by obtaining this set of PCEs, the problem of random eigenvalue calculation is solved by the use of PCE as a byproduct. This problem has been addressed in some recent works, \eg \citet{PichleraMetamodelNaturalfrequency2012}. 
Since the number of random outputs for this set is not very large, PCE can be applied to each of the \textit{selected frequencies} separately, \ie for $i=1,2,\cdots,n_u\times n_y$ and $j=1,2,\cdots \times n_{sf} $
\begin{equation}
\label{eqn:PCE:freq}
\hat{{\mathcal{F}}}_i(j)=\sum_{\ua \in \ca^{M,p}} f_{\ua}^i(j){{\psi}_{\ve{\alpha}}({\ve{x}})}.
\end{equation}

The second set of PCE is for the system response at each individual \textit{scaled frequency}. To this end, let $\tilde{\bf H}(\cn^{ref}) \in \mathbb{C}^{N_{ED} \times (n_\omega \times n_u \times n_y)}$, defined in Algorithm \ref{alg:method:scale}, be a matrix of trajectories at \textit{scaled frequencies}. 
Since the FRFs have complex-valued responses, whereas the PCEs are defined for real-valued functions only\footnote{Limited literature is available on the use of PCE for complex-valued functions, see \eg \citet{soize2004physical}.}, separate PCEs need to be performed for real and imaginary parts of the FRFs. Therefore, the matrix $\ve{\mathcal{G}} = \{\ve{\mathcal{G}}^\mathfrak{R}, \ve{\mathcal{G}}^\mathfrak{I}\} = \{real(\tilde{\bf H}(\cn^{ref})), imag(\tilde{\bf H}(\cn^{ref}))\} \in \mathbb{R}^{N_{ED} \times (2 \times n_\omega \times n_u \times n_y)}$ is the response matrix for which the PCE should be built. 

The number of random outputs for this set, $N = 2 \times n_\omega \times n_u \times n_y$, can be extremely large. As discussed in Section \ref{PCE:PCA}, the PCEs are therefore applied directly to the principal components of $\mathcal{G}$, yielding:

\begin{equation}
\label{eqn:PCE:rFRF}
\hat{\ve{\mathcal{G}}}^\mathfrak{R}= \bar{\ve{\mathcal{G}}}^\mathfrak{R}+ \sum_{j=1}^{N'} \sum_{\ua \in \ca^{M,p}} (u_{\ua}^\mathfrak{R}{{\psi}_{\ve{\alpha}}({\ve{x}})})_j \ve{v}^{\mathfrak{R}^\text{T}}_j,
\end{equation}  

\begin{equation}
\label{eqn:PCE:iFRF}
\hat{\ve{\mathcal{G}}}^\mathfrak{I}= \bar{\ve{\mathcal{G}}}^\mathfrak{I}+ \sum_{j=1}^{N'} \sum_{\ua \in \ca^{M,p}} (u_{\ua}^\mathfrak{I}{{\psi}_{\ve{\alpha}}({\ve{x}})})_j \ve{v}^{\mathfrak{I}^\text{T}}_j,
\end{equation}  

where $u_{\ua}^\mathfrak{R}$ and $u_{\ua}^\mathfrak{I}$ are respectively the vectors of coefficients of the PCEs made for the real and imaginary parts of the FRFs.

\subsection{Surrogate response prediction}
\label{Method:predict}

To predict the surrogate model response at a new sample point $\ve{x}^{(0)}$, several steps need to be taken to transform the PCE predictors in Eqs. (\ref{eqn:PCE:rFRF}) and (\ref{eqn:PCE:iFRF}) from the \textit{scaled frequency} axis $\nu$ to the original frequency axis $\omega$.  
The matrices $\hat{\ve{\mathcal{G}}}^\mathfrak{R}$ and $\hat{\ve{\mathcal{G}}}^\mathfrak{I}$ are obtained by evaluating the second set of PCEs in Eqs. (\ref{eqn:PCE:rFRF}) and (\ref{eqn:PCE:iFRF}), respectively. Then, the FRFs at the \textit{scaled frequencies} can be obtained at the new sample point by the inverse vectorization of ${\hat{\tilde{\bf{H}}}}(\cn^{ref})=\hat{\ve{\mathcal{G}}}^\mathfrak{R}+j\hat{\ve{\mathcal{G}}}^\mathfrak{I}$ where $j=\sqrt{-1}$. 
To obtain the FRF at the original frequency $\omega$ the following transformation is used,

\begin{equation}
\label{eq:siTrans:sFRF}
\hat{\tilde{\mathcal{H}}}_i(\ve{x}^{(0)},\Omega^{(0)}_i)=\hat{\tilde{\mathcal{H}}}_i(\ve{x}^{(0)},\cn^{ref}) \circ (\mathcal{T}^0_i)^{-1}, \quad i=1,2,\cdots,n_u\times n_y
\end{equation}
where $\mathcal{T}^0_i$ is obtained by evaluating Eq. (\ref{eq:trans:scaling}) at $\hat{\ve{\mathcal{F}}}^{(0)}=\hat{\ve{\mathcal{F}}}(\ve{x}^{(0)}; \omega)$ which is the matrix of \textit{selected frequencies} at the new sample point $\ve{x}^{(0)}$ evaluated by Eq. (\ref{eqn:PCE:freq}). Besides, $\Omega_i^{(0)}$ is a set of discretized frequencies which are non-equidistantly spread over the frequency rage of interest. In order to provide the frequency response at the desired frequencies $\Omega_d$, interpolation is inevitable.
The algorithm for predicting the system response at a new sample point is briefly presented in Algorithm \ref{alg:method:predict}.  

\begin{algorithm}
	\begin{algorithmic}[1]
		\STATE {\bf Input}: $\ve{x}^{(0)} \neq \ve{x}^{(l)}, \quad l=1,2,...,N_{ED}$ and $\mathcal{H}^{ref}, \ve{\mathcal{F}}^{ref}, \bar{\mathcal{G}}$ and $\cn^{ref}$
		\STATE $\hat{\ve{\mathcal{G}}}^\mathfrak{R}$= $\hat{\ve{\mathcal{G}}}^\mathfrak{R}(\ve{x}^{(0)}, \cn^{ref})$ using Eq. (\ref{eqn:PCE:rFRF}).
		\STATE $\hat{\ve{\mathcal{G}}}^\mathfrak{I}$= $\hat{\ve{\mathcal{G}}}^\mathfrak{I}(\ve{x}^{(0)}, \cn^{ref})$ using Eq. (\ref{eqn:PCE:iFRF}).
		\STATE $\hat{\tilde{\bf{H}}}$($\ve{x}^{(0)}, \cn^{ref}$) = $\hat{\ve{\mathcal{G}}}^\mathfrak{R}$+$j \hat{\ve{\mathcal{G}}}^\mathfrak{I}$
		\STATE Construct $\hat{\tilde{\ve{\mathcal{H}}}}$($\ve{x}^{(0)}, \cn^{ref}$) from $\hat{\tilde{\bf{H}}}$($\ve{x}^{(0)}, \cn^{ref}$) by inverse vectorization operation
		\FOR{$i=1$ \TO $n_u \times n_y$}
		\STATE Evaluate $\hat{\mathcal{F}}_i^{0}$=$\hat{\mathcal{F}}_i(\ve{x}^{(0)}; \omega)$ using Eq. (\ref{eqn:PCE:freq})
		\STATE Evaluate $\mathcal{T}^0_i$ using Eq. (\ref{eq:trans:scaling})
		\STATE $\hat{\mathcal{H}}_i(\ve{x}^{(0)},\Omega_i^{(0)})$=$\hat{\tilde{\mathcal{H}}}_i(\ve{x}^{(0)},\cn^{ref}) \circ (\mathcal{T}^0_i)^{-1}$
		\STATE $\hat{\mathcal{H}}_i(\ve{x}^{(0)},\Omega_d)$=interpolate$(\hat{\mathcal{H}}_i(\ve{x}^{(0)},\Omega_i^{(0)}), \Omega_i^{(0)},\Omega_d)$
		\ENDFOR	
		\STATE {\bf Output}: $\hat{\ve{\mathcal{H}}}(\Omega_d)$=$\{\hat{\mathcal{H}}_1(\ve{x}^{(0)},\Omega_d), \hat{\mathcal{H}}_2(\ve{x}^{(0)},\Omega_d), \cdots,\hat{\mathcal{H}}_{n_u\times n_y}(\ve{x}^{(0)},\Omega_d)\}$
	\end{algorithmic}	
	\caption{Predicting system responses}
	\label{alg:method:predict}
\end{algorithm}

\section{Examples}
\label{example}
\subsection{Introduction}
\label{exam:intro}
In this section, the proposed method will be applied to two case studies. The first one is a simple 2-DOF system to illustrate how the method works. The second one is a 6-DOF system with a relatively large (16-dimensional), parameter space. For the sake of readability, only results for one output (the $1^{st}$ output for the 2-DOF and the $6^{th}$ output for the 6-DOF system) are shown for each case while the results for the other outputs are reported for completeness in the Appendices.

To assess the accuracy of the surrogate models quantitatively, the following measure based on the root mean square (rms) error of the vectors is defined.

\begin{equation}
\label{eq:Err:rms}
Error(\bullet) = \frac{\text{rms}((\bullet)^{ex}-(\bullet)^{approx}))}{\text{rms}((\bullet)^{ex})}\times 100,
\end{equation}
in which $(\bullet)$ is the vector of interest. $(\cdot)^{ex}$ and $(\cdot)^{approx}$ represent results obtained by running the true and surrogate models, respectively.  
This error aims at measuring the relative difference between vectorial data, such as one FRF or the mean and standard deviation of several FRFs. 
For the mean and standard deviation of the data, the reference results are obtained by evaluating the true model at 10,000 Monte-Carlo samples and the approximations are calculated by the PCE surrogate at the same 10,000 points. 

\subsection{Simple 2-DOF system \citep{jacquelinpolynomial2015}} 
\begin{figure} [H]
	\centering
	\begin{adjustbox}{max width=0.5\columnwidth}
		\begin{tikzpicture}
		
		\tikzstyle{spring}=[thick,decorate,decoration={zigzag,pre length=0.3cm,post
			length=0.3cm,segment length=6}]
		
		\tikzstyle{damper}=[thick,decoration={markings,  
			mark connection node=dmp,
			mark=at position 0.5 with 
			{
				\node (dmp) [thick,inner sep=0pt,transform shape,rotate=-90,minimum
				width=10pt,minimum height=3pt,draw=none] {};
				\draw [thick] ($(dmp.north east)+(2pt,0)$) -- (dmp.south east) -- (dmp.south
				west) -- ($(dmp.north west)+(2pt,0)$);
				\draw [thick] ($(dmp.north)+(0,-4pt)$) -- ($(dmp.north)+(0,4pt)$);
			}
		}, decorate]
		
		\tikzstyle{ground}=[fill,pattern=north east lines,draw=none,minimum
		width=0.75cm,minimum height=0.3cm]
		
		\node[draw,outer sep=0pt,thick, fill=white!60!yellow] (M1) [minimum width=1.5cm, minimum height=1.5cm] {$m$};
		\node[draw,outer sep=0pt,thick, fill=white!60!yellow] (M2) at (3,0) [minimum width=1.5cm, minimum height=1.5cm] {$m$};
		
		\node (ground1) [ground,anchor=north,yshift=-0.25cm,minimum width=4.8cm] at (M1.south) {};
		\draw (ground1.north east) -- (ground1.north west);
		\draw [thick, fill={cyan}] (M1.south west) ++ (0.2cm,-0.125cm) circle (0.125cm)  (M1.south east) ++ (-0.2cm,-0.125cm) circle (0.125cm);
		
		\node (ground2) [ground,anchor=north,yshift=-0.25cm,minimum width=3cm] at (M2.south) {};
		\draw (ground2.north east) -- (ground2.north west);
		\draw [thick, fill={cyan}] (M2.south west) ++ (0.2cm,-0.125cm) circle (0.125cm)  (M2.south east) ++ (-0.2cm,-0.125cm) circle (0.125cm);
		
		\node (wall1) [ground, rotate=-90, minimum width=2cm,yshift=-2.5cm] {};
		\draw (wall1.north east) -- (wall1.north west);
		
		\draw [spring] ($(wall1.90) + (0,0.5)$) -- ($(M1.west) + (0,0.5)$)
		node [midway,above] {$k$};
		\draw [damper] ($(wall1.90) - (0,0.5)$) -- ($(M1.west) - (0,0.5)$)
		node [midway,above=3] {$c$};
		
		\draw[spring] ($(M1.east) + (0,0.5)$) -- ($(M2.west) + (0,0.5)$) 
		node [midway,above] {$k$};
		\draw[damper] ($(M1.east) - (0,0.5)$) -- ($(M2.west) - (0,0.5)$)
		node [midway,above=3] {$c$};

		\draw[thick, dashed] ($(M1.north west)$) -- ($(M1.north west) + (0,1.1)$);
		\draw[thick, dashed] ($(M2.north west)$) -- ($(M2.north west) + (0,0.6)$);
		\draw[ultra thick, -latex] ($(M2.north west) + (0,0.5)$) -- 
		($(M2.north west) + (1,0.5)$)
		node [midway, below] {$q_2$};
		\draw[ultra thick, -latex] ($(M1.north west) + (0,0.5)$) -- 
		($(M1.north west) + (1,0.5)$)
		node [midway, below] {$q_1$};
		\draw[ultra thick, -latex] ($(M1.north west) + (0,1)$) -- 
		($(M1.north west) + (1.5,1)$)
		node [midway, below] {$f$};
		\end{tikzpicture}
	\end{adjustbox}
	\caption{2-DOF system}
	\label{fig:2DOF:simple}
\end{figure}
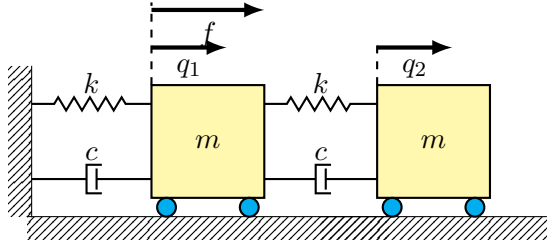

As the first example, the simple 2-DOF system shown in Figure \ref{fig:2DOF:simple} is selected to highlight the steps of the proposed method. In this system, stiffness is assumed to be uncertain 
\begin{equation}
\label{eqn:2DOF:stiffness}
k=\bar{k}(1+\delta_k \xi)
\end{equation}
where $\xi$ is a standard normal random variable. Other properties of the system are listed in Table \ref{tab:2DOF}. The system has one input force $f$ at mass 1, two physical outputs $q_1$ and $q_2$ and thus, two FRFs. The FRFs of the system are obtained in the range of 10 to 35 Hz with a frequency step of 0.01 Hz, as shown in Figure \ref{fig:FRF:2DOF}. The \textit{selected frequencies} are also shown in the figure with red asterisks.

40 points are sampled in the parameter space using Latin Hypercube Sampling (LHS) to form an experimental design (ED) ${\ve{\cx}}$ and the model is evaluated at these points to find the system responses of interest, namely the FRFs, $\ve{\mathcal{H}}$, and the \textit{selected frequencies} $\ve{\mathcal{F}}$.

\begin{table} [H]
	\centering
	\caption{2-DOF system's charactristic}
	\begin{tabular}{l|cccc}
		\hline
		Characteristics & m (kg) & $\bar{k}$ (Nm$^{-1}$) & c (Nm$^{-1}$s$^{-1})$ & $\delta_k$ \\
		\hline
		Value &   1 &  15000 &   1 &  5\% \\
		\hline
	\end{tabular}
	\label{tab:2DOF}
\end{table}

Figure \ref{fig:FRF:2DOF:freq} shows the FRFs of the system evaluated at ${\ve{\cx}}$. To find the transformed FRFs, \ie FRFs at \textit{scaled frequencies} $\nu$, one trajectory was selected randomly as the reference and the others were scaled such that their peaks and valleys were at the same \textit{scaled frequencies} as that of the reference trajectory. The transformed FRFs are shown in Figure \ref{fig:FRF:2DOF:Gfreq} and the corresponding continuous piecewise-linear transformations in Figure \ref{fig:FRF:2DOF:CPL}.

The next step is to find a suitable basis and the associated coefficients for the polynomial chaos expansion. In this case, since the random variable is Gaussian, the basis of the polynomial chaos consists of Hermite polynomials. The LARS algorithm \citep{blatman2011adaptive} is employed here to calculated a sparse PCE with adaptive degree.

\begin{figure}[H]
	\centering
	\includegraphics[width=0.75\columnwidth]{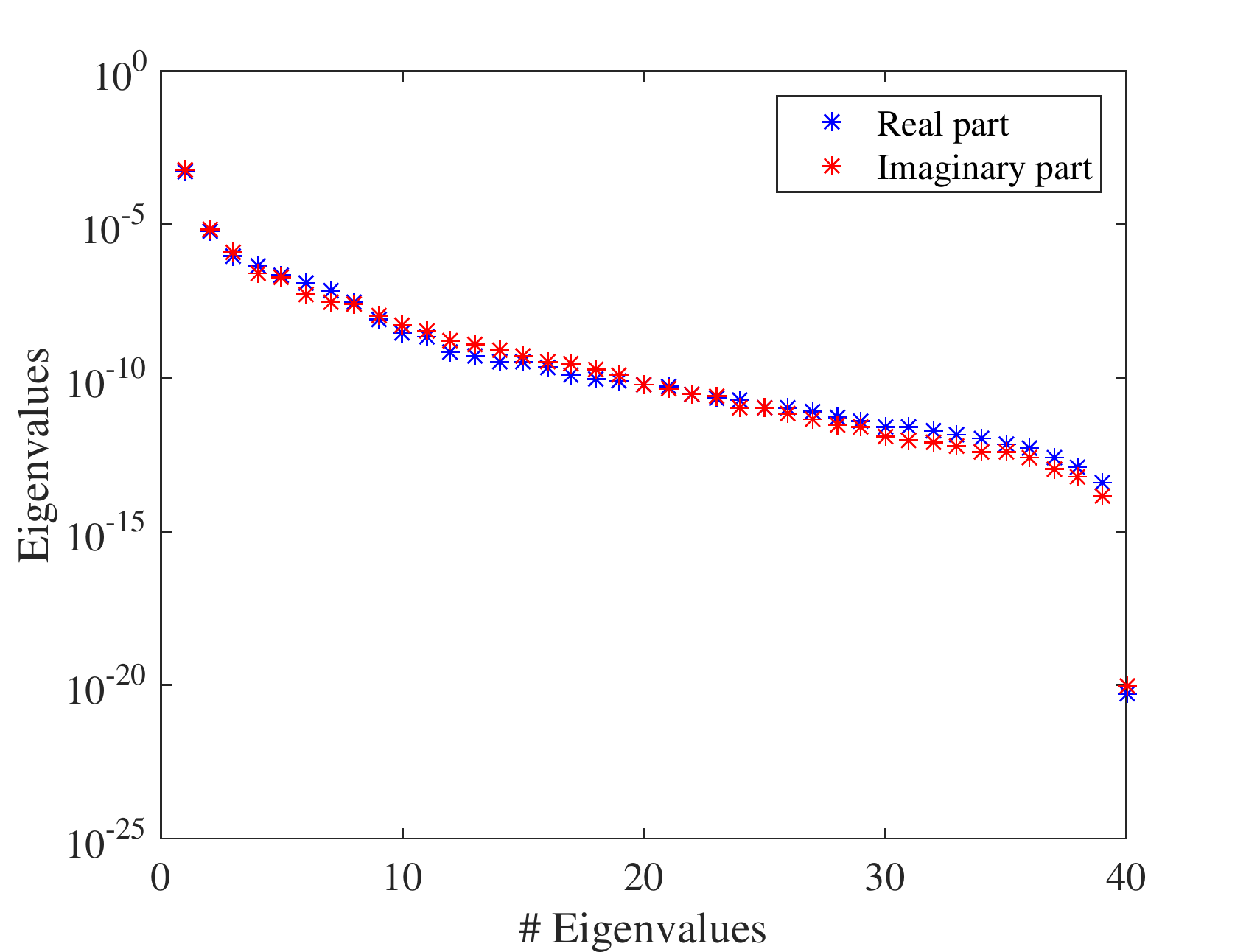}
	\caption{Spectrum of the eigenvalues of the covariance matrix for the 2-DOF system; evaluated for both real part $\ve{\mathcal{G}}^\mathfrak{R}$ and imaginary part $\ve{\mathcal{G}}^\mathfrak{I}$}
	\label{fig:eigval:2DOF}
\end{figure}

\newpage
The first set of expansion consists of 10 PCEs to surrogate the \textit{selected frequencies} $\ve{\mathcal{F}}$, shown in Figure \ref{fig:FRF:2DOF} by red asterisks. As the second set of expansions, PCE is made for the dominant components of $\ve{\mathcal{G}}$ as explained in Section \ref{method:PCA}. To do so, the $\hat{N}$ largest principal components are selected such that the sum of their associated eigenvalues amounts to $99\%$ of the sum of all the eigenvalues, \ie $\sum_{i=1}^{\hat{N}}\lambda_i = 0.99 \sum_{i=1}^{{N}}\lambda_i $ in which $\lambda_i$'s are the eigenvalues of the covariance matrix of either $\ve{\mathcal{G}}^\mathfrak{R}$ or $\ve{\mathcal{G}}^\mathfrak{I}$. By this truncation, the number of random outputs is reduced from 2501 $\times$ 2 $\times$ 2 to 6 components, namely 3 components for the real part and 3 components for the imaginary part. The spectra of the eigenvalues of the covariance matrix of the $\ve{\mathcal{G}}^\mathfrak{R}$ and $\ve{\mathcal{G}}^\mathfrak{I}$ are displayed in Figure \ref{fig:eigval:2DOF}. All the PCEs used to make this surrogate model, including those for the \textit{selected frequencies} and those for the dominant components have orders between 3 and 8.

\begin{figure}[H]
	\centering
	\begin{subfigure}[b]{1\columnwidth}
		\centering
		\includegraphics[width=.7\columnwidth]{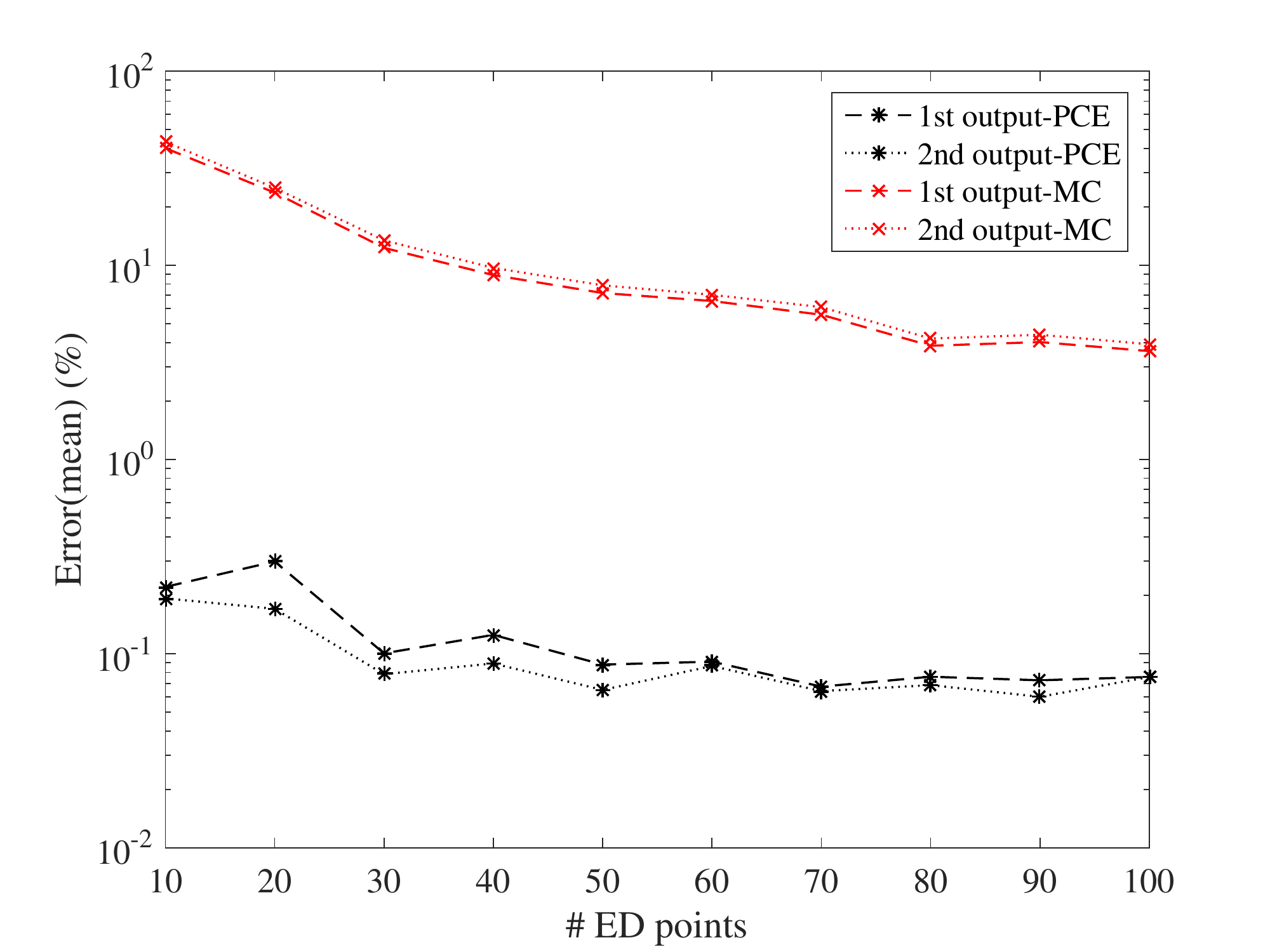}
		\caption{Convergence plot of the mean value of the FRFs}
		\label{fig:2DOF:PCE:convmean}
	\end{subfigure}
	\\
	\begin{subfigure}[b]{1\columnwidth}
		\centering
		\includegraphics[width=.7\columnwidth]{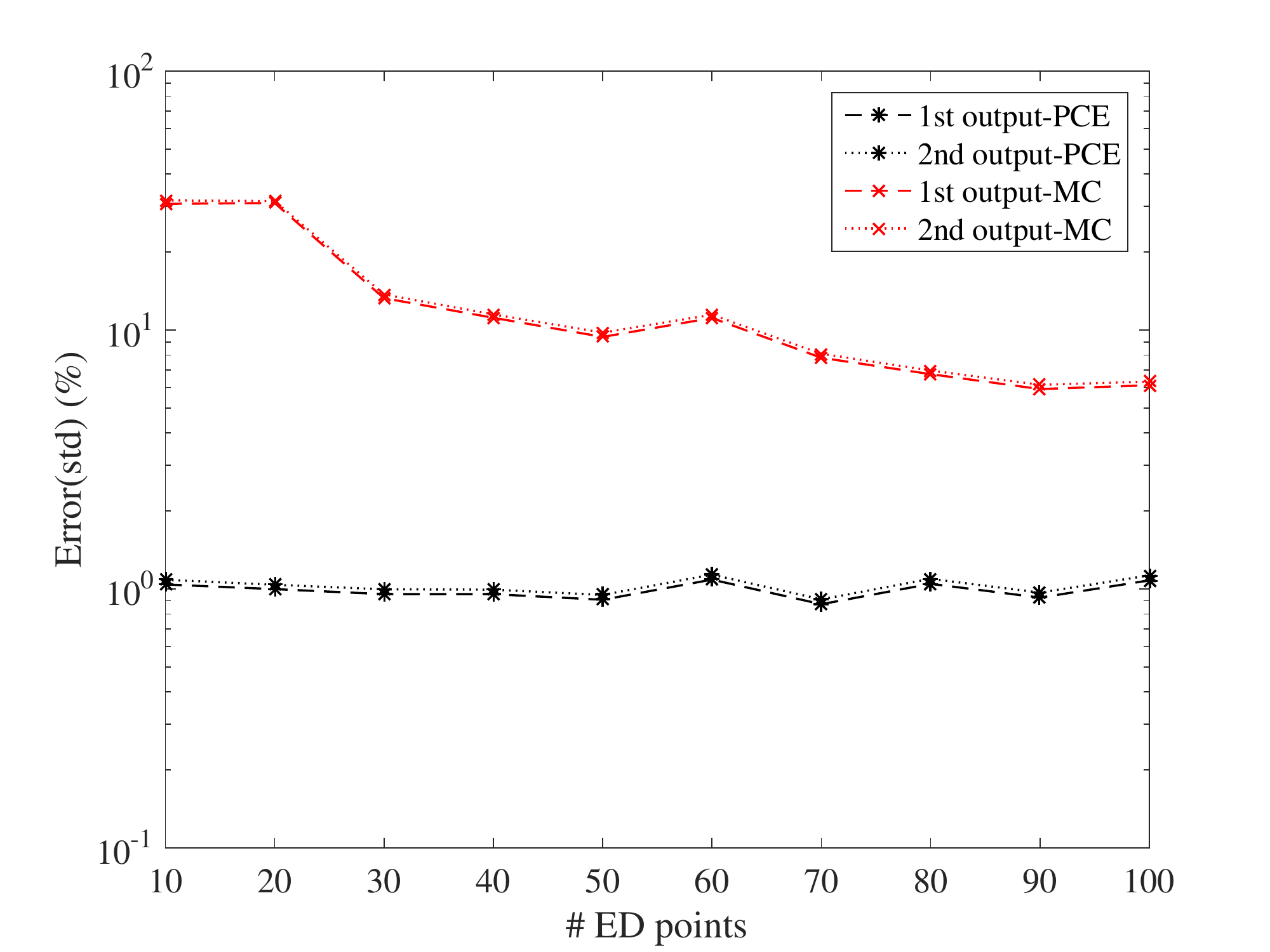}
		\caption{Convergence plot of the standard deviation of the FRFs}
		\label{fig:2DOF:PCE:convstd}
	\end{subfigure}
	\caption{Convergence plot of the statistics of the FRFs obtained by the PCE ($\ast$) and the true model ($\times$) with increasing ED size. The reference results were obtained by 10,000 Monte-Carlo simulations of the true model. }
	\label{fig:FRF:2DOF:conv} 
\end{figure}

The efficiency of the proposed approach is assessed by comparing the prediction accuracy of the surrogate model on a large reference validation set (10,000 samples calculated with the full model). PCE estimate of the mean and standard deviation of the surrogate model are compared to their Monte-Carlo estimators on experimental designs of increasing size. The resulting convergence curves are given in Figure \ref{fig:FRF:2DOF:conv}. 

They indicate that the PCE converges faster to the reference results for the mean and standard deviation. Their estimates are approximately two and one order of magnitude more accurate than those from the MC estimators. In addition, one can conclude that 40 points are enough for the ED in this example, since for larger sizes the accuracy does not improve significantly. 

It is worth mentioning that the coefficient of variation (COV) of the parameters and the level of damping are among the criteria that can affect the size of the ED. Therefore, larger COV and lower levels of damping are not obstacles for the proposed method provided a sufficiently large ED is used.

\begin{figure}[H]
	\centering
	\begin{subfigure}[b]{.5\columnwidth}
		\centering
		\includegraphics[width=1\columnwidth]{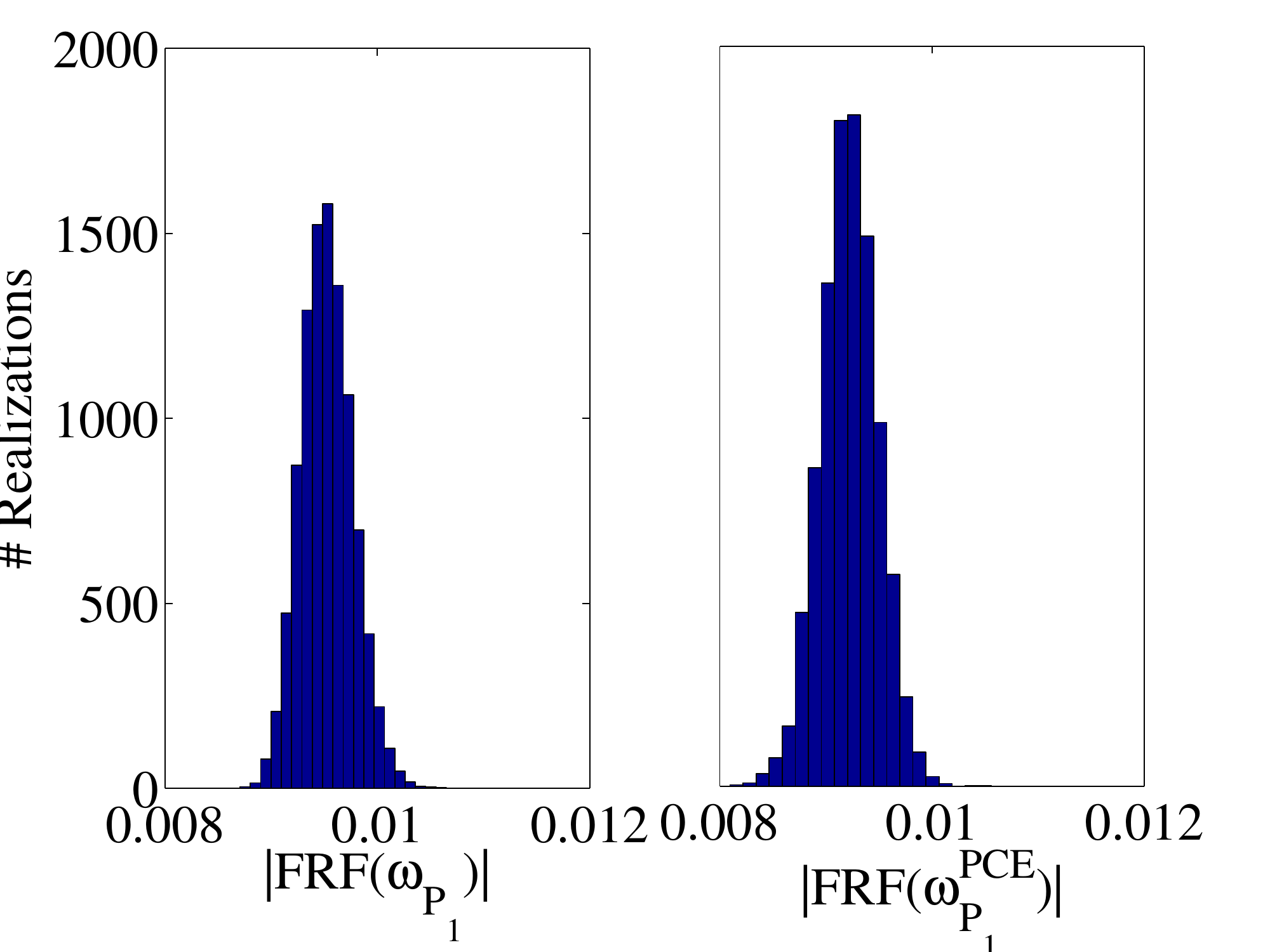}
		\caption{First resonant frequency}
		\label{fig:2DOF:Amp:freq1}
	\end{subfigure}
	\begin{subfigure}[b]{.5\columnwidth}
		\centering
		\includegraphics[width=1\columnwidth]{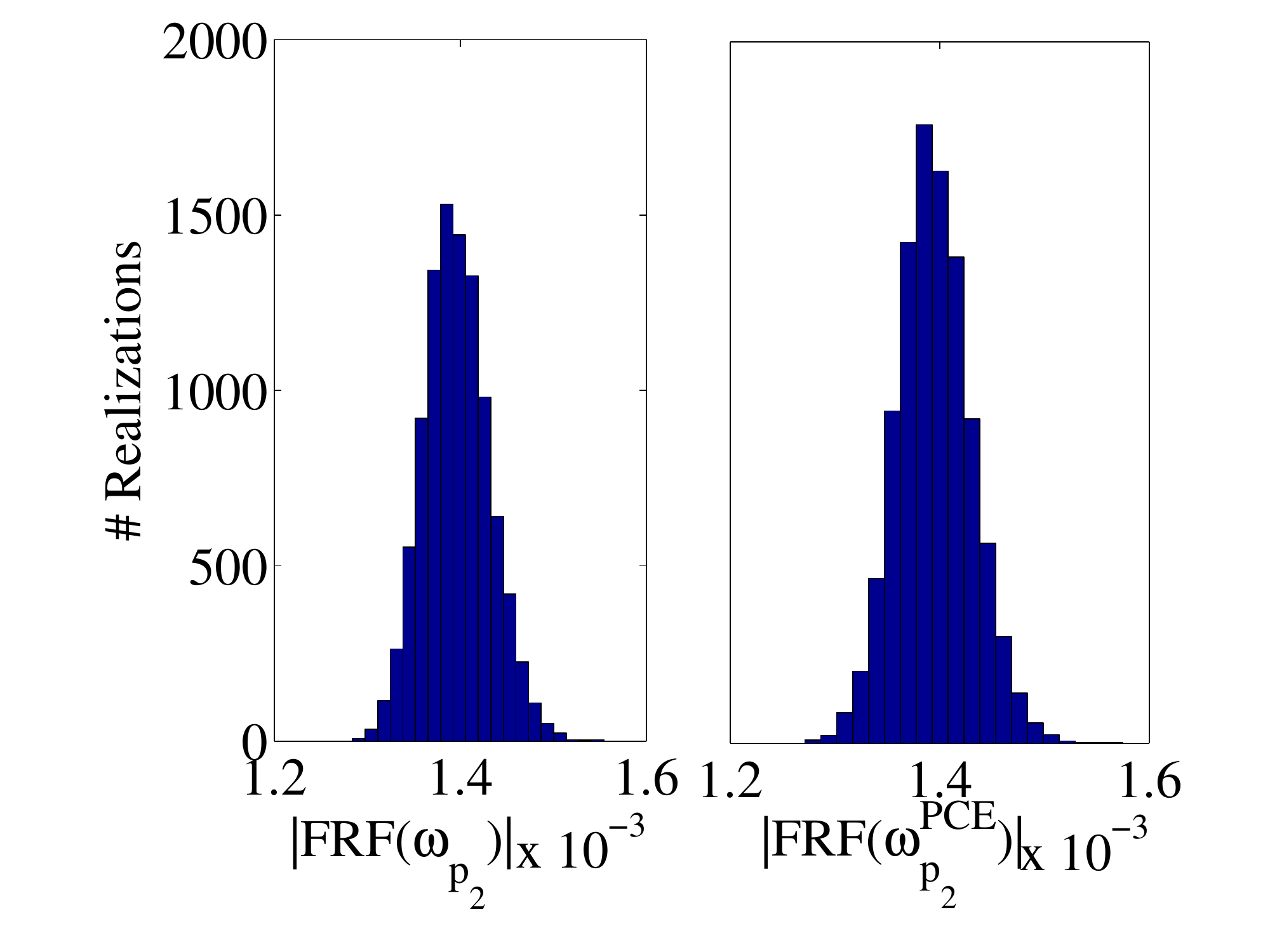}
		\caption{Second resonant frequency}
		\label{fig:2DOF:Amp:freq2}
	\end{subfigure}
	\caption{Histogram of the amplitude of the FRF at the resonant frequencies, obtained by evaluating the true and surrogate models on the 10,000 validation points.}
	\label{fig:2DOF:Amp:freqs}
\end{figure}

\begin{figure}[H]
	\centering 
	\begin{subfigure}[b]{.5\columnwidth}
		\centering
		\includegraphics[width=1\columnwidth]{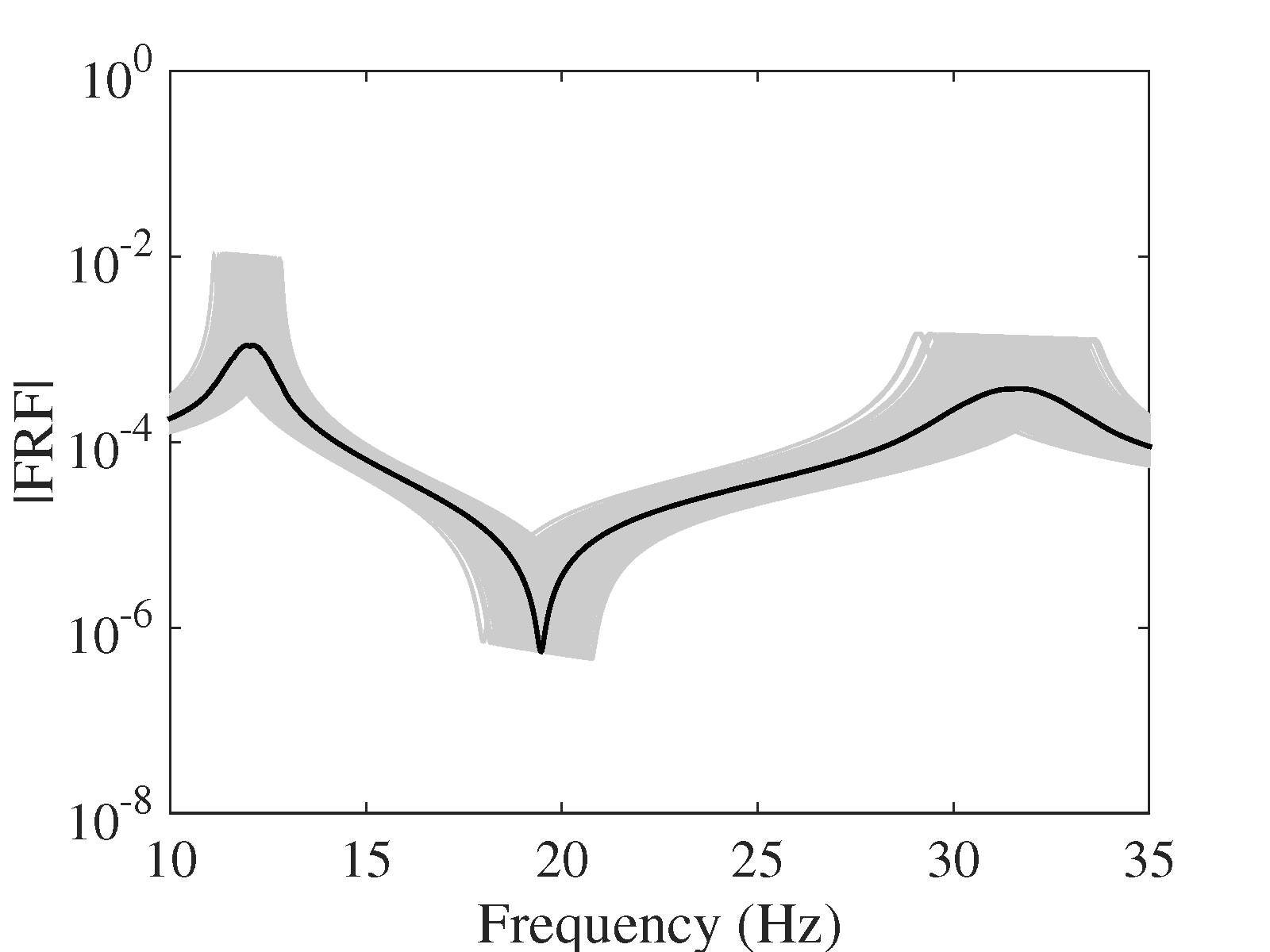}
		\caption{First system output- True model}
		\label{fig:2DOF:envelope:FRF:MC1}
	\end{subfigure}
	\begin{subfigure}[b]{.5\columnwidth}
		\centering
		\includegraphics[width=1\columnwidth]{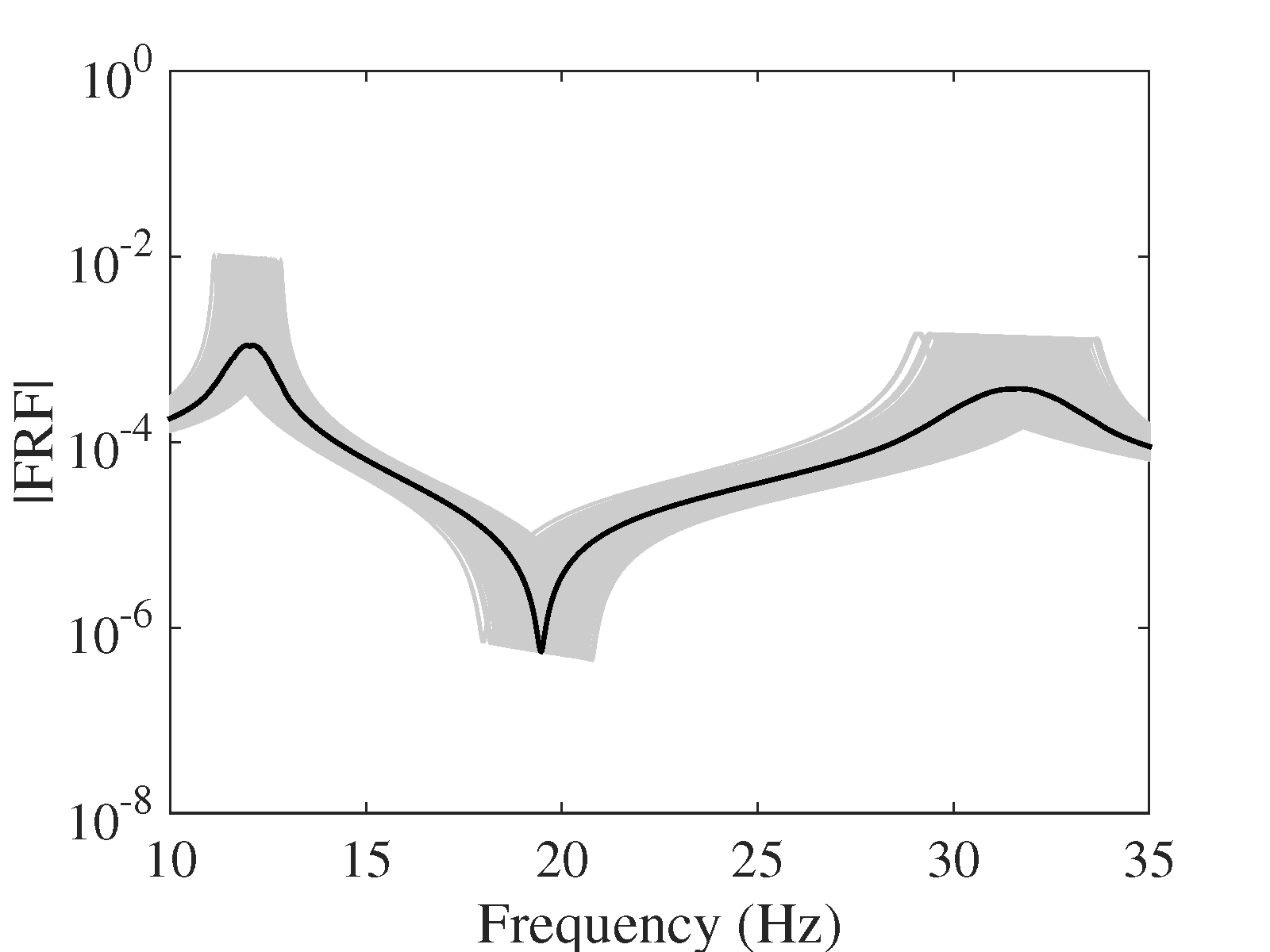}
		\caption{First system output- Surrogate model}
		\label{fig:2DOF:envelope:FRF:PCE1}
	\end{subfigure}
	
	
	\caption{All the FRFs obtained by evaluating the true and the surrogate model at 10,000 MC samples.} 
	\label{fig:FRF:2DOF:envelope}
\end{figure}

\begin{figure}[H]
	\centering
	\begin{subfigure}[b]{.5\columnwidth}
		\centering
		\includegraphics[width=1\columnwidth]{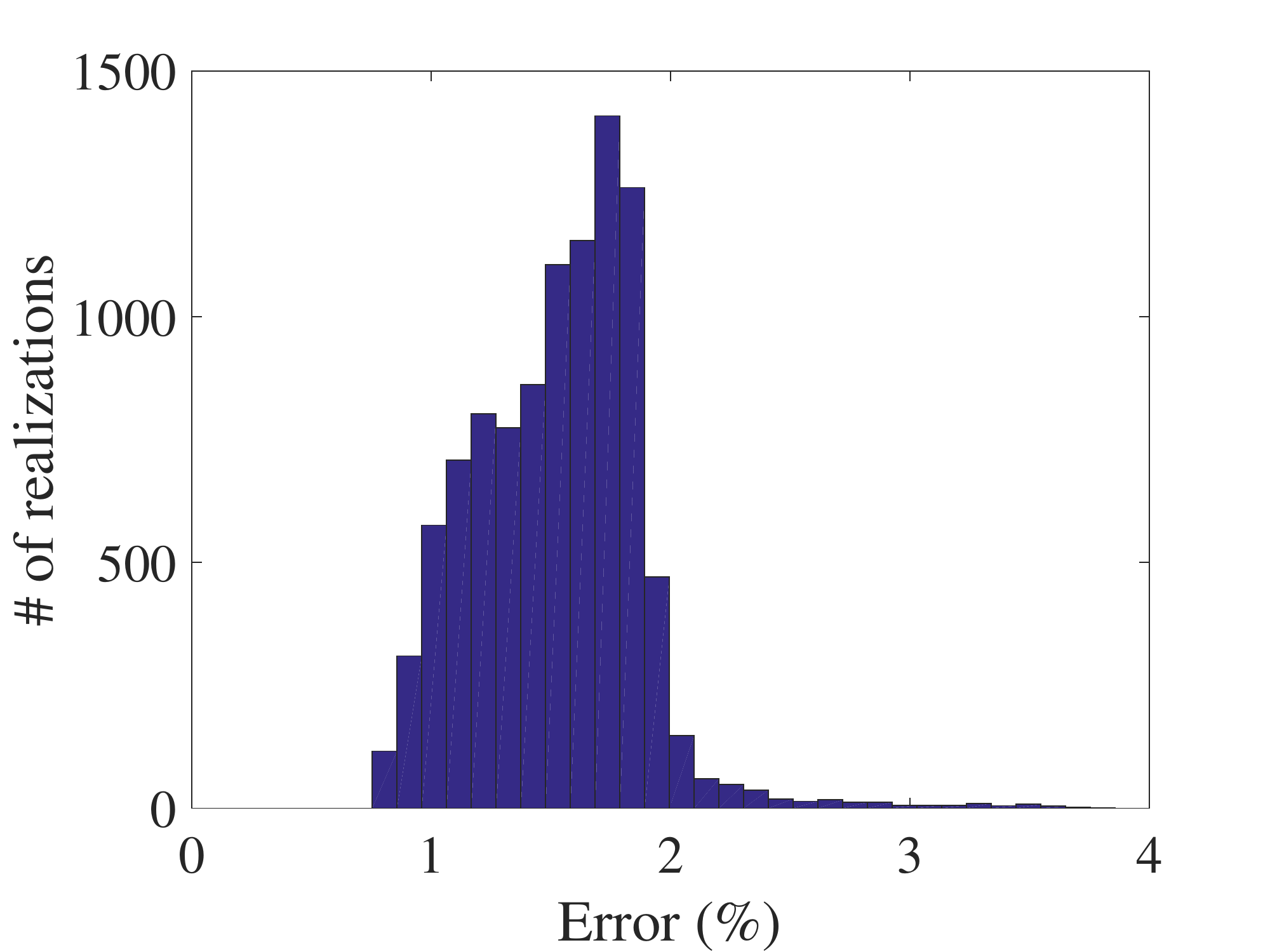}
		\caption{First system output}
		\label{fig:2DOF:error:frf:out1}
	\end{subfigure}
	\begin{subfigure}[b]{0.5\columnwidth}
		\centering
		\includegraphics[width=1\columnwidth]{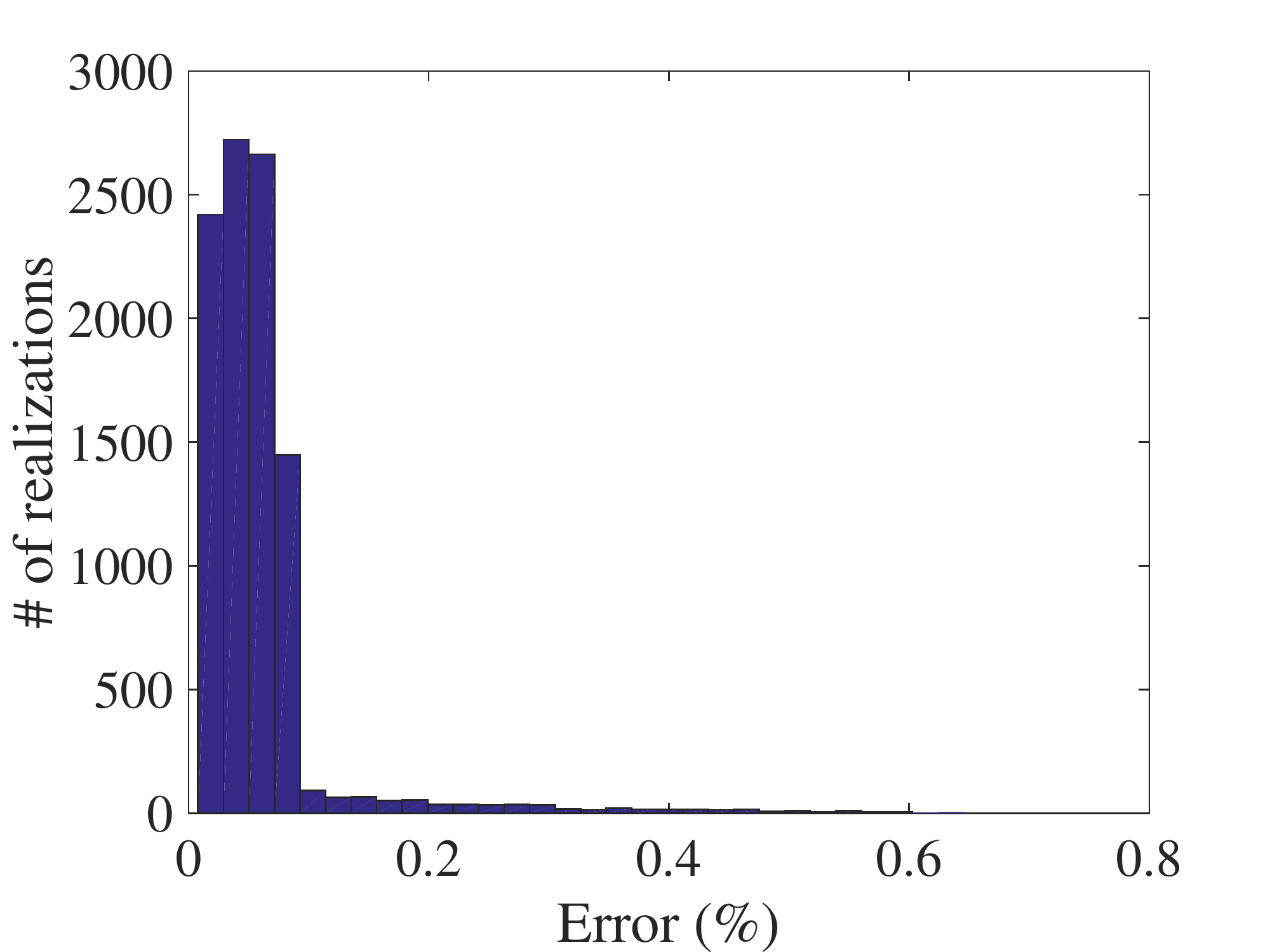}
		\caption{Second system output}
		\label{fig:2DOF:error:frf:out2}
	\end{subfigure}
	\caption{Error of the FRF predicted by PCE surrogate model, evaluated by Eq. (\ref{eq:Err:rms}).} 
	\label{fig:2DOF:error:frf}
\end{figure}

The 10,000 model evaluations used to produce the convergence curves in Figure \ref{fig:FRF:2DOF:conv} are also used to provide a detailed validation of the performance of the PCE surrogates on various quantities of interests.
The results are presented in the following figures.
In Figure \ref{fig:2DOF:predict:eigval}, the \textit{selected frequencies} obtained by the PCE are shown versus the true ones. The results show that the PCE model accurately predicts the \textit{selected frequencies}.
Since the amplitudes at the resonant frequencies are the most sensitive parts of the FRFs and contains most of crucial information about the system, it is one of the most interesting parts of the FRFs for the researcher. Figure \ref{fig:2DOF:Amp:freqs} shows the histograms of these amplitudes obtained by the true and the surrogate models at at all 10,000 validation points. Moreover, 
in Figure \ref{fig:FRF:2DOF:envelope}, the whole FRFs at the validation points are depicted and compared. Their associated means are also shown by black lines. The results indicate accurate prediction of the amplitude at the resonant frequencies as well as the whole FRFs. A quantitative accuracy analysis for the whole FRF was done by using Eq. (\ref{eq:Err:rms}) and its histogram presented in Figure \ref{fig:2DOF:error:frf} confirms the high accuracy of the surrogate model.

\begin{figure}[H]
	\centering
	\begin{subfigure}[b]{0.5\columnwidth}
		
		\includegraphics[width=1\columnwidth]{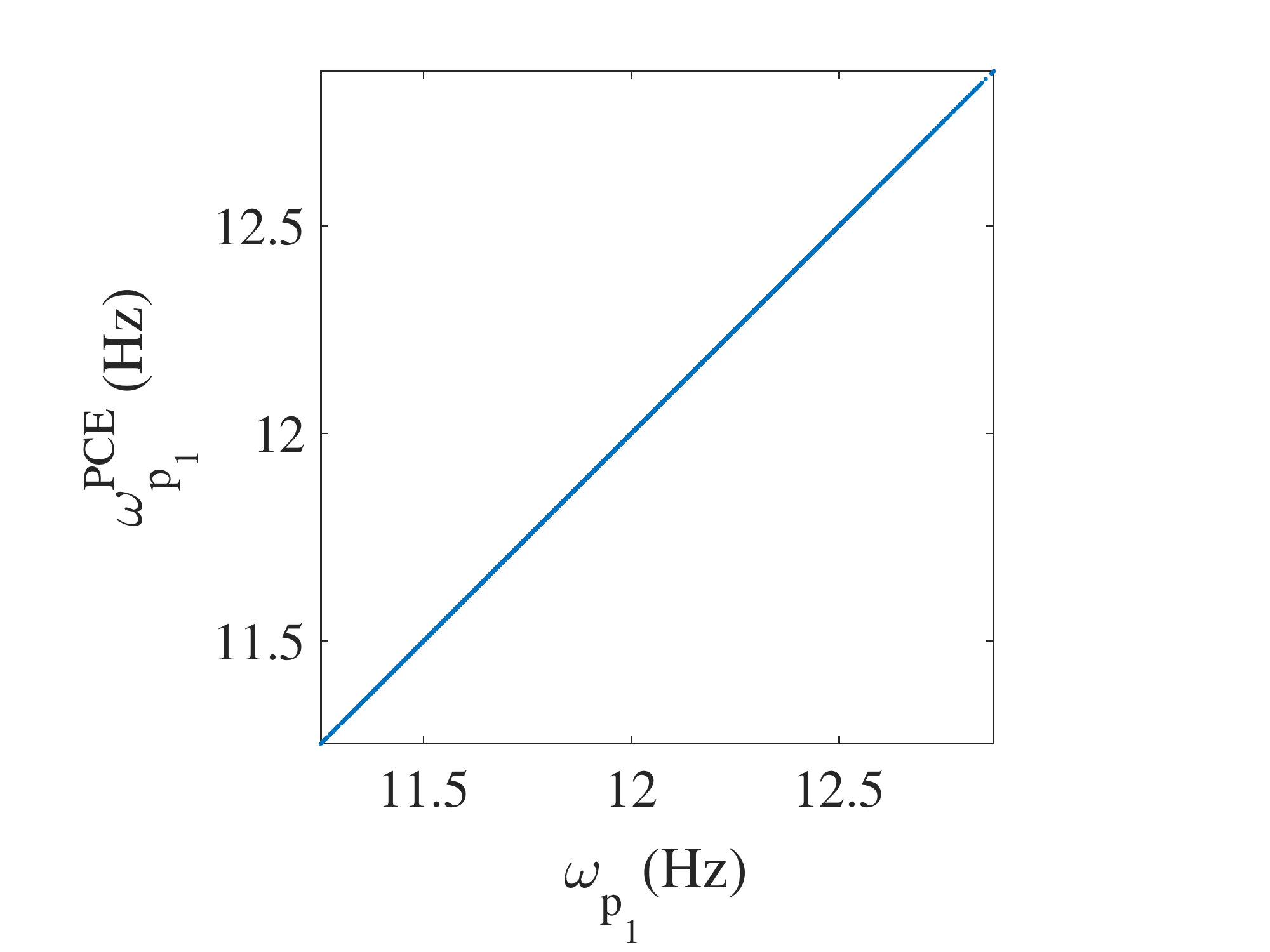}
		\label{fig:2DOF:eig:1}
	\end{subfigure}
	\begin{subfigure}[b]{0.5\columnwidth}
		
		\includegraphics[width=1\columnwidth]{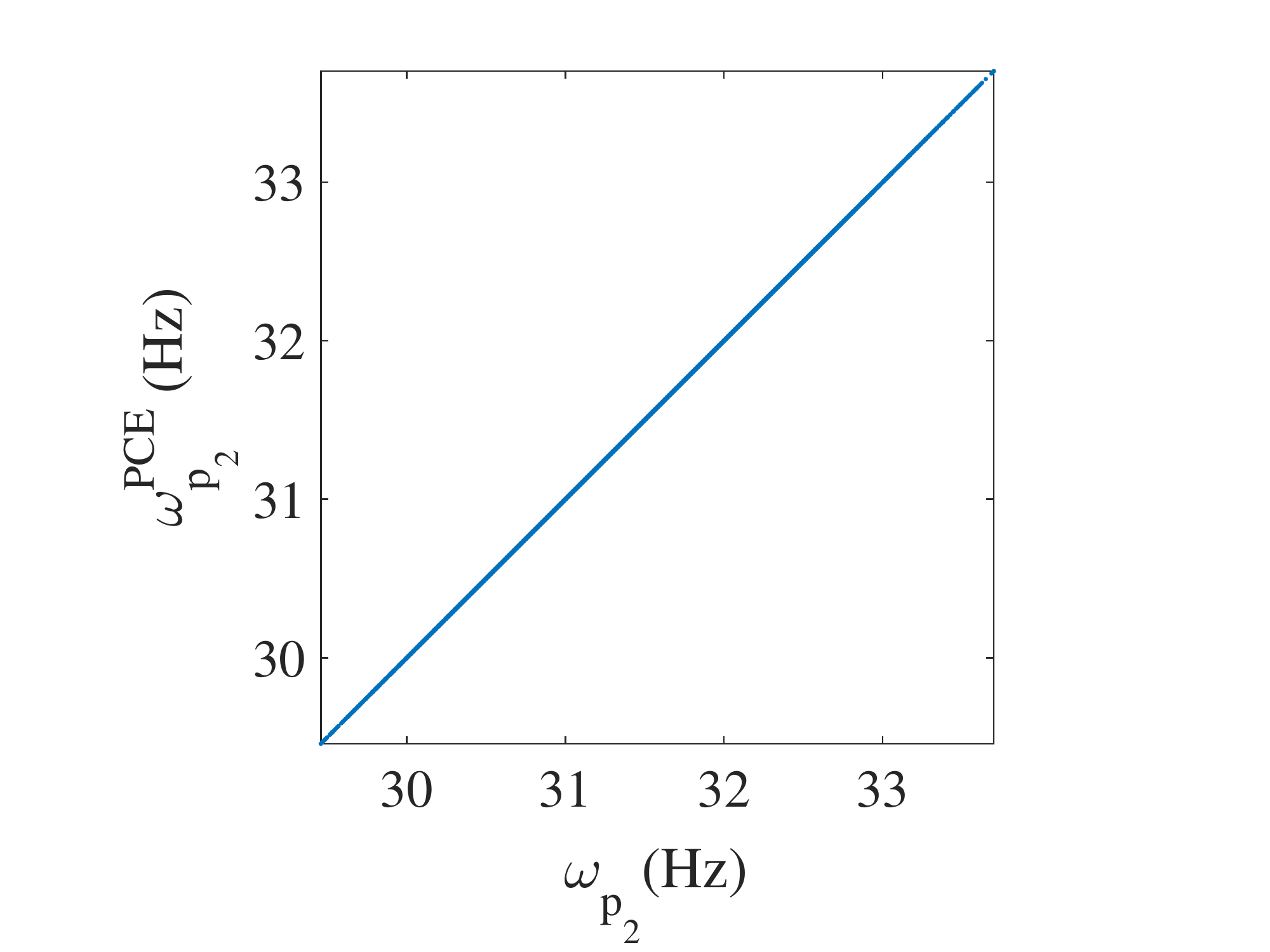}
		\label{fig:2DOF:eig:2}
	\end{subfigure}
	\\
	\begin{subfigure}[b]{0.5\columnwidth}
		
		\includegraphics[width=1\columnwidth]{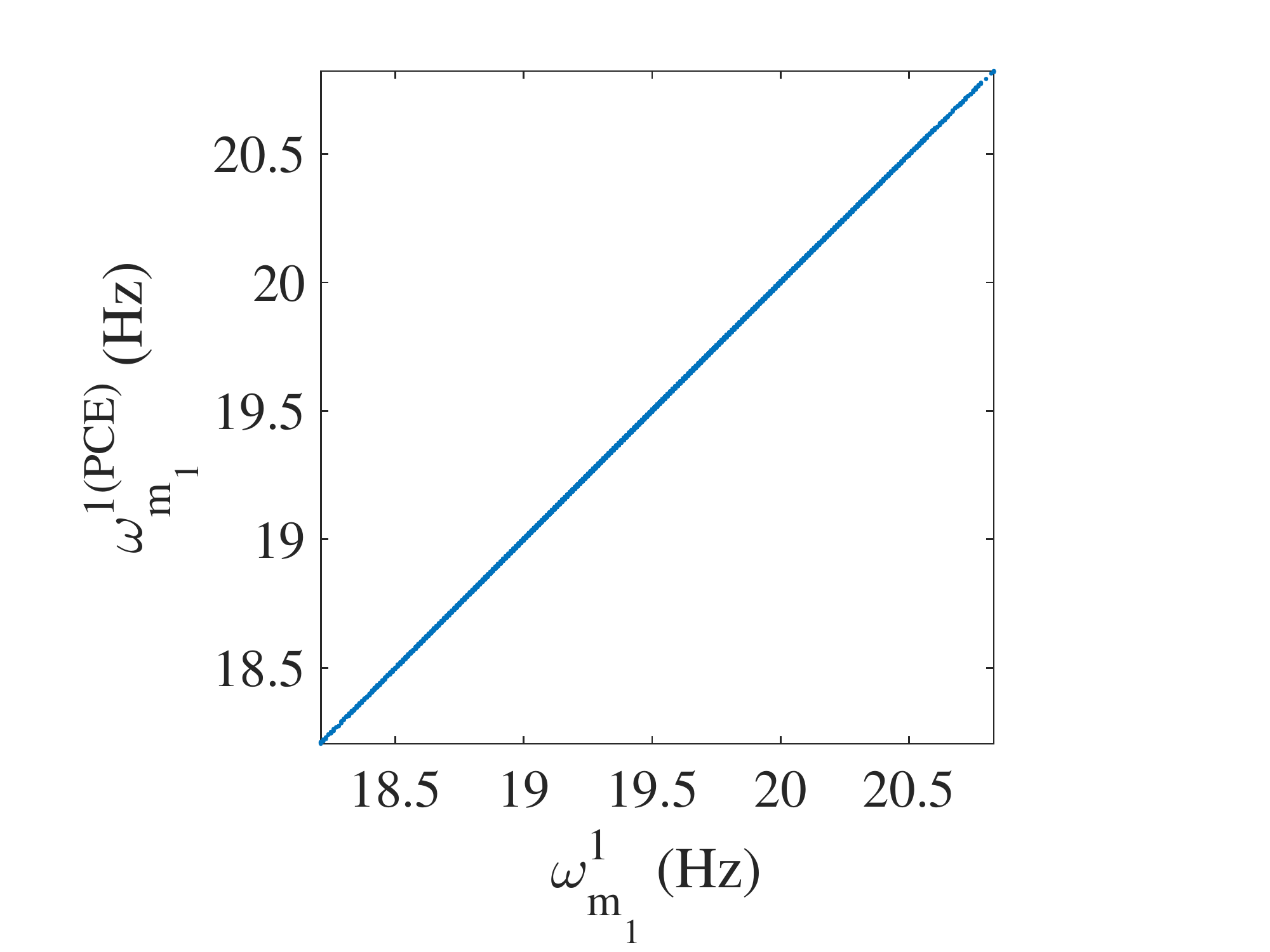}
		\label{fig:2DOF:min:1}
	\end{subfigure}
	\begin{subfigure}[b]{0.5\columnwidth}
		
		\includegraphics[width=1\columnwidth]{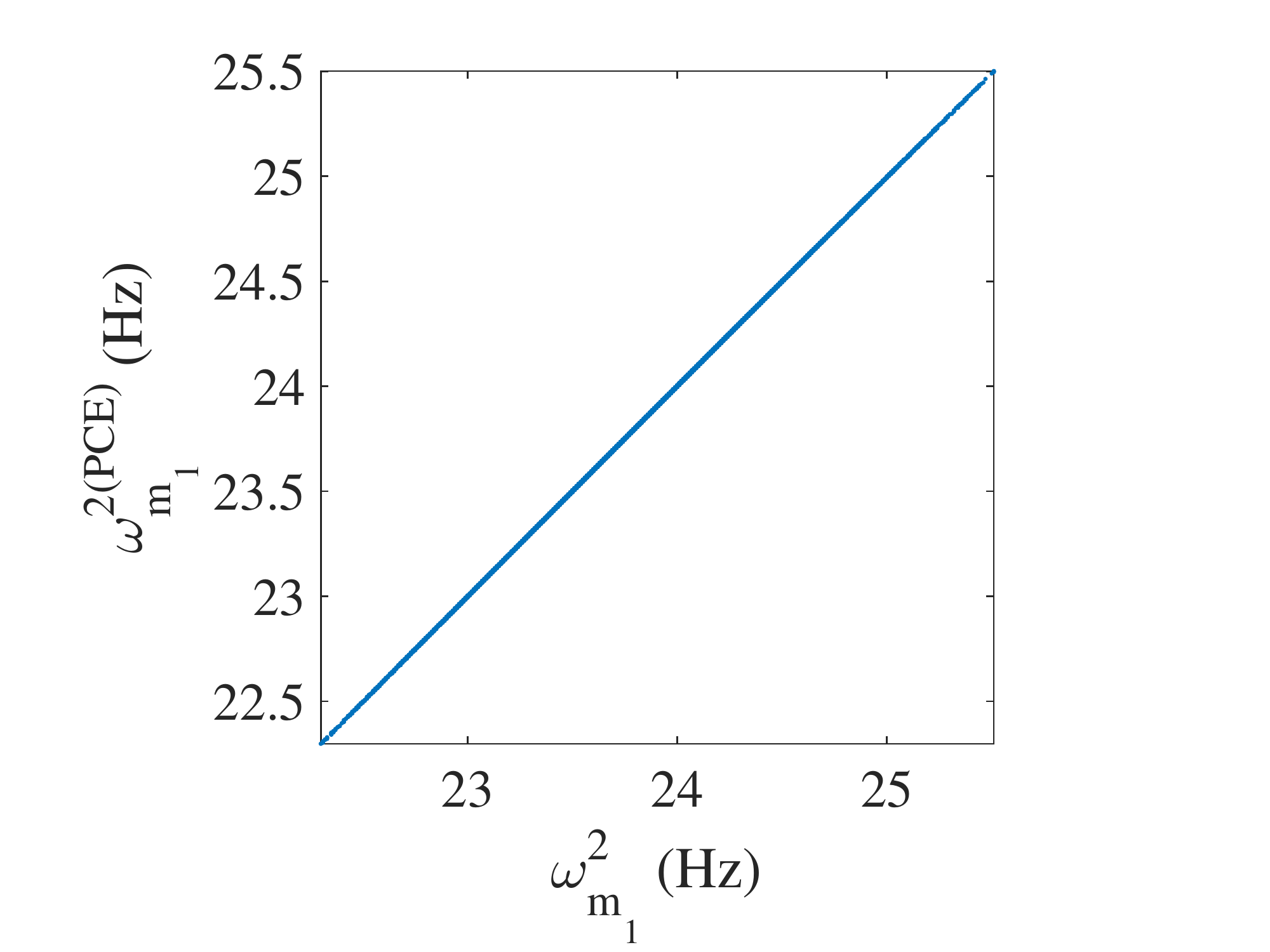}
		\label{fig:2DOF:min:2}
	\end{subfigure}
	\caption{Selected frequencies predicted by the surrogate  model versus by the true model. upper row: eigenfrequencies, lower row: frequencies where minimum amplitude occurs, see matrix (\ref{eq16}) for notations.}
	\label{fig:2DOF:predict:eigval}
\end{figure}

%
%
%
%



\begin{figure}[H]
	\centering
	\begin{subfigure}[b]{.5\columnwidth}
		\centering
		\includegraphics[width=1\columnwidth]{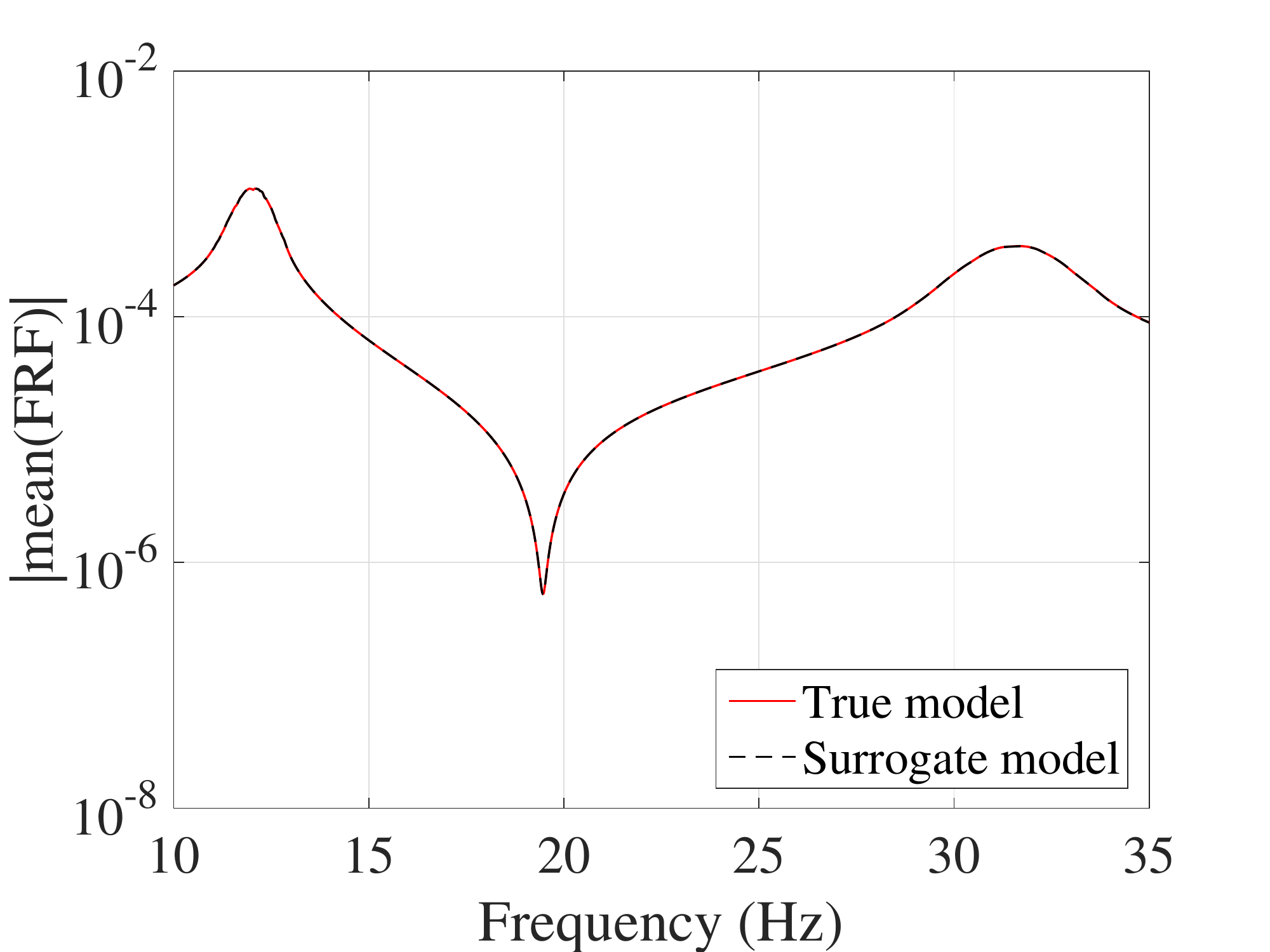}
		\caption{Magnitude of the mean}
		\label{fig:2DOF:FRF:mean1}
	\end{subfigure}
	\begin{subfigure}[b]{.5\columnwidth}
		\centering
		\includegraphics[width=1\columnwidth]{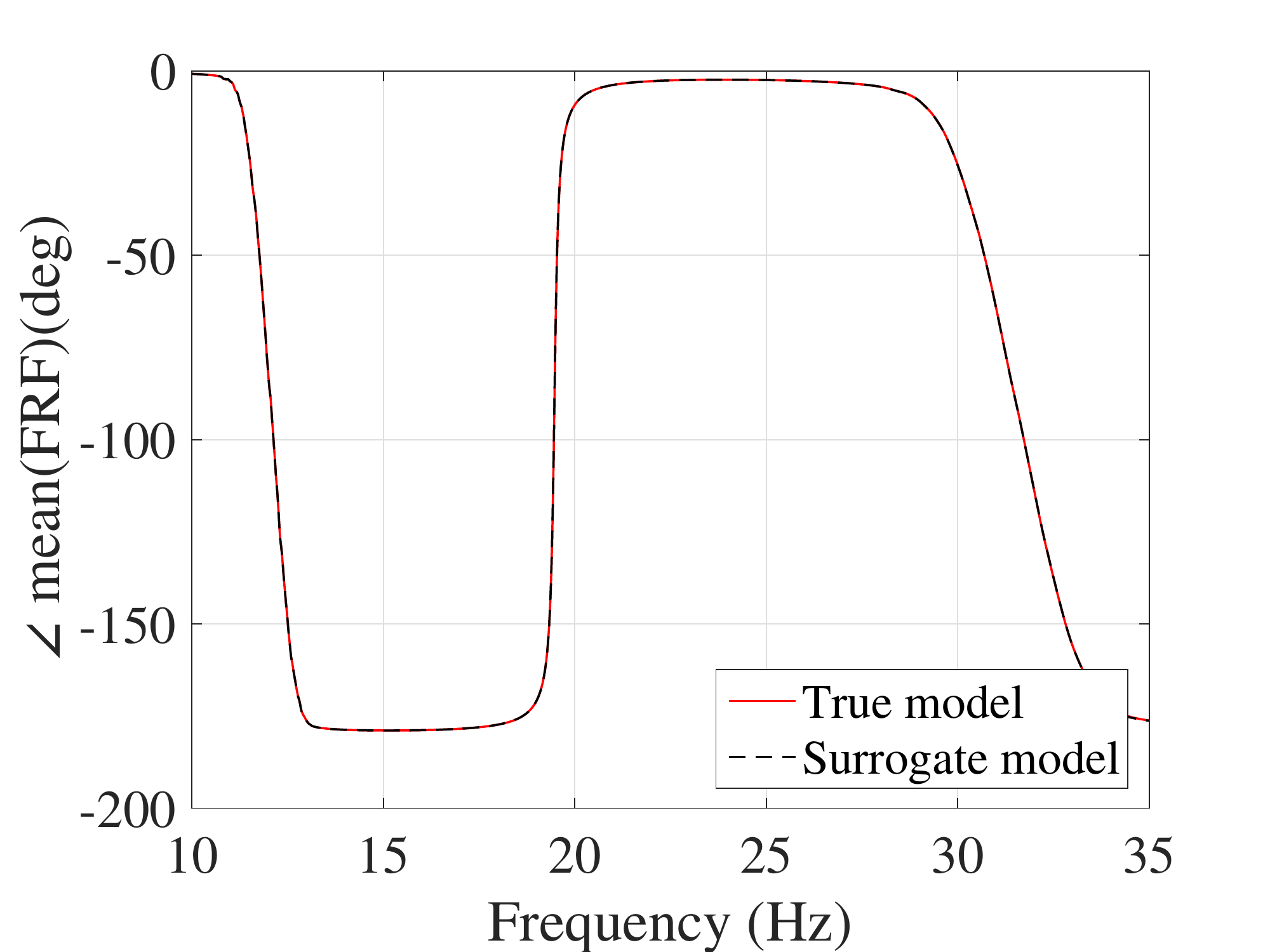}
		\caption{phase of the mean}
		\label{fig:2DOF:FRF:mean1:phase}
	\end{subfigure}%
	\\
	\begin{subfigure}[b]{.5\columnwidth}
		\centering
		\includegraphics[width=1\columnwidth]{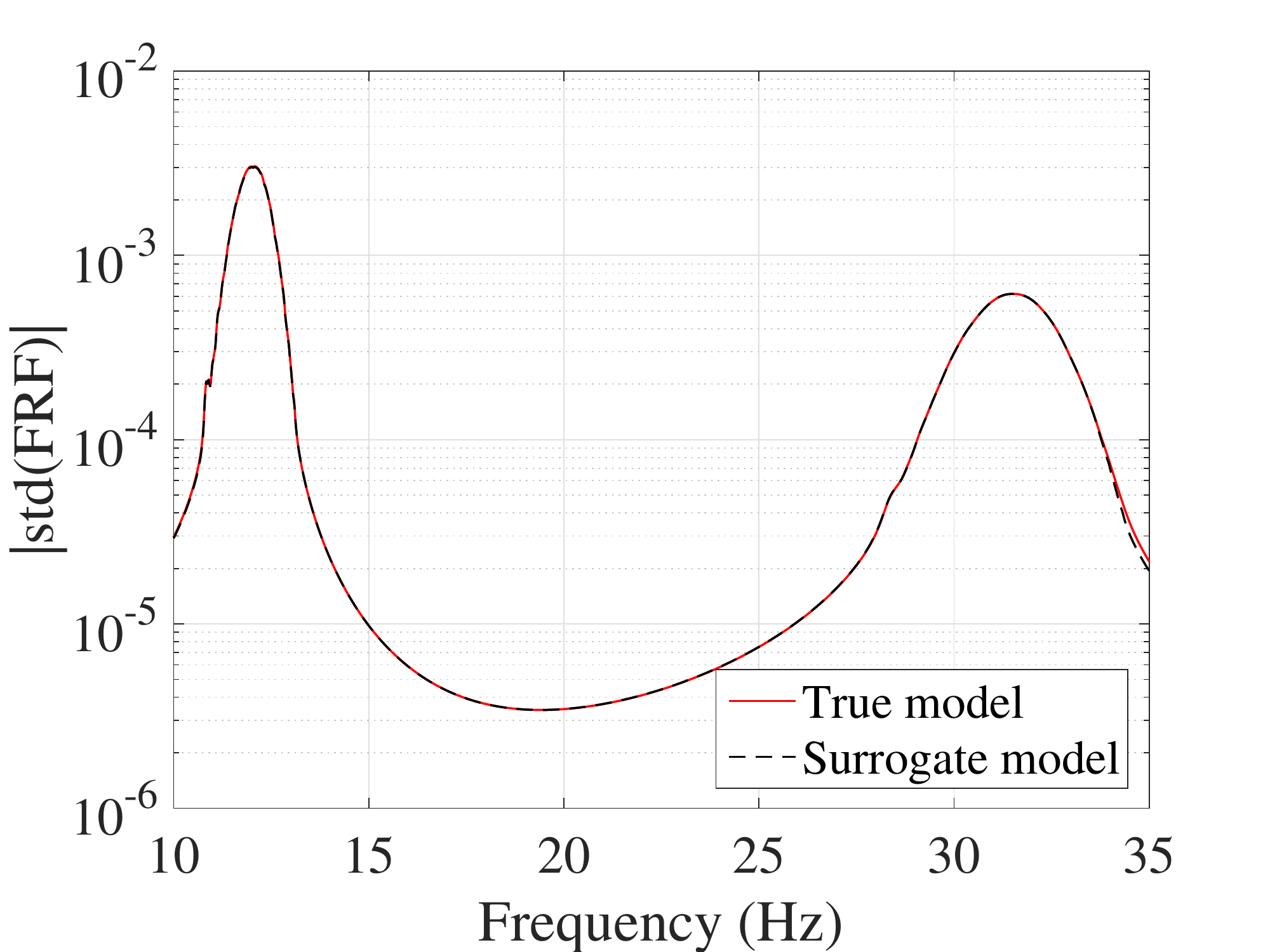}
		\caption{Magnitude of the standard deviation}
		\label{fig:2DOF:FRF:std1}
	\end{subfigure}
	\begin{subfigure}[b]{.5\columnwidth}
		\centering
		\includegraphics[width=1\columnwidth]{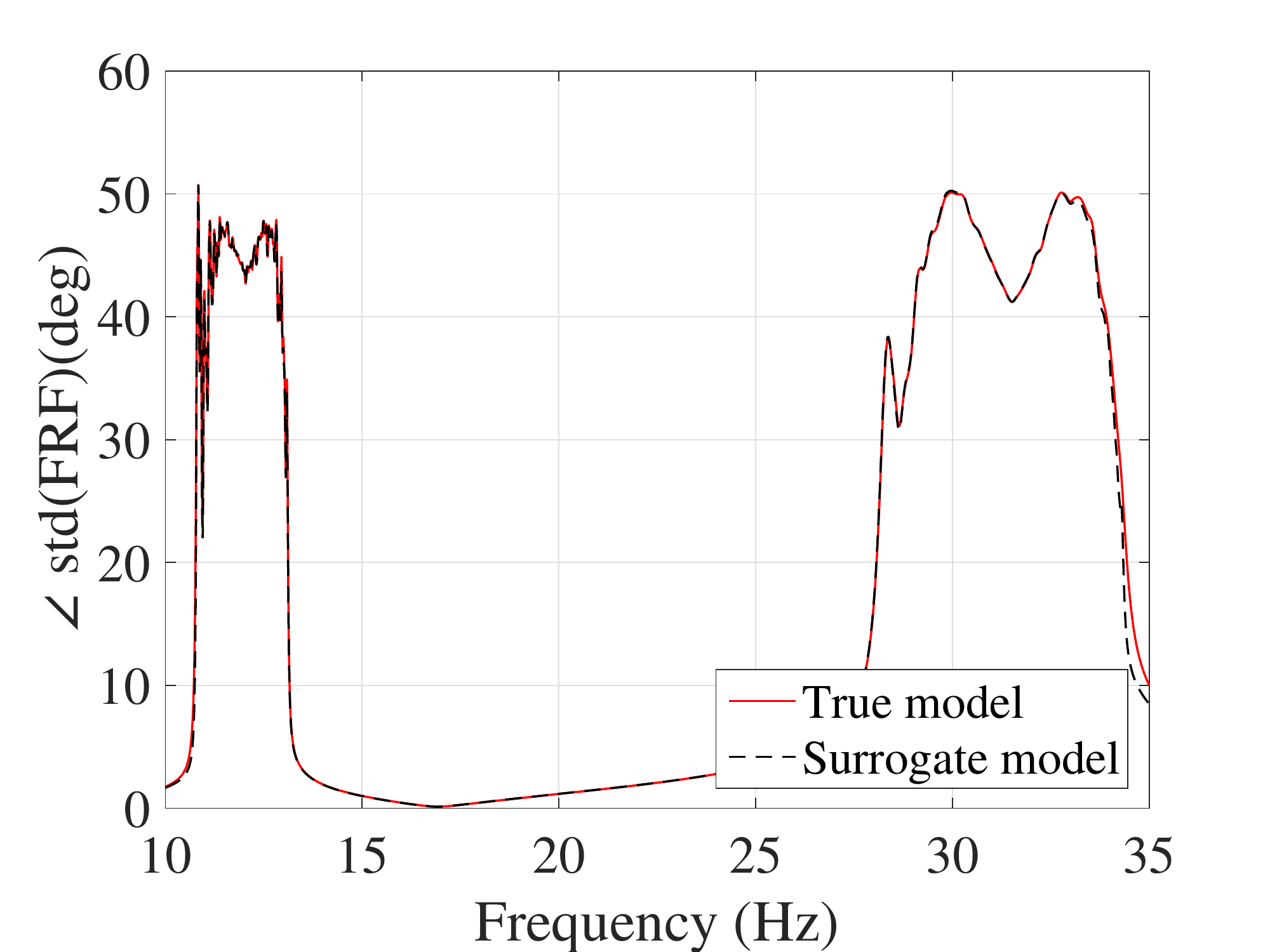}
		\caption{Phase of the standard deviation}
		\label{fig:2DOF:FRF:std1:phase}
	\end{subfigure}%
	
	
	\caption{Mean and standard deviation of the FRFs evaluated over 10,000 sample points, by the true model (red) and by the surrogate model (black).} 
	\label{fig:FRF:2DOF:out1}
\end{figure}

Another accuracy test is given by the comparison between the first two moments of the FRFs. The mean  and standard deviation of the trajectories obtained by the true model and the surrogate model are compared in Figure \ref{fig:FRF:2DOF:out1} for the $1^{st}$ output and in \ref{app:2DOF:stat:out2} for the $2^{nd}$ output. They reveal the accuracy of the proposed surrogate model in predicting the first two moments of the FRFs.

\begin{figure}[H]
	\centering
	\begin{subfigure}[b]{0.47\columnwidth}
		\centering
		\includegraphics[width=1\columnwidth]{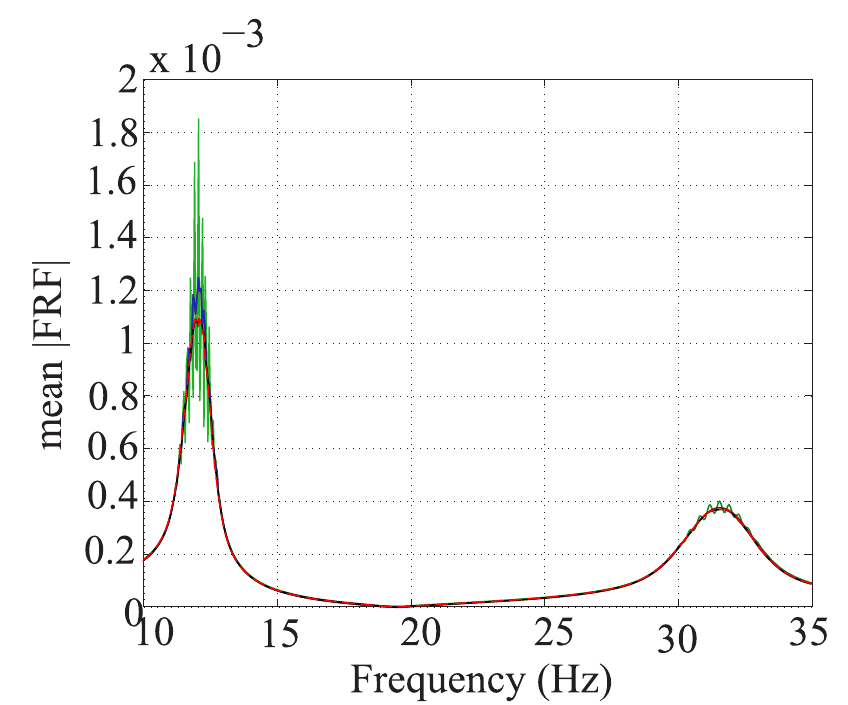}
		\caption{Mean of the FRFs} 
		\label{fig:2DOF:FRF:mean:compinone}
	\end{subfigure}
	\begin{subfigure}[b]{0.47\columnwidth}
		\centering
		\includegraphics[width=1\columnwidth]{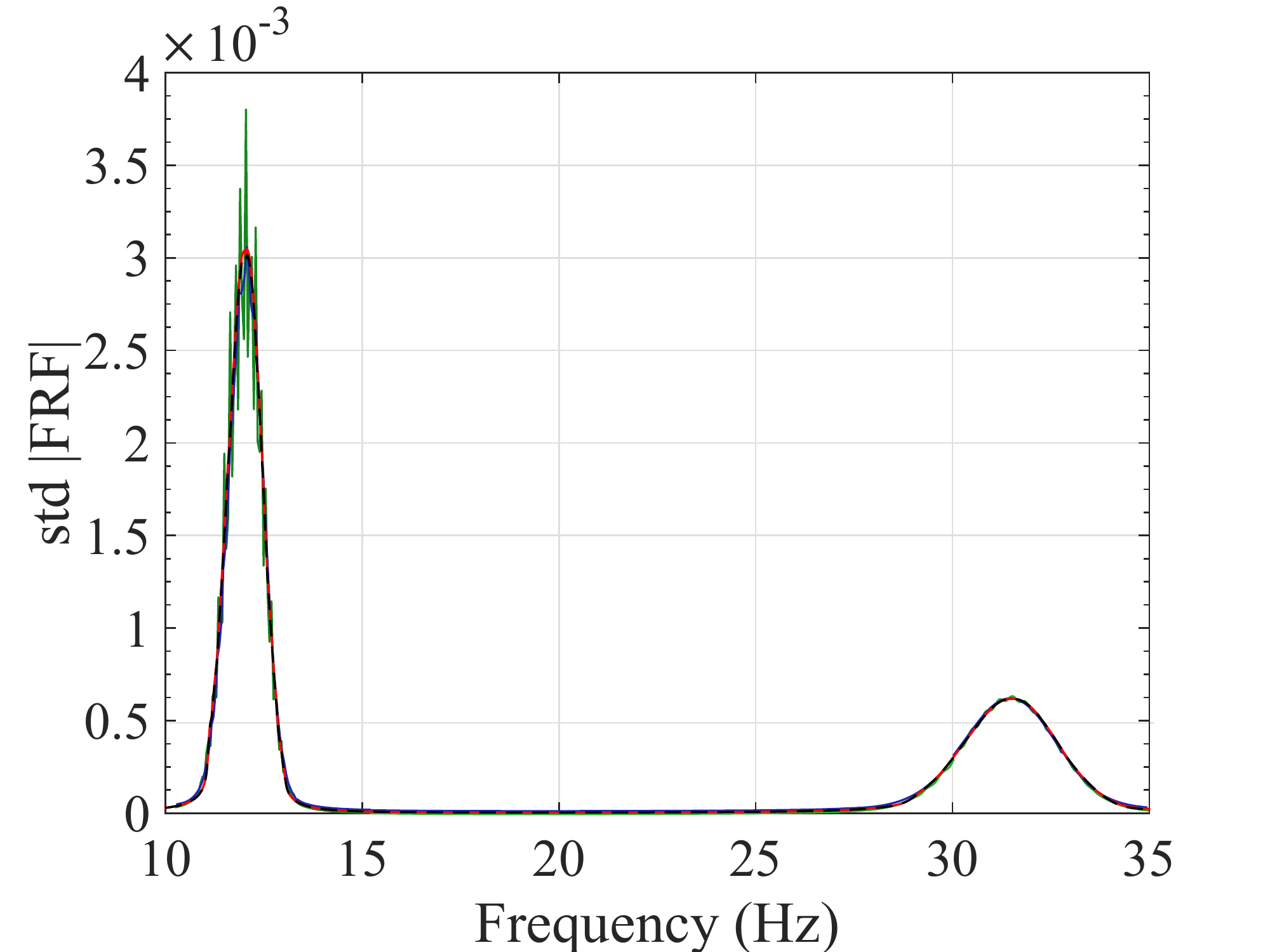}
		\caption{Standard deviation of the FRFs} 
		\label{fig:2DOF:FRF:std:compinone}
	\end{subfigure}
	\caption{Comparison between the methods to estimate the statistics of the FRFs at the first mass by PCE. Green line: Direct use of PCE with order 50. Shown in \citet{Jacquelin2015144, jacquelinpolynomial2015}, blue line: PCE with order 50 and Aitken's transformation. Proposed in \citet{jacquelinpolynomial2015}, black line: the proposed stochastic transformation method, red line: reference result.}
	\label{fig:FRF:2DOF:compinone}
\end{figure}

To show the feasibility of the proposed method to estimate the statistics of the FRFs, the results obtained here are compared to their counterparts in two of the most recent works available in the literature. The first study \citep{Jacquelin2015144} directly uses high-order PCE for estimating the first two moments of the FRF, whereas the second method \citep{jacquelinpolynomial2015} proposes to use Aitken's transformation in conjunction with PCEs,  
Both methods use PCEs of order 50 and tend to produce spurious peaks around the resonance region. The use of Aitken's transformation slightly improves convergence. Their results for the mean and standard deviation are shown in Figures \ref{fig:2DOF:FRF:mean:compinone} and \ref{fig:2DOF:FRF:std:compinone}, respectively.
For comparison, the results from our approach in Figures \ref{fig:2DOF:FRF:mean1} and \ref{fig:2DOF:FRF:std1} are reproduced in Figure \ref{fig:FRF:2DOF:compinone} with a scaling similar to the other panels.
They indicate that the stochastic frequency transformation approach proposed here, significantly improves the estimation accuracy of the PCE surrogate, as no spurious peaks are visible in this case.

\begin{figure}[H]
	\centering
	\begin{subfigure}[b]{.5\columnwidth}
		\centering
		\includegraphics[width=1\columnwidth]{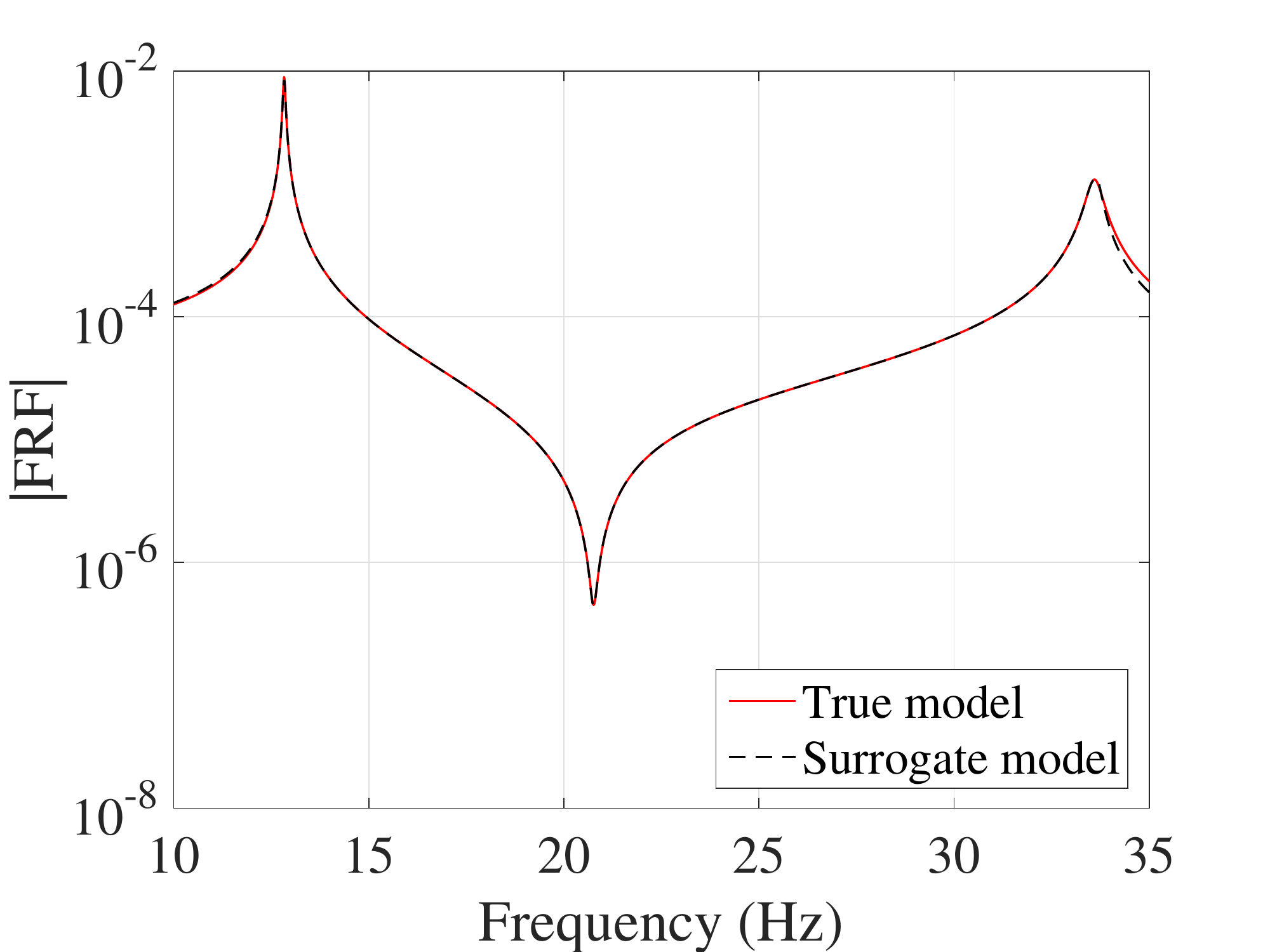}
		\caption{First system output}
		\label{fig:2DOF:predict:worstFRF:out1}
	\end{subfigure}
	\begin{subfigure}[b]{.5\columnwidth}
		\centering
		\includegraphics[width=1\columnwidth]{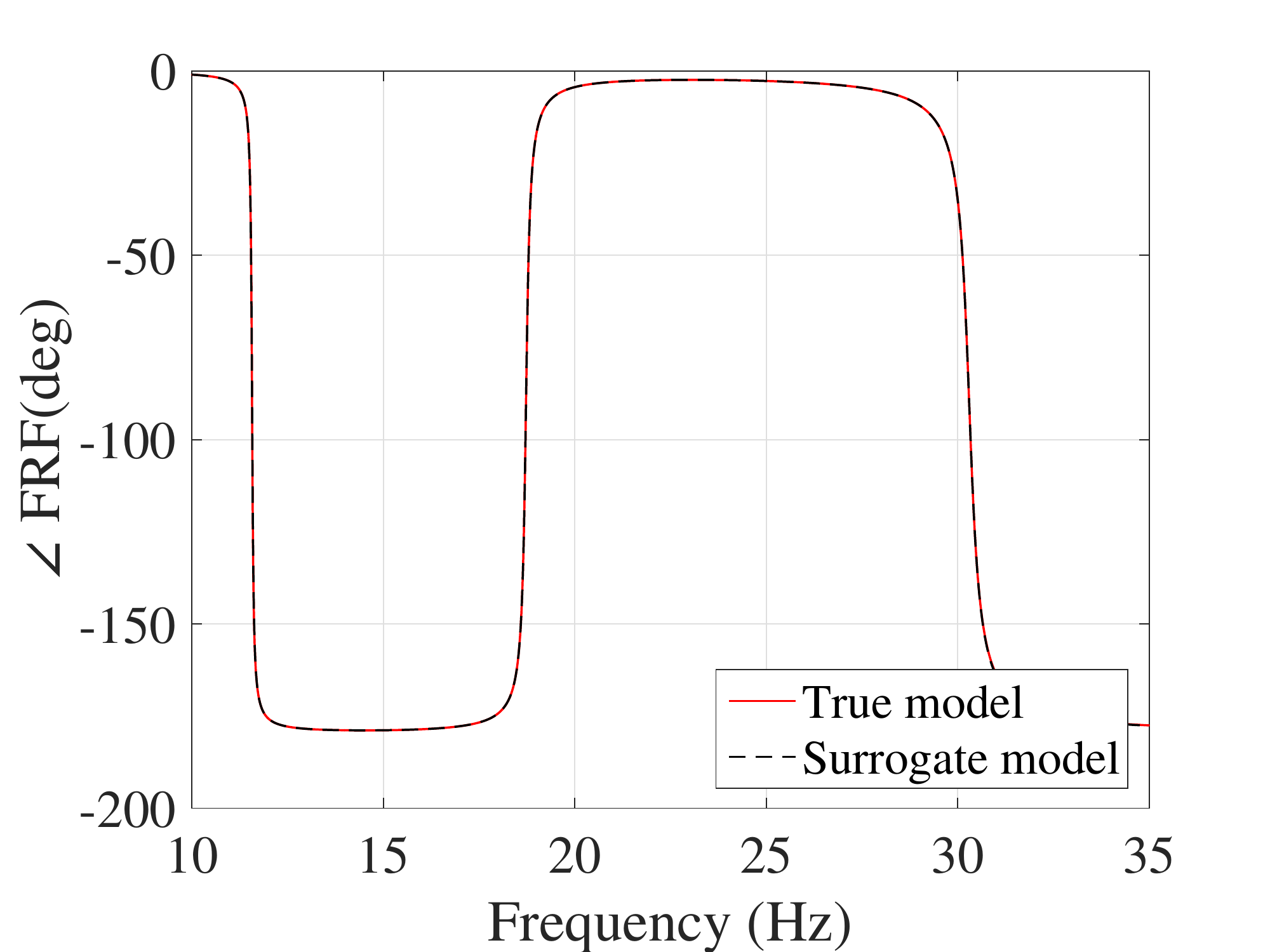}
		\caption{First system output}
		\label{fig:2DOF:predict:worstFRF:out1:phase}
	\end{subfigure}
	
	\caption{Worst case FRF prediction among 10,000 sample points, true (in red) and predicted (in black) FRF of the system.} 
	\label{fig:FRF:2DOF:predict:worst}
\end{figure}

As far as individual comparison between the true and the predicted FRF are concerned, the worst predicted FRF, the one with the maximum error, is presented in Figure \ref{fig:FRF:2DOF:predict:worst}.
They indicate that even for the worst-case, the presented approach results in prediction of the FRFs with excellent accuracy.


\subsection{6-DOF system: large parameter space}

\begin{figure} [H]
	\centering
	\begin{adjustbox}{max width=0.8\columnwidth}
		\begin{tikzpicture}
		
		\tikzstyle{spring}=[thick,decorate,decoration={zigzag,pre length=0.3cm,post
			length=0.3cm,segment length=6}]
		
		\tikzstyle{damper}=[thick,decoration={markings,  
			mark connection node=dmp,
			mark=at position 0.5 with 
			{
				\node (dmp) [thick,inner sep=0pt,transform shape,rotate=-90,minimum
				width=10pt,minimum height=3pt,draw=none] {};
				\draw [thick] ($(dmp.north east)+(2pt,0)$) -- (dmp.south east) -- (dmp.south
				west) -- ($(dmp.north west)+(2pt,0)$);
				\draw [thick] ($(dmp.north)+(0,-4pt)$) -- ($(dmp.north)+(0,4pt)$);
			}
		}, decorate]
		
		\tikzstyle{ground}=[fill,pattern=north east lines,draw=none,minimum
		width=0.75cm,minimum height=0.3cm]
		
		\node[draw,outer sep=0pt,thick, fill=white!60!yellow] (M1) [minimum width=1cm, minimum height=4.5cm] {$m_1$};
		\node[draw,outer sep=0pt,thick, fill=white!60!yellow] (M2) at (2.5,1.5) [minimum width=1cm, minimum height=1.5cm] {$m_2$};
		\node[draw,outer sep=0pt,thick, fill=white!60!yellow] (M3) at (2.5,-1.5) [minimum width=1cm, minimum height=1.5cm] {$m_3$};
		\node[draw,outer sep=0pt,thick, fill=white!60!yellow] (M6) at (5,0) [minimum width=1cm, minimum height=4.5cm] {$m_6$};
		\node[draw,outer sep=0pt,thick, fill=white!60!yellow] (M5) at (7.5,1.5) [minimum width=1cm, minimum height=1.5cm] {$m_5$};
		\node[draw,outer sep=0pt,thick, fill=white!60!yellow] (M4) at (7.5,-1.5) [minimum width=1cm, minimum height=1.5cm] {$m_4$};
		
		\node (ground1) [ground,anchor=north,yshift=-0.25cm,minimum width=4.8cm] at (M1.south) {};
		\draw (ground1.north east) -- (ground1.north west);
		\draw [thick, fill={cyan}] (M1.south west) ++ (0.2cm,-0.125cm) circle (0.125cm)  (M1.south east) ++ (-0.2cm,-0.125cm) circle (0.125cm);
		
		\node (ground2) [ground,anchor=north,yshift=-0.25cm,minimum width=4.8cm] at (M3.south) {};
		\draw (ground2.north east) -- (ground2.north west);
		\draw [thick, fill={cyan}] (M3.south west) ++ (0.2cm,-0.125cm) circle (0.125cm)  (M3.south east) ++ (-0.2cm,-0.125cm) circle (0.125cm);
		
		\node (ground3) [ground,anchor=north,yshift=-0.25cm,minimum width=4.8cm] at (M6.south) {};
		\draw (ground3.north east) -- (ground3.north west);
		\draw [thick, fill={cyan}] (M6.south west) ++ (0.2cm,-0.125cm) circle (0.125cm)  (M6.south east) ++ (-0.2cm,-0.125cm) circle (0.125cm);
		
		\node (ground4) [ground,anchor=north,yshift=-0.25cm,minimum width=4.8cm] at (M4.south) {};
		\draw (ground4.north east) -- (ground4.north west);
		\draw [thick, fill={cyan}] (M4.south west) ++ (0.2cm,-0.125cm) circle (0.125cm)  (M4.south east) ++ (-0.2cm,-0.125cm) circle (0.125cm);
		
		\node (wall1) [ground, rotate=-90, minimum width=2cm,yshift=-2.5cm] {};
		\draw (wall1.north east) -- (wall1.north west);
		
		\node (wall2) [ground, xshift=7.5cm, yshift=1.5cm, rotate=90, minimum width=2cm,yshift=-2.5cm] {};
		\draw (wall2.north east) -- (wall2.north west);
		
		\node (wall3) [ground, xshift=7.5cm, yshift=-1.5cm, rotate=90, minimum width=2cm,yshift=-2.5cm] {};
		\draw (wall3.north east) -- (wall3.north west);
		
		\draw [spring] ($(wall1.90)+ (0,0.35)$) -- ($(M1.west)+ (0,0.35)$)
		node [midway,above] {$k_1$};
		\draw [damper] ($(wall1.90) - (0,0.35)$) -- ($(M1.west) - (0,0.35)$)
		node [midway,above=3] {$c_1$};
		
		\draw[spring] ($(M1.east) + (0,1.85)$) -- ($(M2.west)+(0,0.35)$) 
		node [midway,above] {$k_2$};
		\draw[damper] ($(M1.east) + (0,1.15)$) -- ($(M2.west)-(0,0.35)$) 
		node [midway,above=3] {$c_2$};
		
		\draw [spring] ($(M1.east) + (0,-1.15)$) -- ($(M3.west)+(0,0.35)$)
		node [midway,above] {$k_3$};
		\draw [damper] ($(M1.east) + (0,-1.85)$) -- ($(M3.west)-(0,0.35)$)
		node [midway,above=3] {$c_3$};
		
		\draw [spring] ($(M1.east)+(0,0.35)$) -- ($(M6.west)+(0,0.35)$)
		node [midway,above] {$k_4$};
		\draw [damper] ($(M1.east) - (0,0.35)$) -- ($(M6.west)-(0,0.35)$)
		node [midway,above=3] {$c_4$};

		\draw [spring] ($(M2.east)+ (0,0.35)$) -- ($(M6.west) + (0,1.85)$)
		node [midway,above] {$k_5$};
		\draw [damper] ($(M2.east)-(0,0.35)$) -- ($(M6.west) + (0,1.15)$)
		node [midway,above=3] {$c_5$};
		
		\draw [spring] ($(M3.east)+(0,0.35)$) -- ($(M6.west) - (0,1.15)$)
		node [midway,above] {$k_6$};
		\draw [damper] ($(M3.east)-(0,0.35)$) -- ($(M6.west) - (0,1.85)$)
		node [midway,above=3] {$c_6$};
		
		\draw [spring] ($(M6.east) + (0,-1.15)$) -- ($(M4.west)+(0,0.35)$)
		node [midway,above] {$k_7$};
		\draw [damper] ($(M6.east) + (0,-1.85)$) -- ($(M4.west)-(0,0.35)$)
		node [midway,above=3] {$c_7$};
		
		\draw [spring] ($(M6.east) + (0,1.85)$) -- ($(M5.west)+(0,0.35)$)
		node [midway,above] {$k_8$};
		\draw [damper] ($(M6.east) + (0,1.15)$) -- ($(M5.west)-(0,0.35)$)
		node [midway,above=3] {$c_8$};
		
		\draw [spring] ($(M5.east)+(0,0.35)$) -- ($(wall2.90)+(0,0.35)$)
		node [midway,above] {$k_9$};
		\draw [damper] ($(M5.east)-(0,0.35)$) -- ($(wall2.90)-(0,0.35)$)
		node [midway,above=3] {$c_9$};
		
		\draw [spring] ($(M4.east)+(0,0.35)$) -- ($(wall3.90)+(0,0.35)$)
		node [midway,above] {$k_{10}$};
		\draw [damper] ($(M4.east)-(0,0.35)$) -- ($(wall3.90)-(0,0.35)$)
		node [midway,above=3] {$c_{10}$};
		
		\draw[ultra thick, -latex] ($(M6.east)$) -- ($(M6.east)+(1.5,0) $)
		node [midway, below] {$f$};
		\end{tikzpicture}
	\end{adjustbox}
	\caption{The 6-DOF system}
	\label{fig:6DOF}
\end{figure}
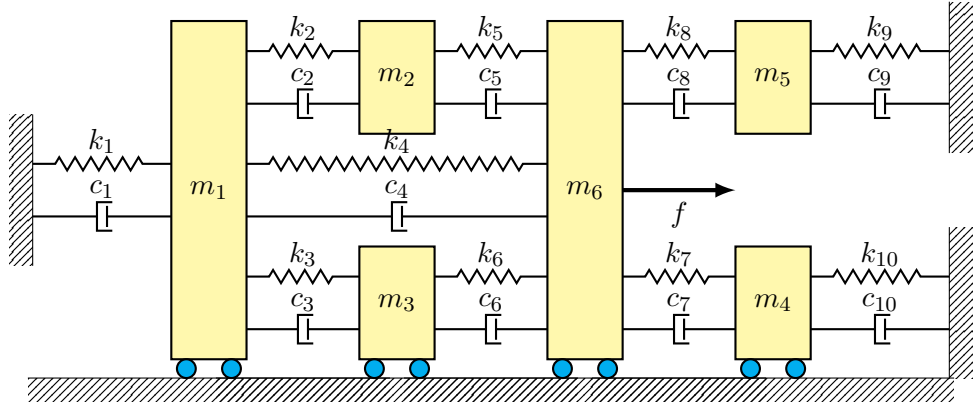

The second example is chosen to illustrate the application of the proposed method to a problem with a relatively large parameter space. The system, shown in Figure \ref{fig:6DOF}, consists of 10 springs and 6 masses which are modeled by random variables with lognormal distributions. Their mean values are listed in Table \ref{tab:6DOF:variables}. The uncertainty on the springs (resp. masses) has a COV = 10\% (resp. COV = 5\%). The damping matrix is $\ve{V}=0.1\widehat{\ve{M}}$,
where $\widehat{\ve{M}}$ is the matrix of the mean value of the system masses. 
Table \ref{tab:6DOF:damping} provides its corresponding mean of modal dampings evaluated over 10,000 samples.

The system has one input force at mass 6 and 6 system outputs, one for each mass. The FRF of the system is evaluated at a frequency range from 1 to 25 rad/s with the step of 0.01$\pi$ rad/s.

In this example, the ED consists of 400 points sampled from the parameter space using LHS. 
The marginal distributions of the input vectors $\ve{X}$ consists of lognormal distributions. Therefore, the chosen PCE basis consists of Hermite polynomials on the reduced variable ${\ve{Z}}=\ln({\ve{X}})$. Eq. (\ref{eq:PCE:infinite}) thus, can be written as $$\ve{Y}=\cm(\ve{X} )=\sum_{\ua \in \ca^{M,p}} {{{\tilde{u}}_{\ua}}} \, {{\psi}_{\ve{\alpha}}(\ln({\ve{X}}))}.$$  

The LARS algorithm has been employed to build sparse PCEs with adaptive degree for both the \textit{selected frequencies} and the principal components of the scaled FRF.

For the second set of PCEs, PCA has been performed and the dominant components are selected such that $\sum_{i=1}^{\hat{N}}\lambda_i = 0.999 \sum_{i=1}^{{N}}\lambda_i$. This truncation reduced the number of random outputs from 761 $\times$ 6 $\times$ 2 to 102 components. Since the dimension of the input parameter space is large, to reduce the unknown coefficients of the PCEs and avoid the curse of dimensionality, a hyperbolic truncation with $q$-norm of 0.7 was used before the LARS algorithm. Besides, only polynomials up to rank 2 were selected here (\ie polynomials that depend at most on 2 of the 16 parameters). It should be mentioned that all the PCEs used for the surrogate model have eventually maximum degrees less than 10.



\begin{table} [H]
	\centering
	\caption{The 6-DOF system's variables}
	\begin{tabular}{lccc}
		\hline
		\multicolumn{2}{r}{Variables} & mean & Coeff. of variation (\%) \\
		\hline
		\multirow{6}{*}{Masses (Kg)}  & $m_1$ & 50 & 5 \\
		& $m_2$ & 35 & 5 \\
		& $m_3$ & 12 & 5 \\
		& $m_4 $& 33 & 5 \\
		& $m_5$ & 100 & 5 \\
		& $m_6$ & 45 & 5 \\
		\hline
		\multirow{10}{*}{Stiffnesses (N/m)} & $k_1$ & 3000 & 10 \\
		& $k_2$ & 1725 & 10 \\
		& $k_3$ & 1200 & 10 \\
		& $k_4$ & 2200 & 10 \\
		& $k_5$ & 1320 & 10 \\
		& $k_6$ & 1330 & 10 \\
		& $k_7$ & 1500 & 10 \\
		& $k_8$ & 2625 & 10 \\
		& $k_9$ & 1800 & 10 \\
		& $k_{10}$ & 850 & 10 \\
		\hline 
	\end{tabular}
	\label{tab:6DOF:variables}
\end{table}



\begin{table} [H]
	\centering
	\caption{Mean of modal dampings of the 6-DOF system evaluated over 10,000 samples}
	\begin{tabular}{l|cccccc}
		\hline
		\multicolumn{1}{c|}{\multirow{2}{*}{Damping} }
		& \multicolumn{6}{c}{Mean of modal dampings (\%)}  \\  
		\cline{2-7}
		\multicolumn{1}{c|}{} & $1^{st} $ & $2^{nd}$ & $3^{rd}$ & $4^{th}$ & $5^{th}$ & $6^{th}$\\
		\hline
		$\ve{V}=0.1\widehat{\ve{M}}$ & $1.30$ & $0.72$ & $0.52$ & $0.44$ & $0.33$ & $0.30$ \\
		\hline			
	\end{tabular}
	\label{tab:6DOF:damping}
\end{table}

\begin{figure}[H]
	\centering
	\begin{subfigure}[b]{.5\columnwidth}
		\centering
		\includegraphics[width=1\columnwidth]{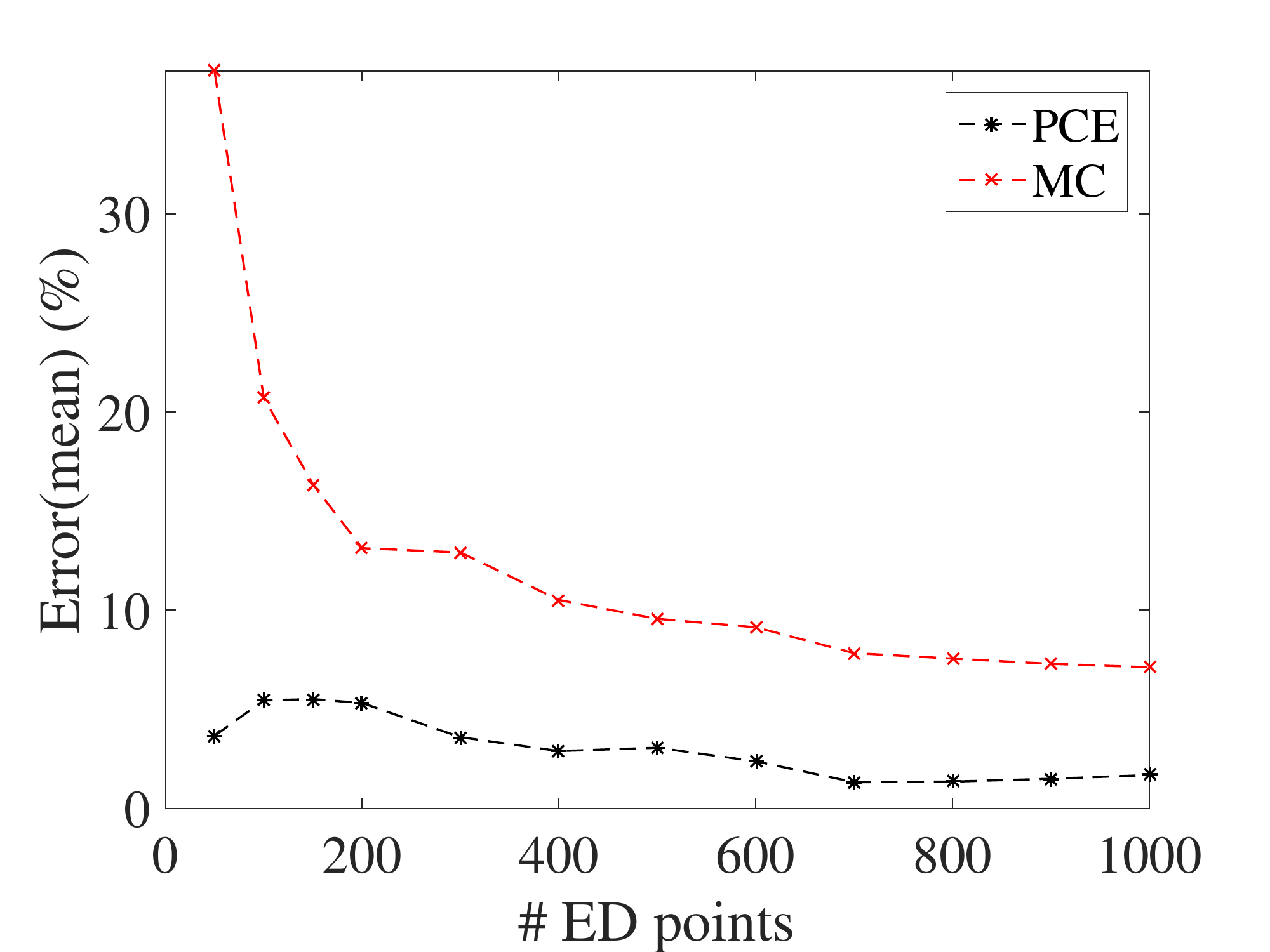}
		\caption{Mean}
		\label{fig:6DOF:FRF:mean:conv:out6:p1D}
	\end{subfigure}
	\begin{subfigure}[b]{.5\columnwidth}
		\centering
		\includegraphics[width=1\columnwidth]{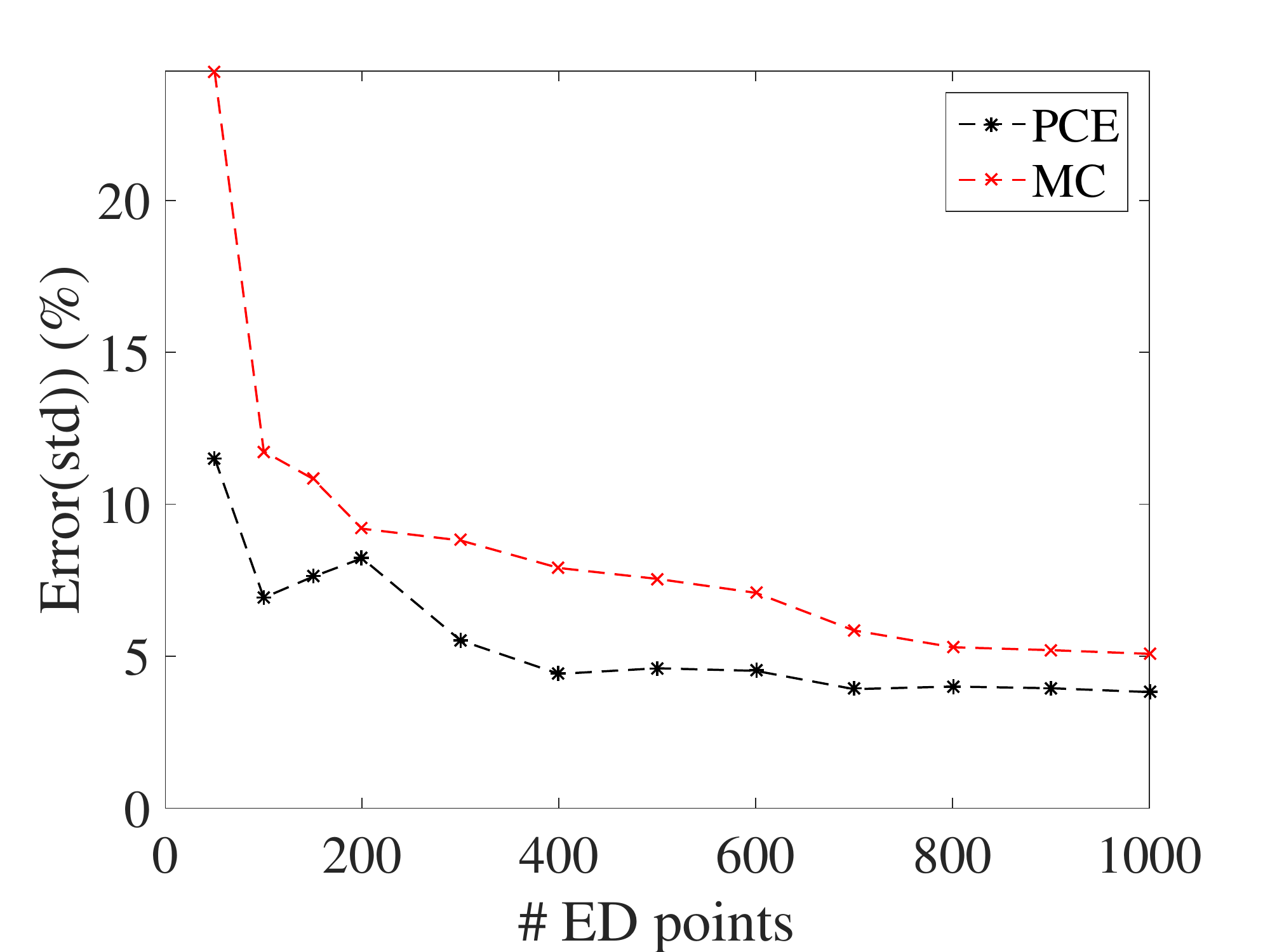}
		\caption{Standard deviation}
		\label{fig:6DOF:FRF:std:conv:out6:p1D}
	\end{subfigure}
	\caption{Convergence plot of the first two moments of the FRFs of the 6-DOF system calculated at $6^{th}$ mass by the PCE (black $\ast$) and the true model (red $\times$) by enlarging the experimental design. The reference results were obtained by 10,000 Monte-Carlo simulation of the true model.}
	\label{fig:6DOF:conv:p1D}
\end{figure}

The efficiency of the proposed method is assessed by comparing the PCE estimates of the first two moments of the surrogate model with the plain Monte-Carlo estimators on experimental designs of increasing size. The reference validation set is obtained by 10,000 points sampled from the parameter space by LHS at which the full model is evaluated. 
The results are shown in Figure \ref{fig:6DOF:conv:p1D} for the mean and standard deviation at $6^{th}$ output. The results for the other outputs are presented in \ref{app:6DOF:conv}.
They indicate that both mean and standard deviation evaluated by the surrogate model converge faster than those of the Monte-Carlo simulations. Besides, it can be inferred that 400 points are enough for the ED. 

\begin{figure}[H]
	\centering
	\begin{subfigure}[b]{.5\columnwidth}
		\centering
		\includegraphics[width=1\columnwidth]{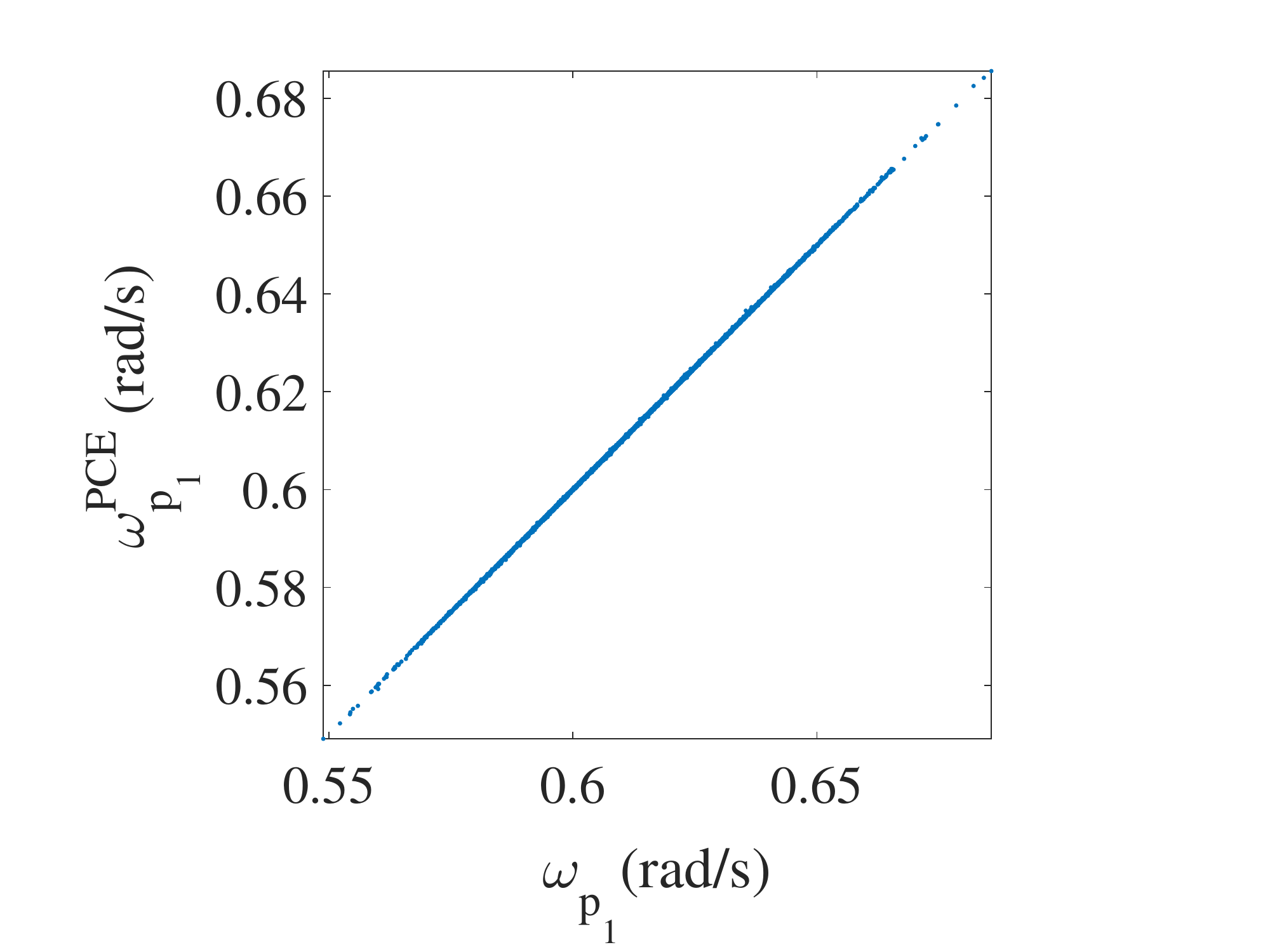}
		\caption{The best predicted eigenfrequency}
		\label{fig:6DOF:eigenvalue:1st:p1D}
	\end{subfigure}
	\begin{subfigure}[b]{.5\columnwidth}
		\centering
		\includegraphics[width=1\columnwidth]{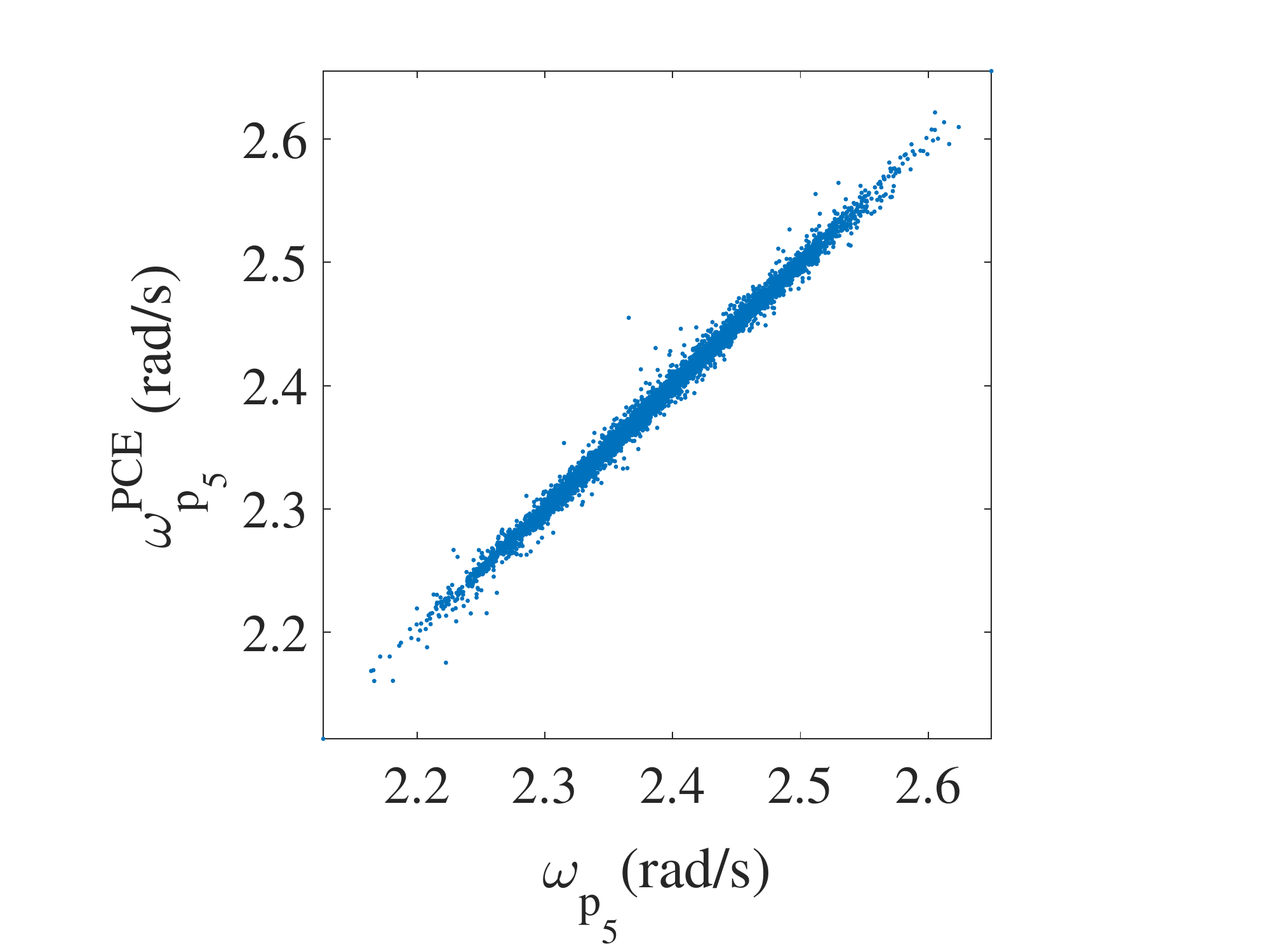}
		\caption{The worst predicted eigenfrequency}
		\label{fig:6DOF:eigenvalue:5st:p1D}
	\end{subfigure}
	\caption{The eigenfrequencies predicted by the surrogate model versus obtained by the true model, evaluated at 10,000 MC samples.} 
	\label{fig:6DOF:eigval:PCEvsMC:p1D}
\end{figure}

In order to assess the accuracy of the surrogate model in estimating various quantities of interests, the same 10,000 points used as the reference validation set to study the convergence are used here.  
Figure \ref{fig:6DOF:eigval:PCEvsMC:p1D} illustrates two of the predicted \textit{selected frequencies} versus the true ones, namely the best and worst predicted eigenfrequency, so that the accuracy of the surrogate model in this step can be inferred. While the overall accuracy is very good for all frequencies, it tends to degrade somewhat at higher frequencies. 

\begin{figure}[H]
	\centering
	\begin{subfigure}[b]{.5\columnwidth}
		\centering
		\includegraphics[width=1\columnwidth]{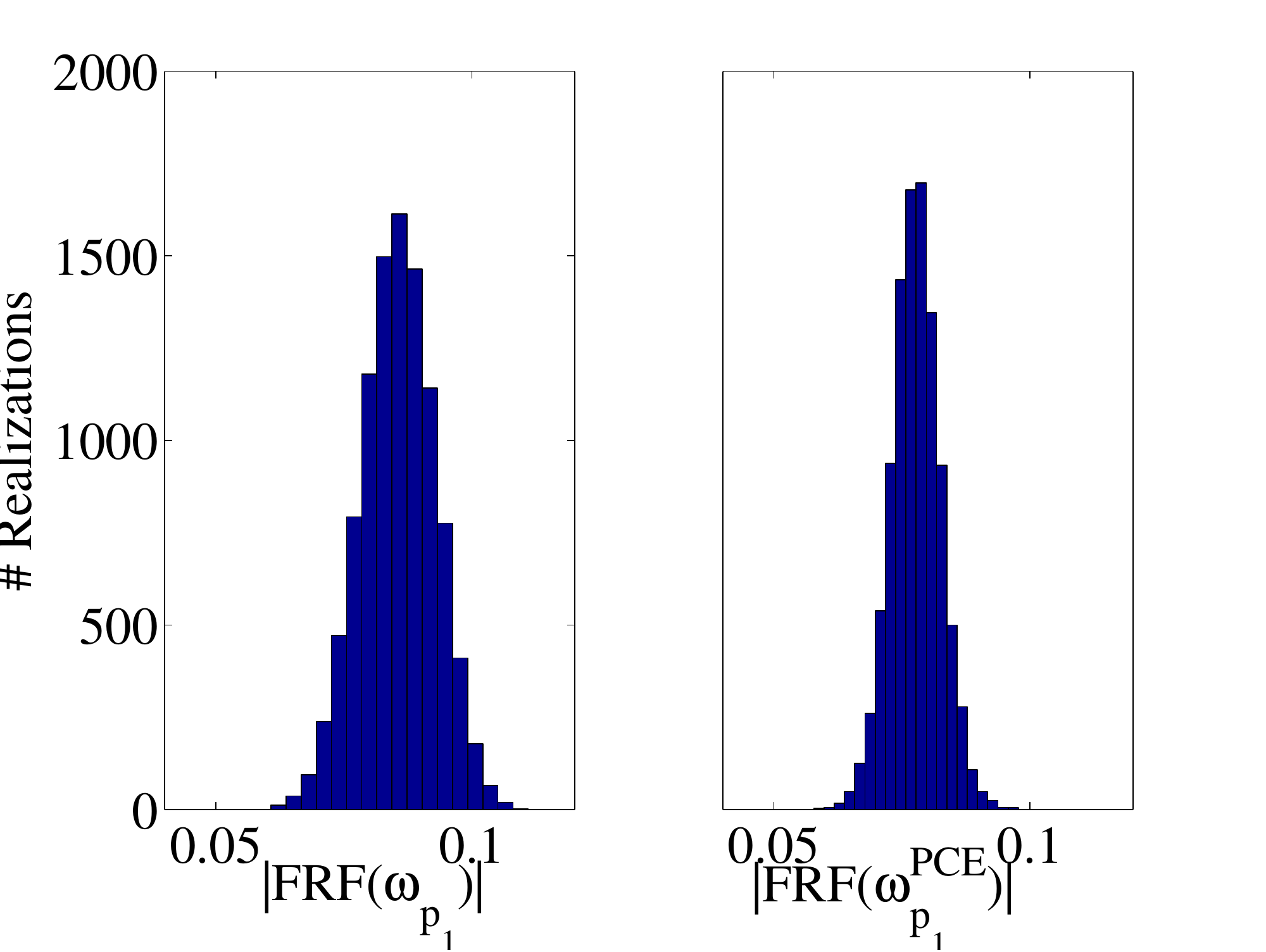}
		\caption{First resonant frequency}
		\label{fig:6DOF:Amp:freq1}
	\end{subfigure}
	\begin{subfigure}[b]{.5\columnwidth}
		\centering
		\includegraphics[width=1\columnwidth]{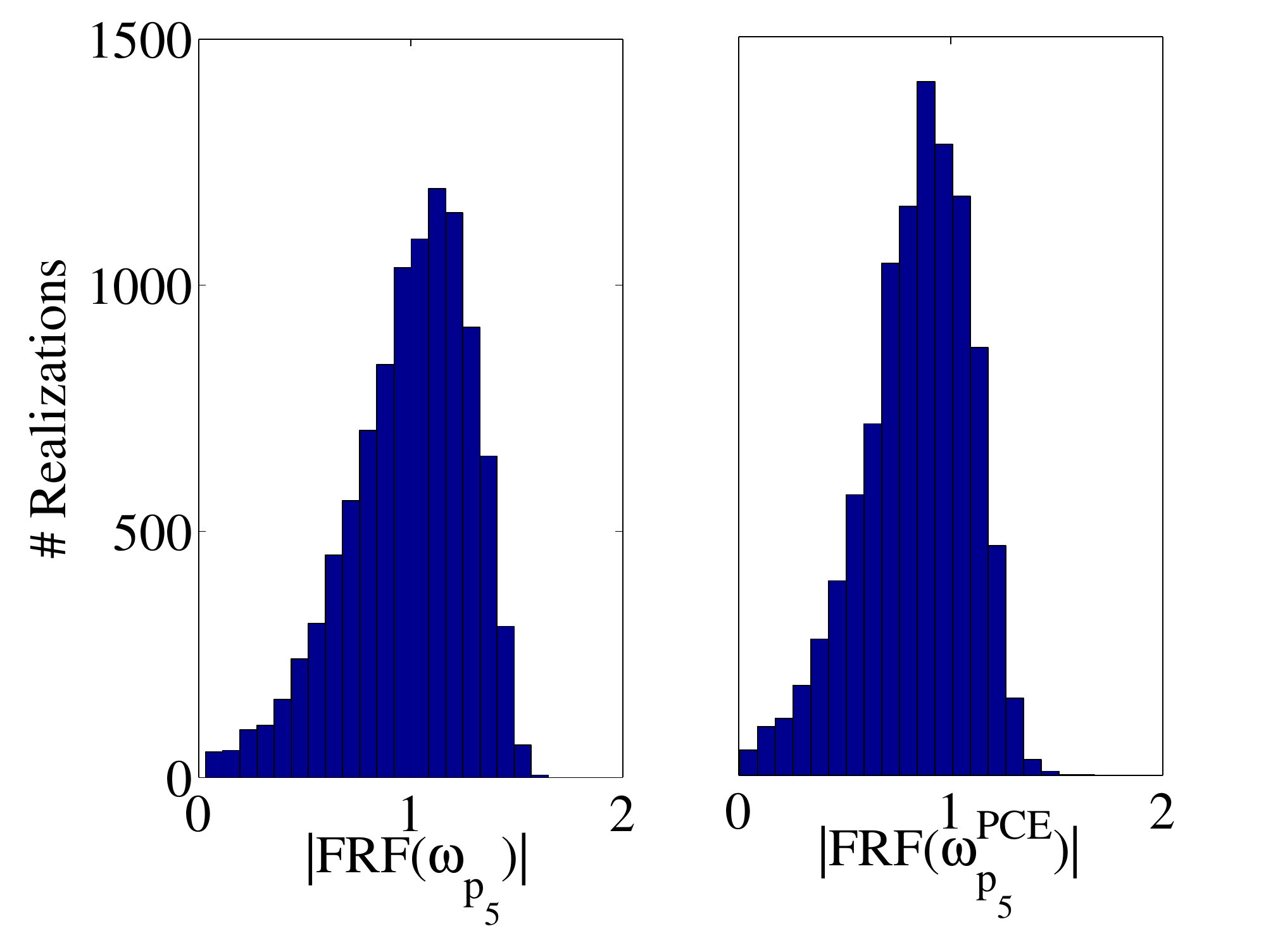}
		\caption{Fifth resonant frequency}
		\label{fig:6DOF:Amp:freq6}
	\end{subfigure}
	\caption{Histogram of the amplitude of the FRF at the first and fifth resonant frequencies, obtained by evaluating the true and surrogate models on the 10,000 MC samples.}
	\label{fig:6DOF:Amp:freqs}
\end{figure}

\begin{figure}[H]
	\centering 
	\begin{subfigure}[b]{.5\columnwidth}
		\centering
		\includegraphics[width=1\columnwidth]{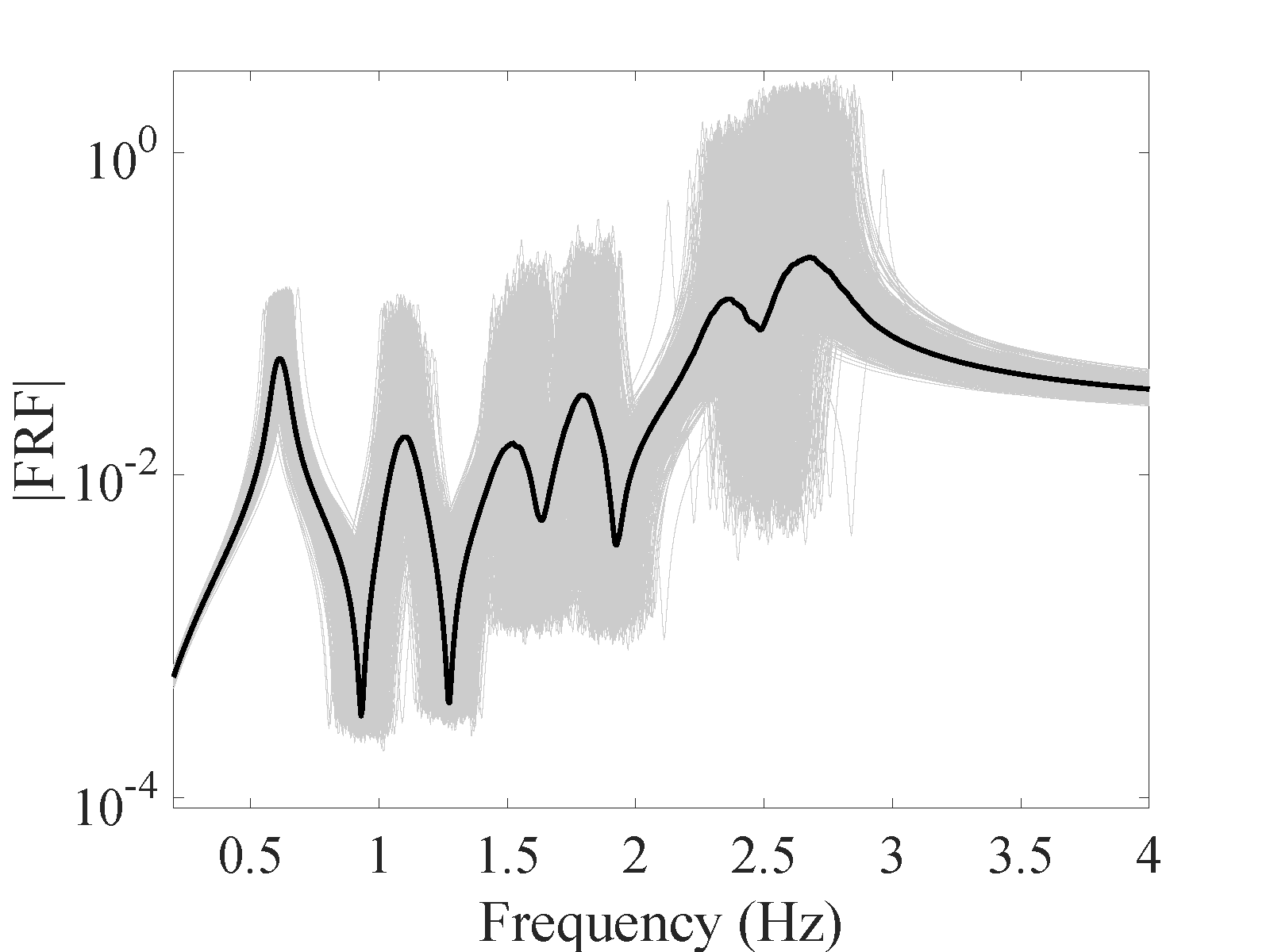}
		\caption{$6^{th}$ system output- True model}
		\label{fig:6DOF:envelope:FRF:out6}
	\end{subfigure}
	\begin{subfigure}[b]{.5\columnwidth}
		\centering
		\includegraphics[width=1\columnwidth]{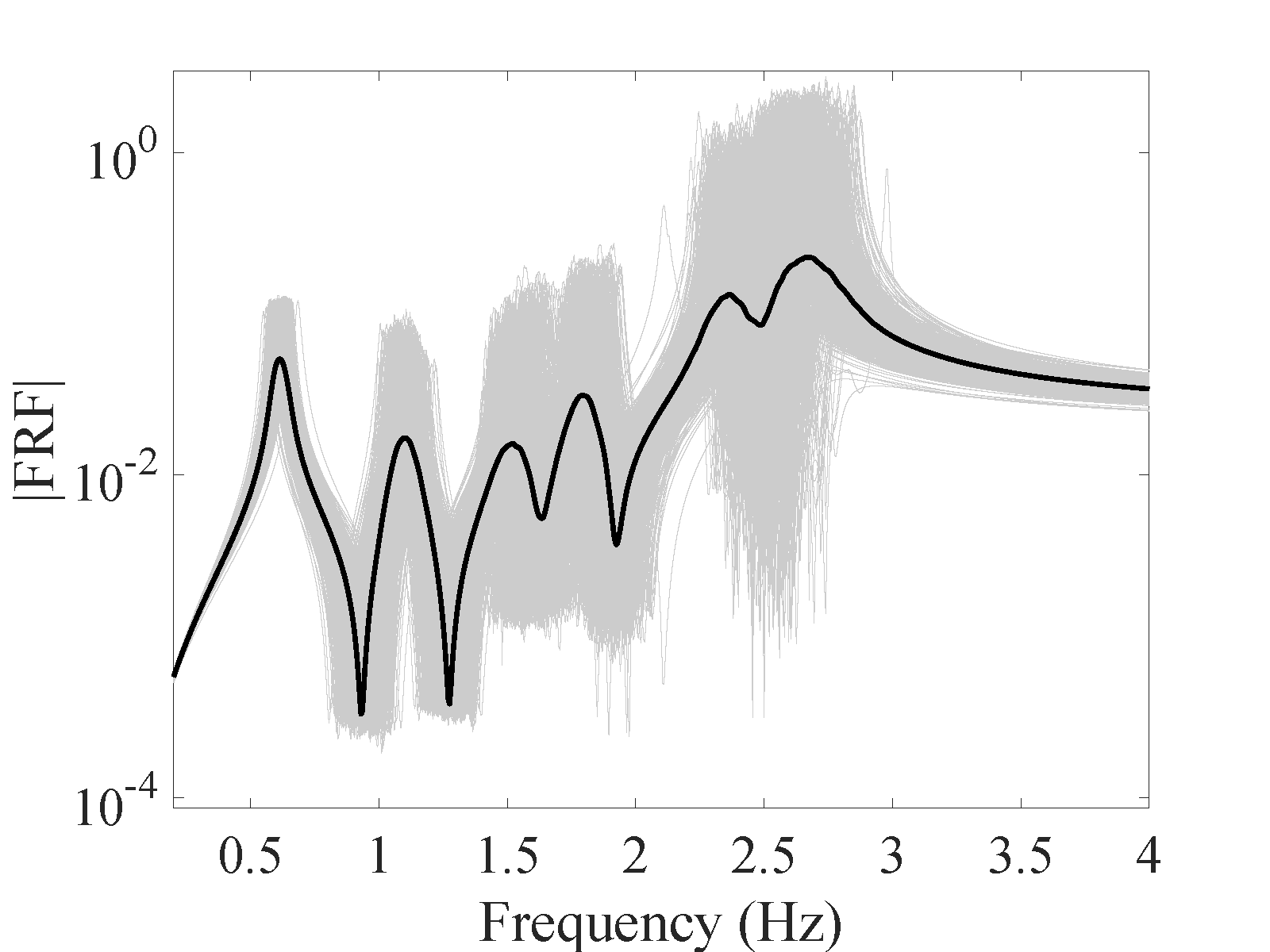}
		\caption{$6^{th}$ system output- Surrogate model}
		\label{fig:6DOF:envelope:FRF:out6:PCE}
	\end{subfigure}
	\caption{FRFs at $6^{th}$ mass obtained by evaluating the true and the surrogate model at 10,000 MC samples.} 
	\label{fig:FRF:6DOF:envelope}
\end{figure}

\begin{figure}[H]
	\centering
	\begin{subfigure}[b]{0.5\columnwidth}
		\centering
		\includegraphics[width=1\columnwidth]{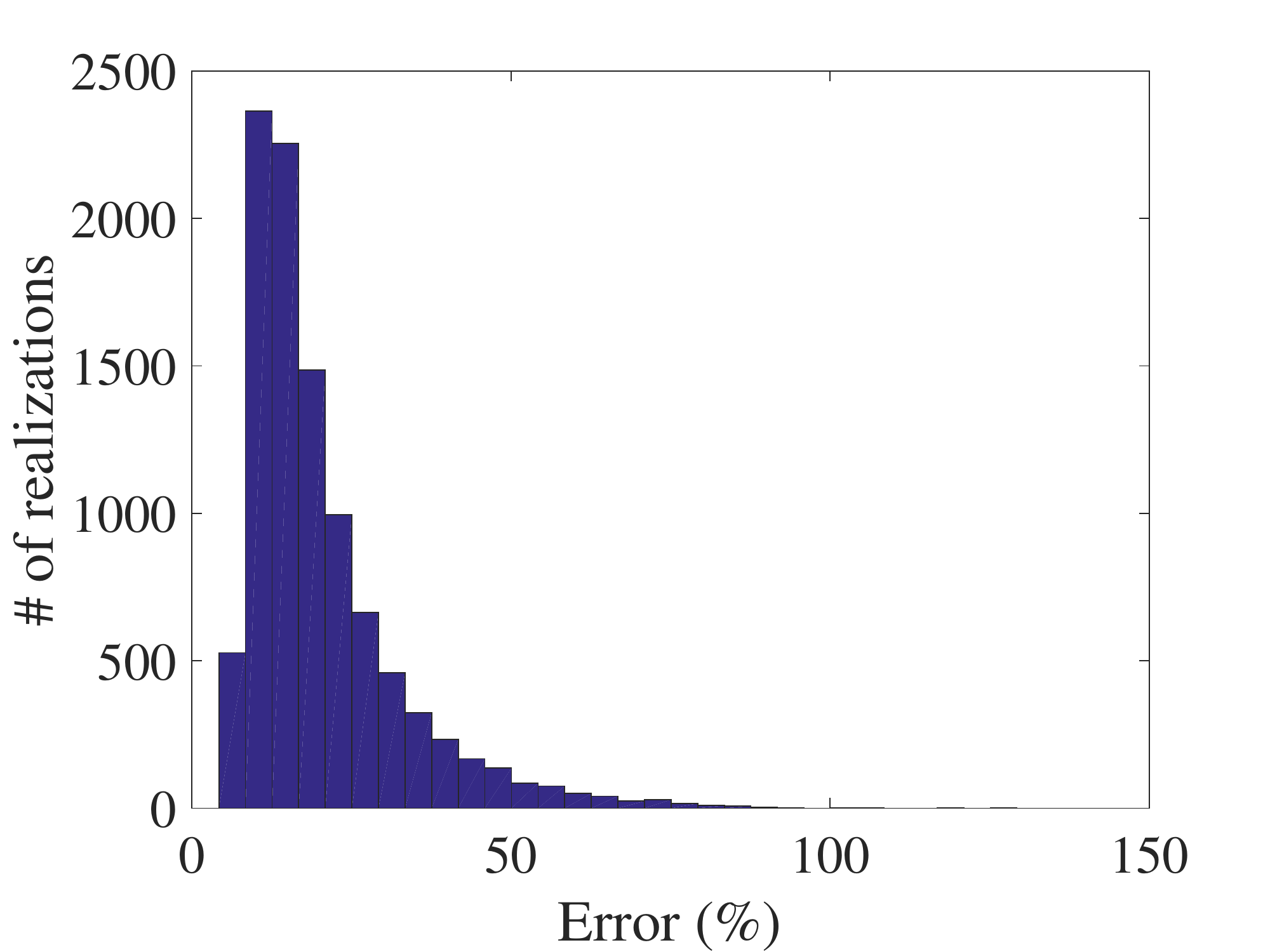}
		\caption{First output}
		\label{fig:6DOF:FRF:error:out1:p1D}
	\end{subfigure}
	\begin{subfigure}[b]{0.5\columnwidth}
		
		\includegraphics[width=1\columnwidth]{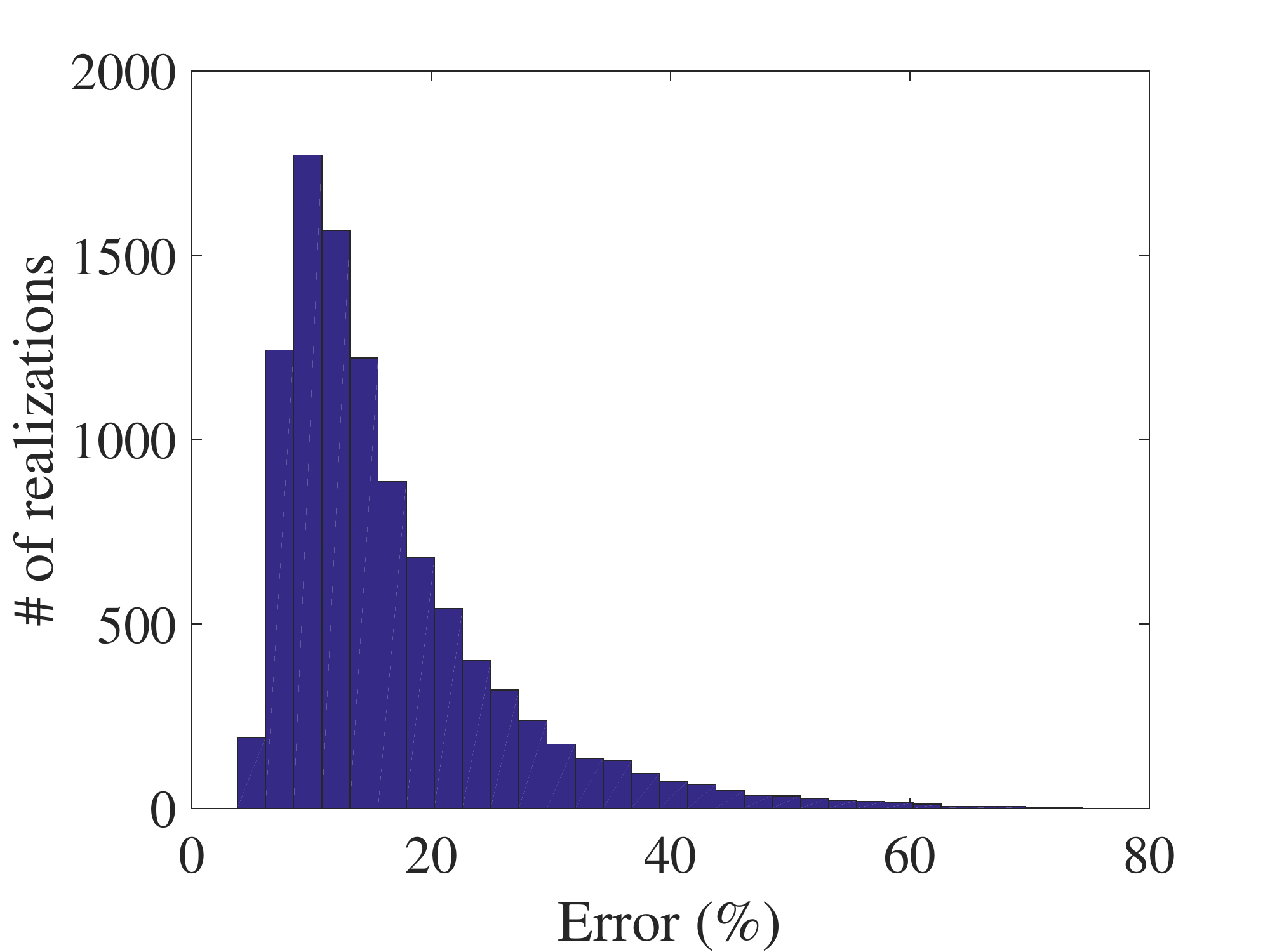}
		\caption{Sixth output}
		\label{fig:6DOF:FRF:error:out6:p1D}
	\end{subfigure}
	\caption{Error of the FRFs predicted by the surrogate model, evaluated at 10,000 MC samples by Eq. (\ref{eq:Err:rms}).}
	\label{fig:6DOF:error:frf:p1D}
\end{figure}

Besides, at all the validation points the FRFs are calculated by both the true model and the surrogate model. The variation of the amplitudes at the first and fifth resonant frequencies are shown as histograms in Figure \ref{fig:6DOF:Amp:freqs}. Plots of the individual FRFs are reported in Figure  \ref{fig:FRF:6DOF:envelope} for $6^{th}$ output. In order to assess the error quantitatively, each response of the surrogate model has been compared with the corresponding one of the true model in the root-mean-square sense. This error is evaluated using Eq. (\ref{eq:Err:rms}) and the corresponding results are presented in Figure \ref{fig:6DOF:error:frf:p1D}. They indicate the high accuracy of the proposed surrogate model in predicting the FRFs.

\begin{figure}[H]
	\centering
	\begin{subfigure}[b]{.5\columnwidth}
		\centering
		\includegraphics[width=1\columnwidth]{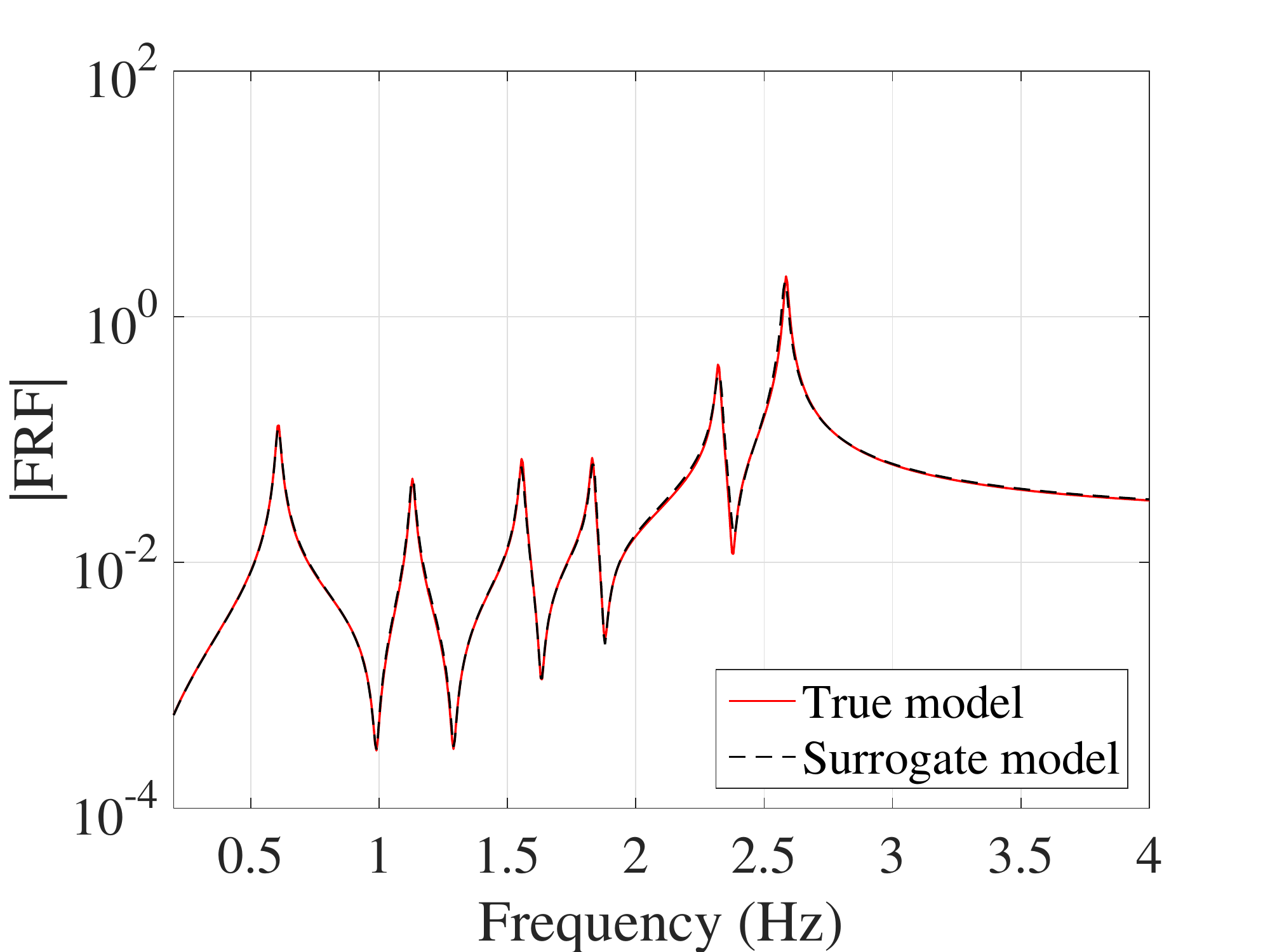}
		\caption{Typical FRF prediction}
		\label{fig:6DOF:FRF:typ:out1}
	\end{subfigure}
	\begin{subfigure}[b]{.5\columnwidth}
		\centering
		\includegraphics[width=1\columnwidth]{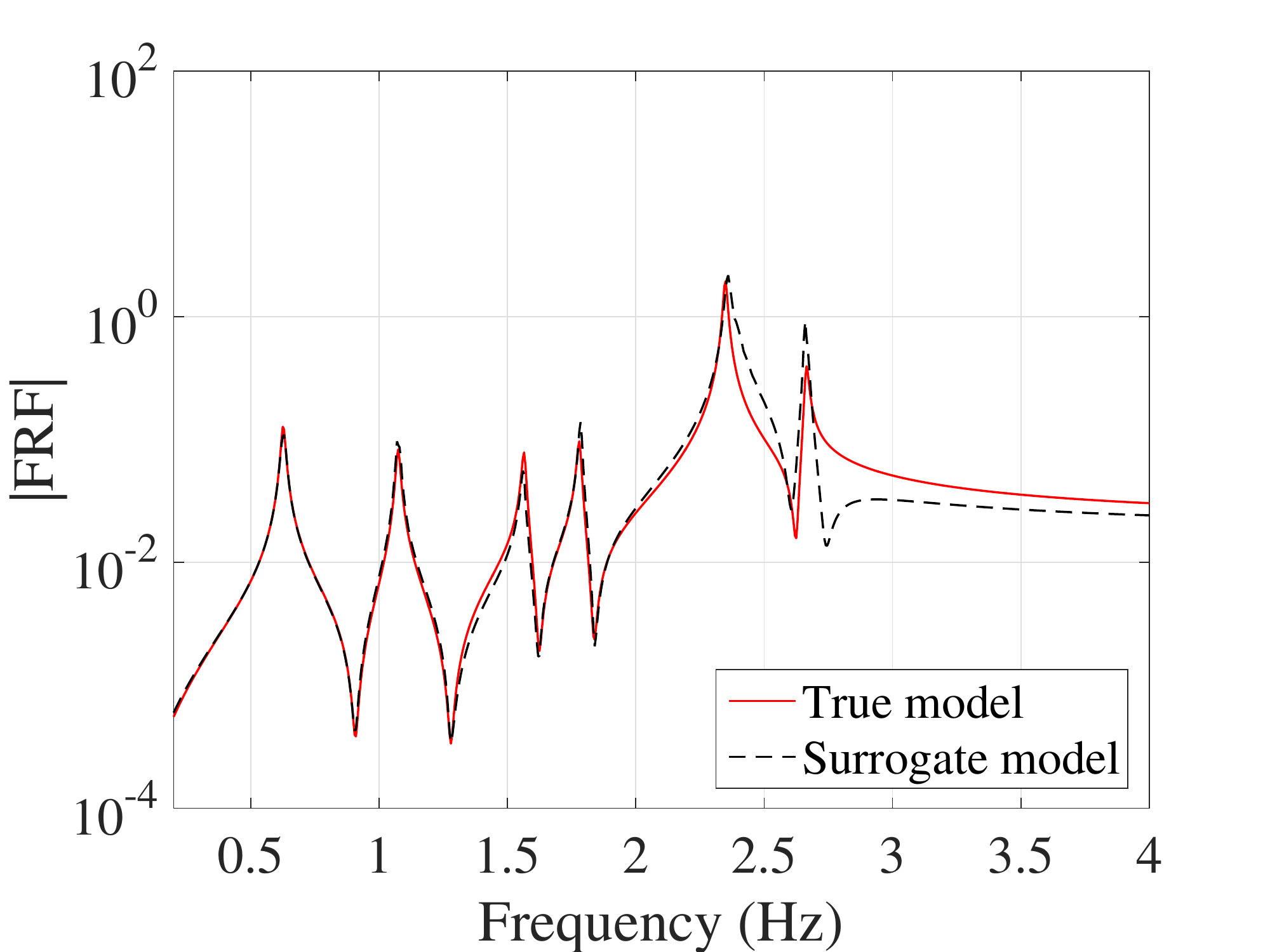}
		\caption{Worst FRF prediction}
		\label{fig:6DOF:FRF:typ:out6}
	\end{subfigure}
	\caption{Two samples of the FRFs predicted by the surrogate model at $6^{th}$ output, evaluated by the true model (red line) and the surrogate model (black line).}
	\label{fig:FRF:6DOF:predict:individual:p1D}
\end{figure}

As an individual comparison between the true FRFs and predicted by the surrogate model, two cases are considered: one case with an average and one with the maximum overall error, are selected and their $6^{th}$ output are demonstrated in Figure \ref{fig:FRF:6DOF:predict:individual:p1D}. The other outputs are presented in \ref{app:6DOF:individual:FRF}.

\begin{figure}[H]
	\centering
	\begin{subfigure}[b]{.5\columnwidth}
		\centering
		\includegraphics[width=1\columnwidth]{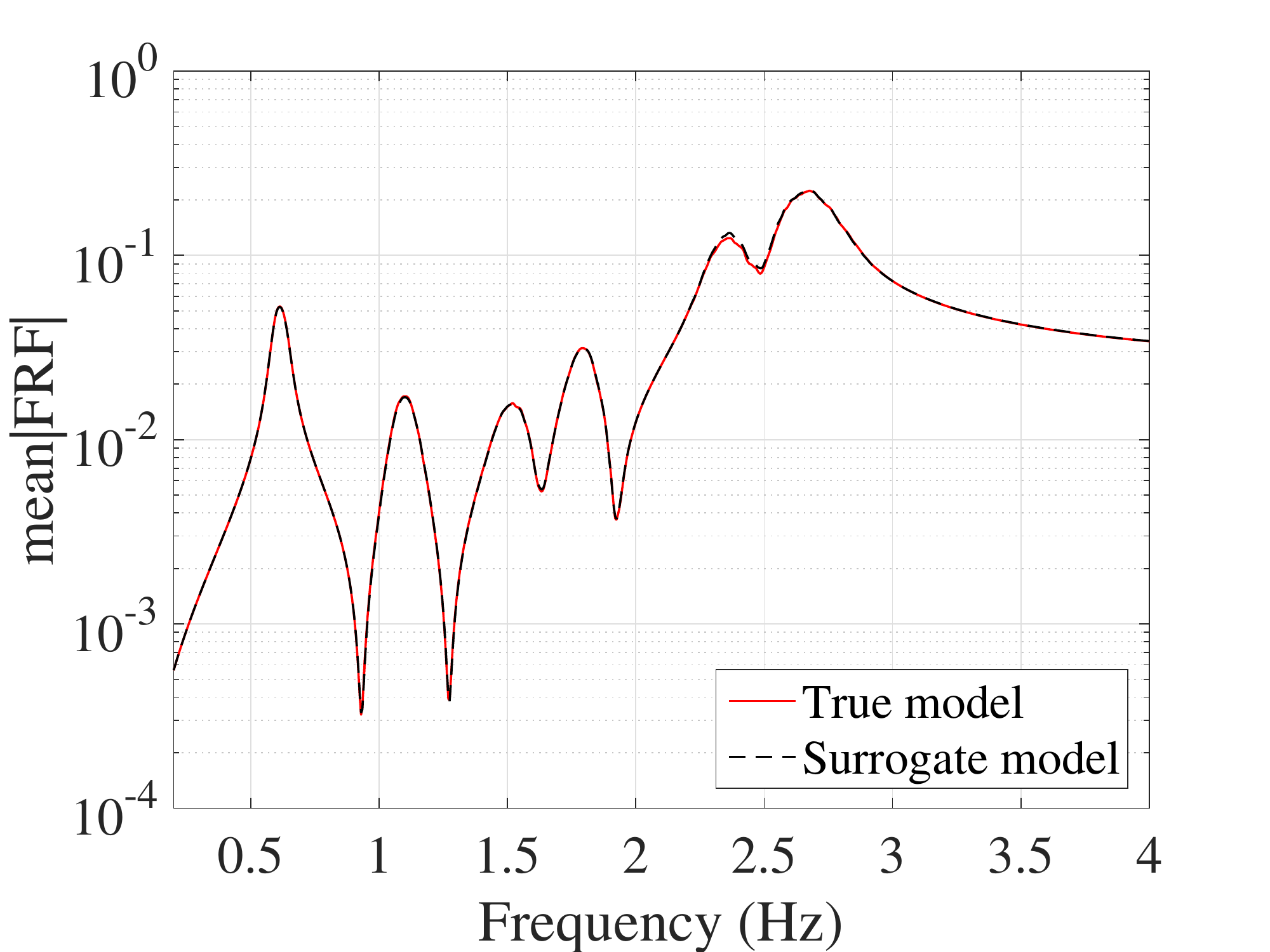}
		\caption{Magnitude of the mean}
		\label{fig:6DOF:FRF:mean:out6:abs}
	\end{subfigure}
	\begin{subfigure}[b]{.5\columnwidth}
		\centering
		\includegraphics[width=1\columnwidth]{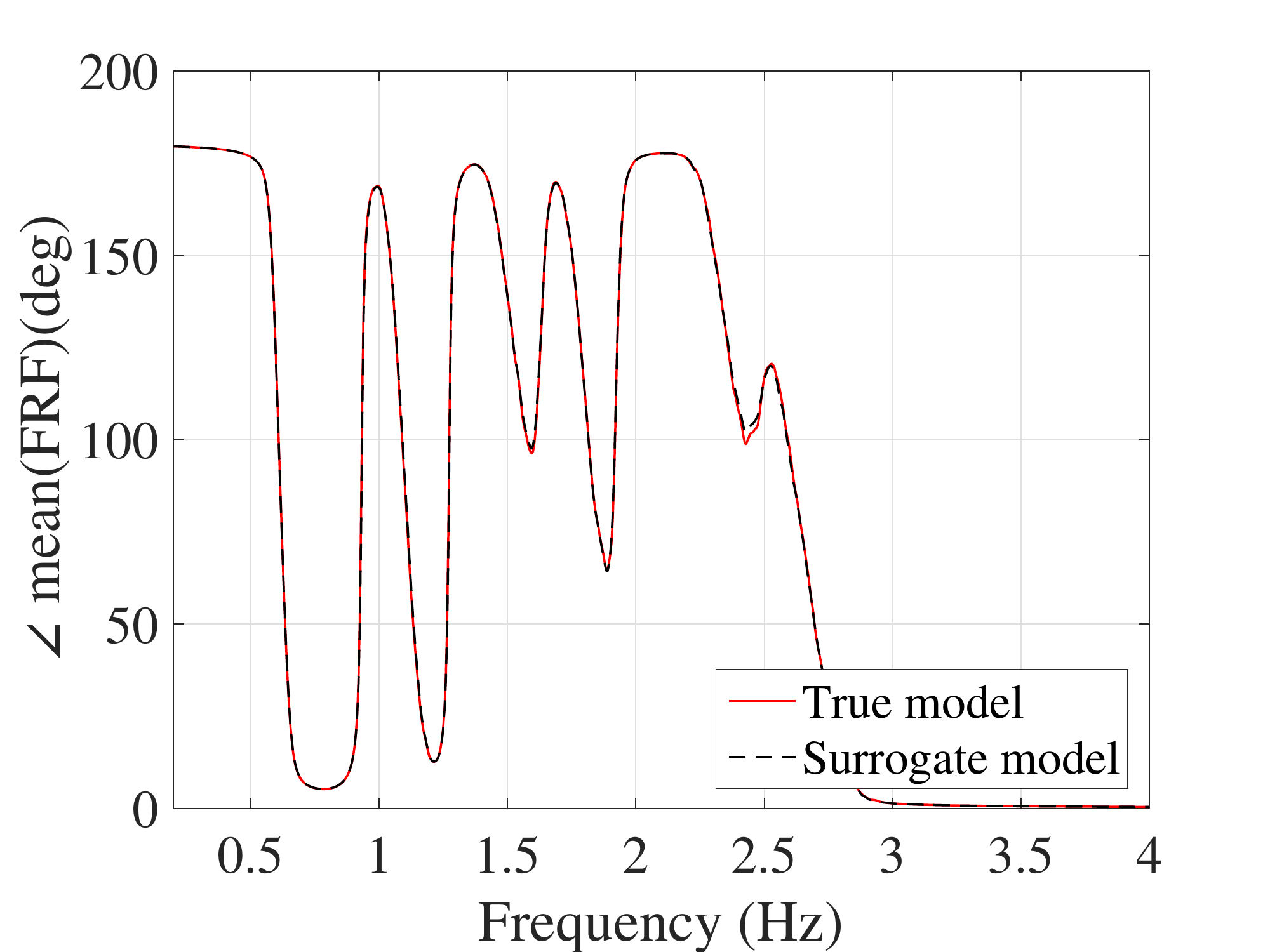}
		\caption{Phase of the mean}
		\label{fig:6DOF:FRF:mean:out6:phase}
	\end{subfigure}
	\\
	\begin{subfigure}[b]{0.5\columnwidth}
		\centering
		\includegraphics[width=1\columnwidth]{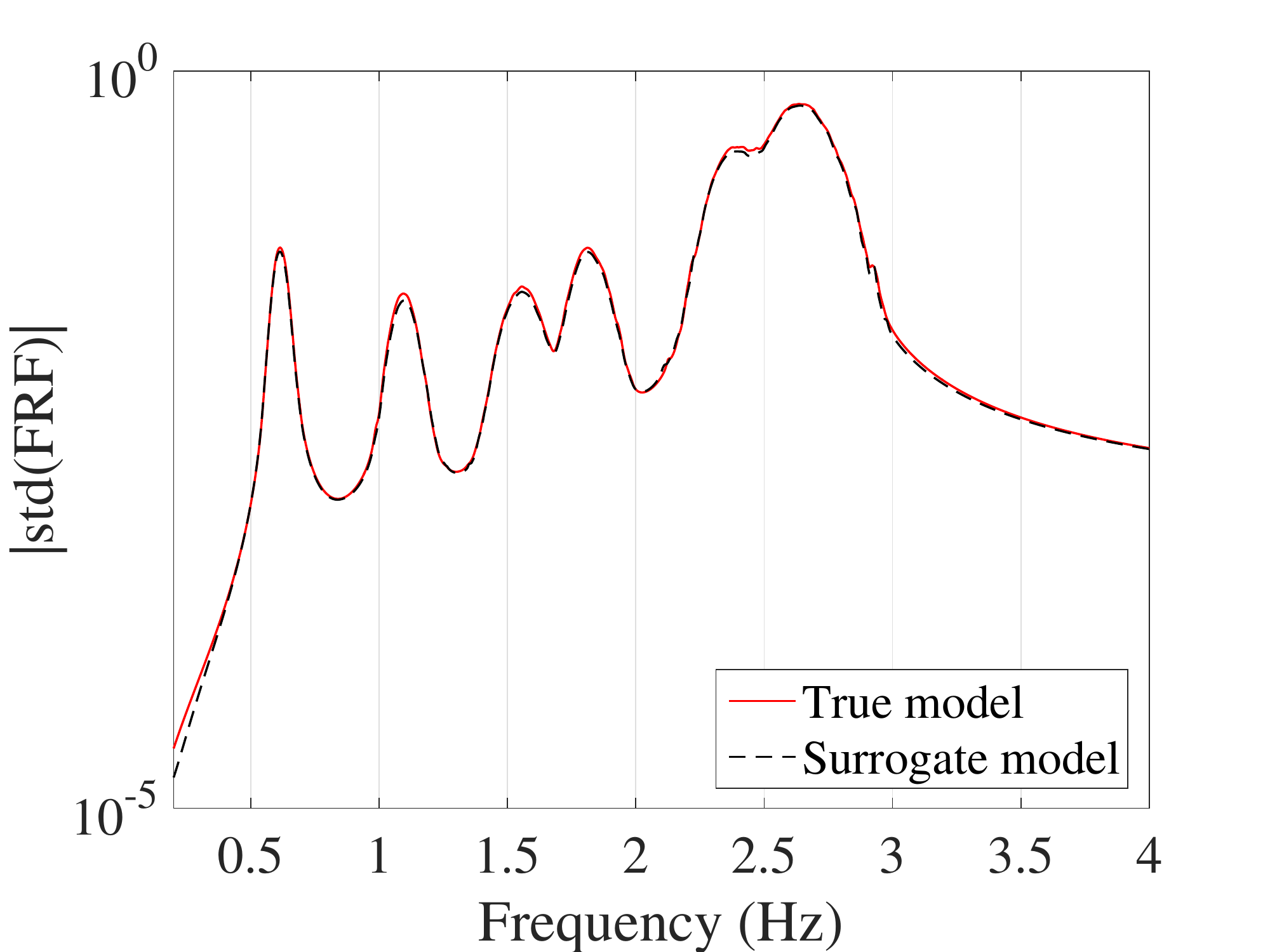}
		\caption{Magnitude of the standard deviation}
		\label{fig:6DOF:FRF:std:out6:abs}
	\end{subfigure}
	\begin{subfigure}[b]{0.5\columnwidth}
		\centering
		\includegraphics[width=1\columnwidth]{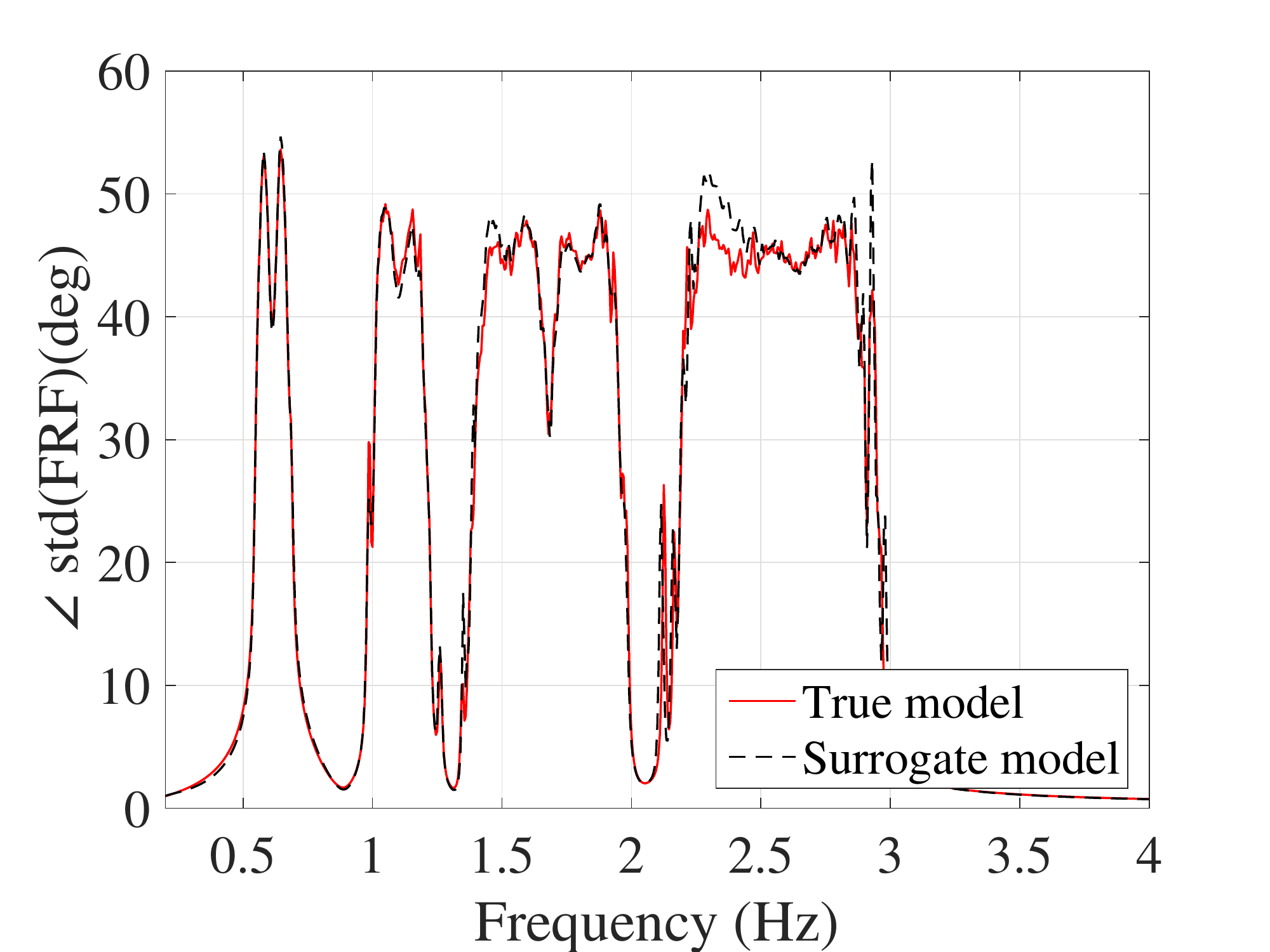}
		\caption{Phase of the standard deviation}
		\label{fig:6DOF:FRF:std:out6:phase}
	\end{subfigure}
	\caption{Mean and standard deviation of the FRF of the 6-DOF system at $6^{th}$ output, evaluated at 10,000 MC sample points by the true model (red line) and the surrogate model (black line).}
	\label{fig:6DOF:frf:stat:out6}
\end{figure}

The mean and standard deviations of the FRFs were compared with the reference ones. The results for $6^{th}$ output are plotted in Figure \ref{fig:6DOF:frf:stat:out6}. The other outputs are plotted in \ref{app:6DOF:moments:FRF}. They indicate that although both the mean and the standard deviation match with their reference, the standard deviation presents minor mismatch at the peaks.



\begin{figure}[H]
	\centering
	\begin{subfigure}[b]{.5\columnwidth}
		\centering
		\includegraphics[width=1\columnwidth]{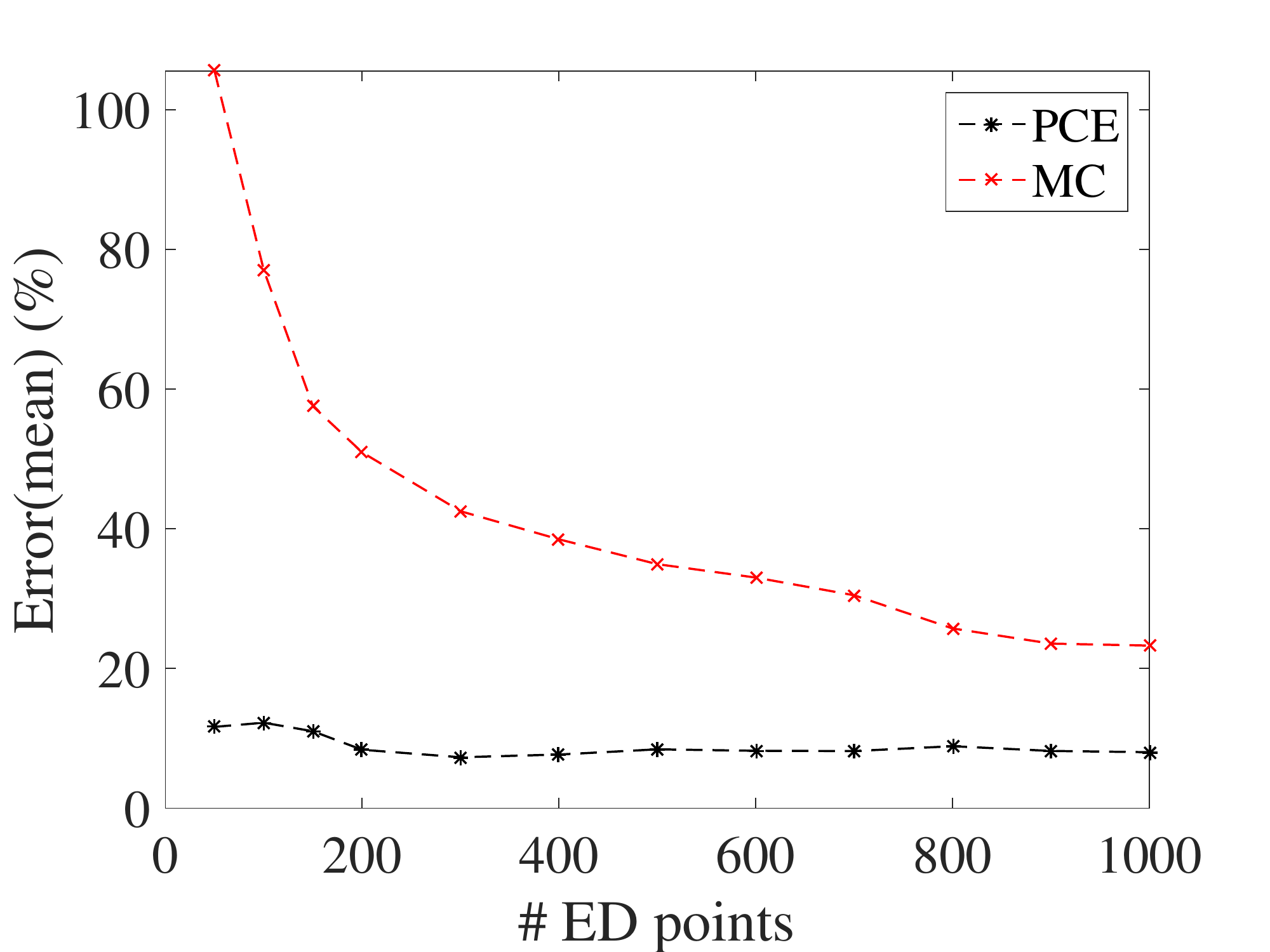}
		\caption{Mean}
		\label{fig:6DOF:FRF:mean:conv:out6:p01D}
	\end{subfigure}
	\begin{subfigure}[b]{.5\columnwidth}
		\centering
		\includegraphics[width=1\columnwidth]{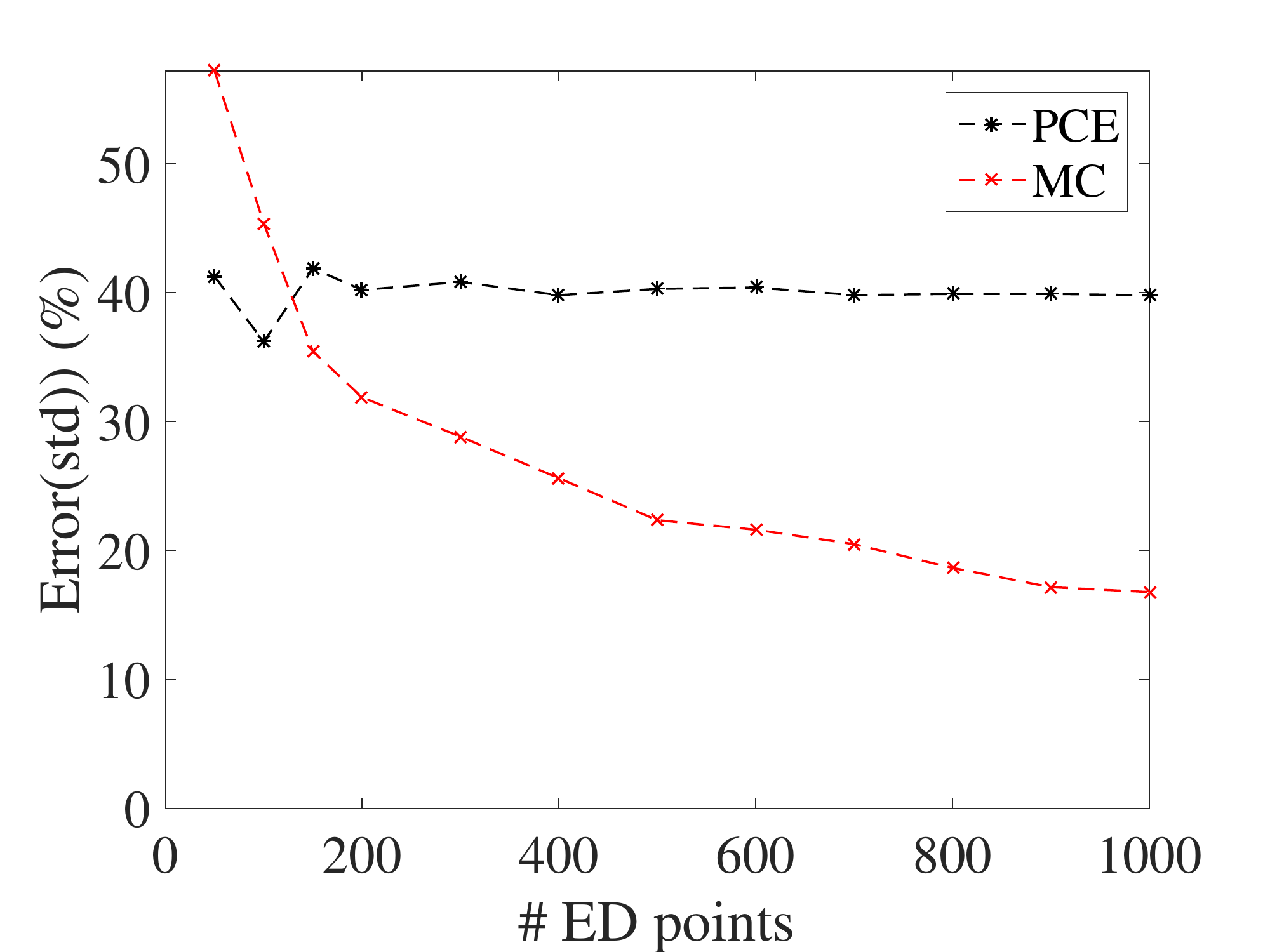}
		\caption{Standard deviation}
		\label{fig:6DOF:FRF:std:conv:out6:p01D}
	\end{subfigure}
	\caption{Convergence plot of the first two moments of the FRFs of the 6-DOF system with $\ve{V}=0.01\widehat{\ve{M}}$ calculated at $6^{th}$ mass by the PCE (black $\ast$) and the true model (red $\times$) by enlarging the experimental design. The reference results were obtained by 10,000 Monte-Carlo simulation of the true model.}
	\label{fig:6DOF:conv:p01D}
\end{figure}

\begin{figure}[H]
	\centering
	\begin{subfigure}[b]{0.5\columnwidth}
		\centering
		\includegraphics[width=1\columnwidth]{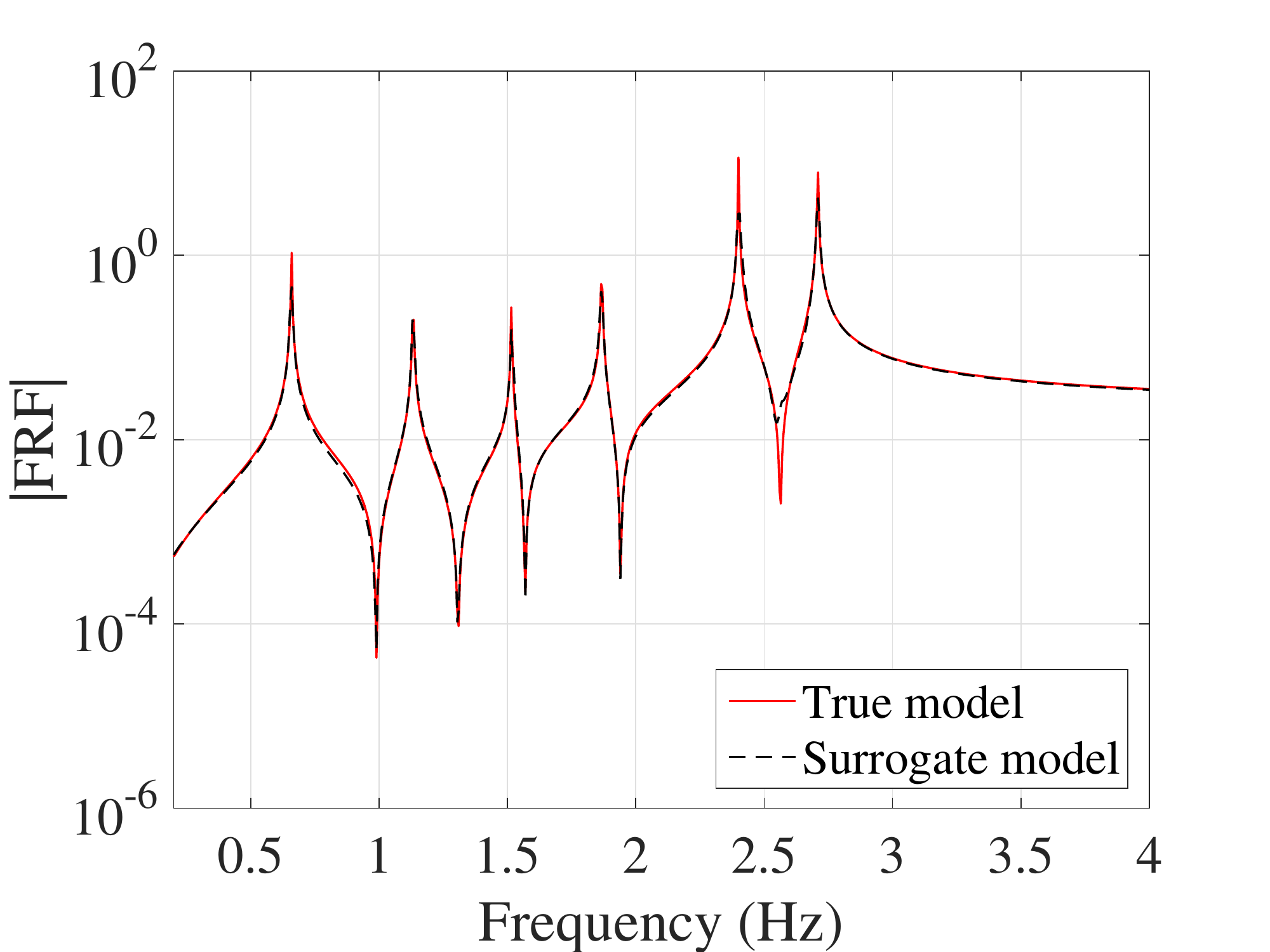}
		\caption{Typical FRF prediction}
		\label{fig:6DOF:FRF:out6:typ:p01D}
	\end{subfigure}
	\begin{subfigure}[b]{0.5\columnwidth}
		\centering
		\includegraphics[width=1\columnwidth]{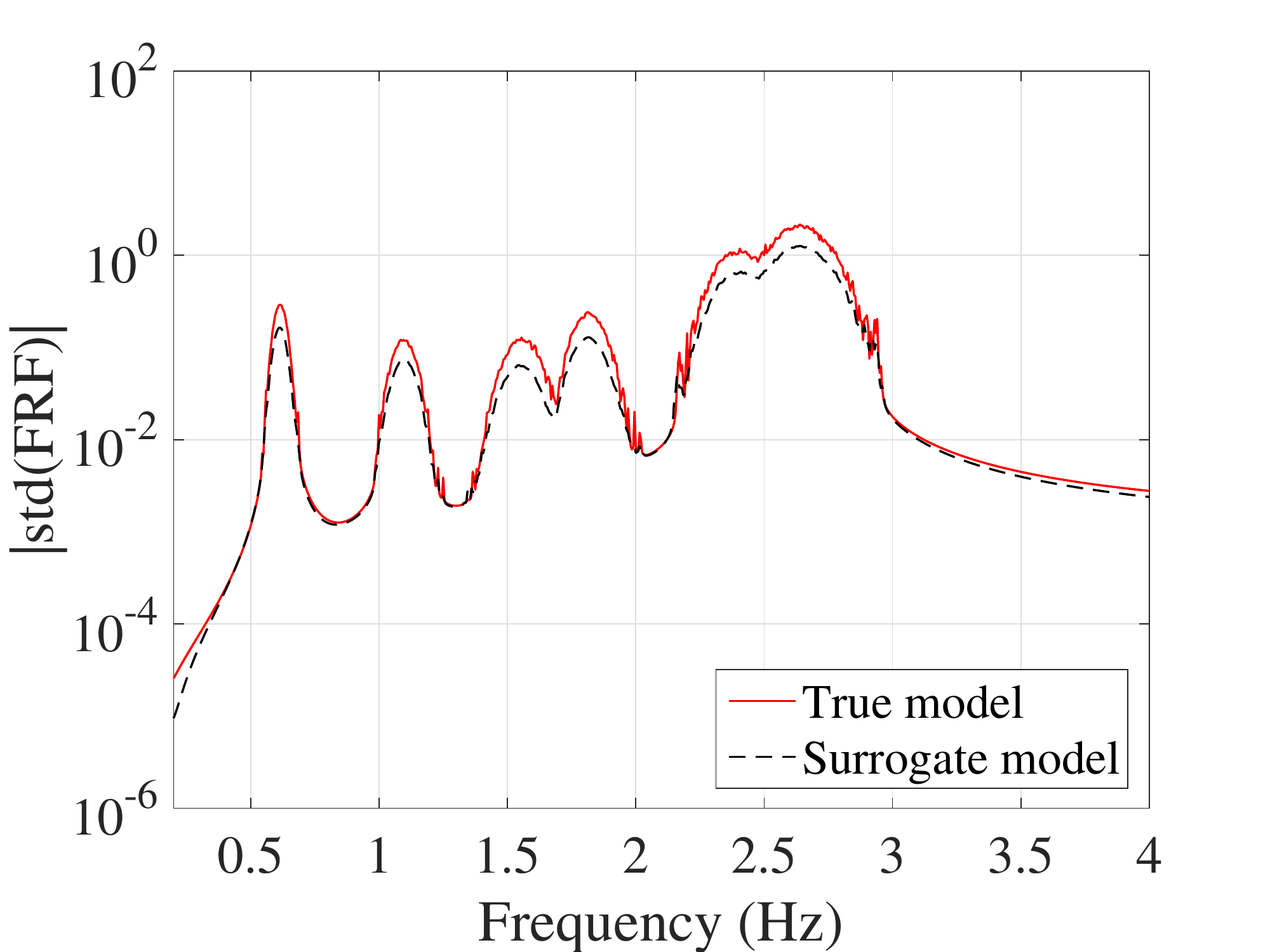}
		\caption{Magnitude of the standard deviation}
		\label{fig:6DOF:FRF:std:out6:abs:p01D}
	\end{subfigure}
	\caption{Analysis of the FRF of the 6-DOF system  with $\ve{V}=0.01\widehat{\ve{M}}$ at $6^{th}$ output, evaluated at 10,000 MC sample points by the true model (red line) and the surrogate model (black line). The surrogate model has been made with 1000 ED points.}
	\label{fig:6DOF:frf:out6:p01D}
\end{figure}

In order to study the effect of the damping level on the accuracy of the proposed method, the study was repeated on the 6-DOF system after setting a much lower damping level. As is shown in Figure \ref{fig:6DOF:conv:p01D}, if the damping is decreased of one order of magnitude $\ve{V}=0.01\widehat{\ve{M}}$, the convergence of the mean response is still improved, whereas the standard deviation shows a more erratic behavior and reaches a plateau of approximately 40\% RMS error. 
To analyze this phenomenon, a surrogate model has been made with a larger experimental design comprised of 1000 points. A typical FRF predicted by this surrogate model is shown in Figure \ref{fig:6DOF:FRF:out6:typ:p01D}. Significant mismatch is observed especially around the peaks, which leads to an overall less accurate estimation of the standard deviation, (see Figure \ref{fig:6DOF:FRF:std:out6:abs:p01D}). Since this error did not change by enriching the ED, it is concluded that the main source for this error must lie somewhere else in the processing chain.
An in-depth analysis revealed that with such low damping levels, peak estimation is inaccurate even when using the full model, if the frequency step is not chosen fine enough. This is one of the main sources of error. Similarly, the interpolation step between pre- and post-processing the frequency axis of the PCE also plays an important role.
Both of the errors could in fact be reduced by reducing the frequency step. Indeed, when in a further investigation we reduced the frequency step to 0.005$\pi$, the error plateau was reduced to 20\%.

\section{Conclusions}

A novel method to build a surrogate model directly for the FRFs of stochastic linear dynamic systems based on sparse PCE has been proposed. To this end, there were two major challenges which have been addressed in this paper: the frequency shifts in the \textit{selected frequencies} of the FRFs, \ie peaks and valleys, due to the uncertainty in the parameters of the system and the non-smooth behavior of the FRFs. These can lead to very high-order PCEs even for the FRFs obtained from cases with 1 or 2 DOFs. We thus propose a stochastic frequency-transformation as a preprocessing step before building PCEs. This transformation scales the FRFs in the frequency horizon so that their \textit{selected frequencie}s become aligned. Although this preprocessing step results in one extra set of PCEs, they do not require any additional full model evaluations.
After the transformation, FRFs are very similar and low-order PCEs can be built for each frequency. This leads to an enormously large number of random outputs. An efficient implementation of principal component analysis has been used to alleviate this issue. Moreover, the problem of curse of dimensionality of PCEs in cases with large parameter space was resolved by employing the LARS algorithm to build sparse PCEs together with adaptive degree.

Successfully applied to two case studies, the proposed method shows its capability of accurately 1) predicting individual FRFs, 2) estimating the mean and standard deviation of the FRFs, and 3) estimating the eigenfrequencies of the system and their statistics. 
In cases of very low damping, significant errors can be observed around the peaks. Interpolation error, both for the full model and for the surrogate model were identified as the main cause. In fact, when the frequency step in the full model was too coarse, both the estimation of the resonant frequencies and that of their amplitudes were significantly inaccurate. This in turn resulted in an inaccurate experimental design and, consequently, in inaccurate surrogate models. Reducing the frequency step has been shown to be effective in  reducing interpolation error for very low-damping applications.



\appendix
\section{ 2-DOF system}
\label{appendix:2DOF}

\subsection{ Statistics of the FRF at $2^{nd}$ output}
\label{app:2DOF:stat:out2}

\begin{figure}[H]
	\centering
	\begin{subfigure}[b]{.5\columnwidth}
		\centering
		\includegraphics[width=1\columnwidth]{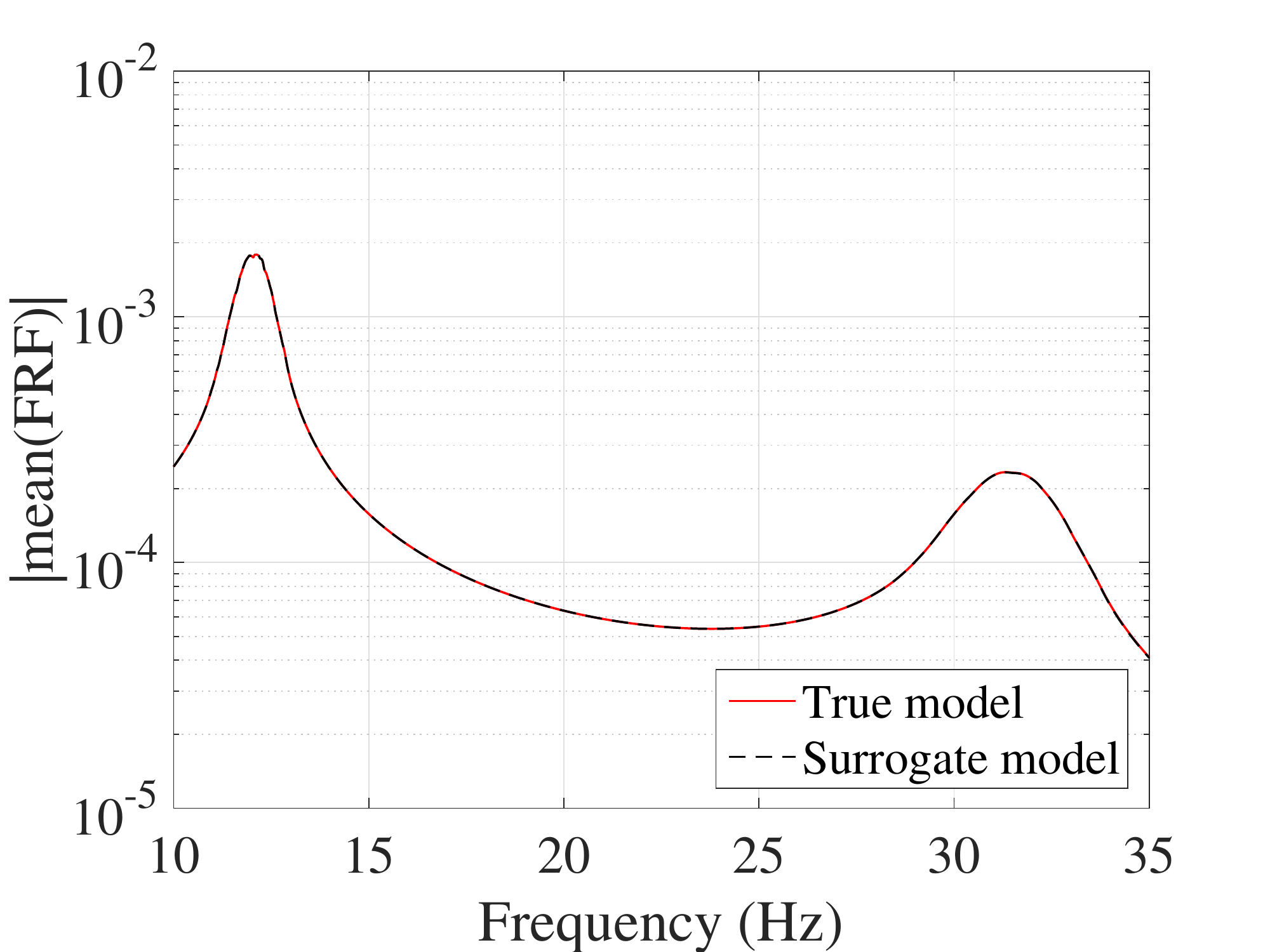}
		\caption{Magnitude of the mean}
		\label{fig:2DOF:FRF:mean2}
	\end{subfigure}
	\begin{subfigure}[b]{.5\columnwidth}
		\centering
		\includegraphics[width=1\columnwidth]{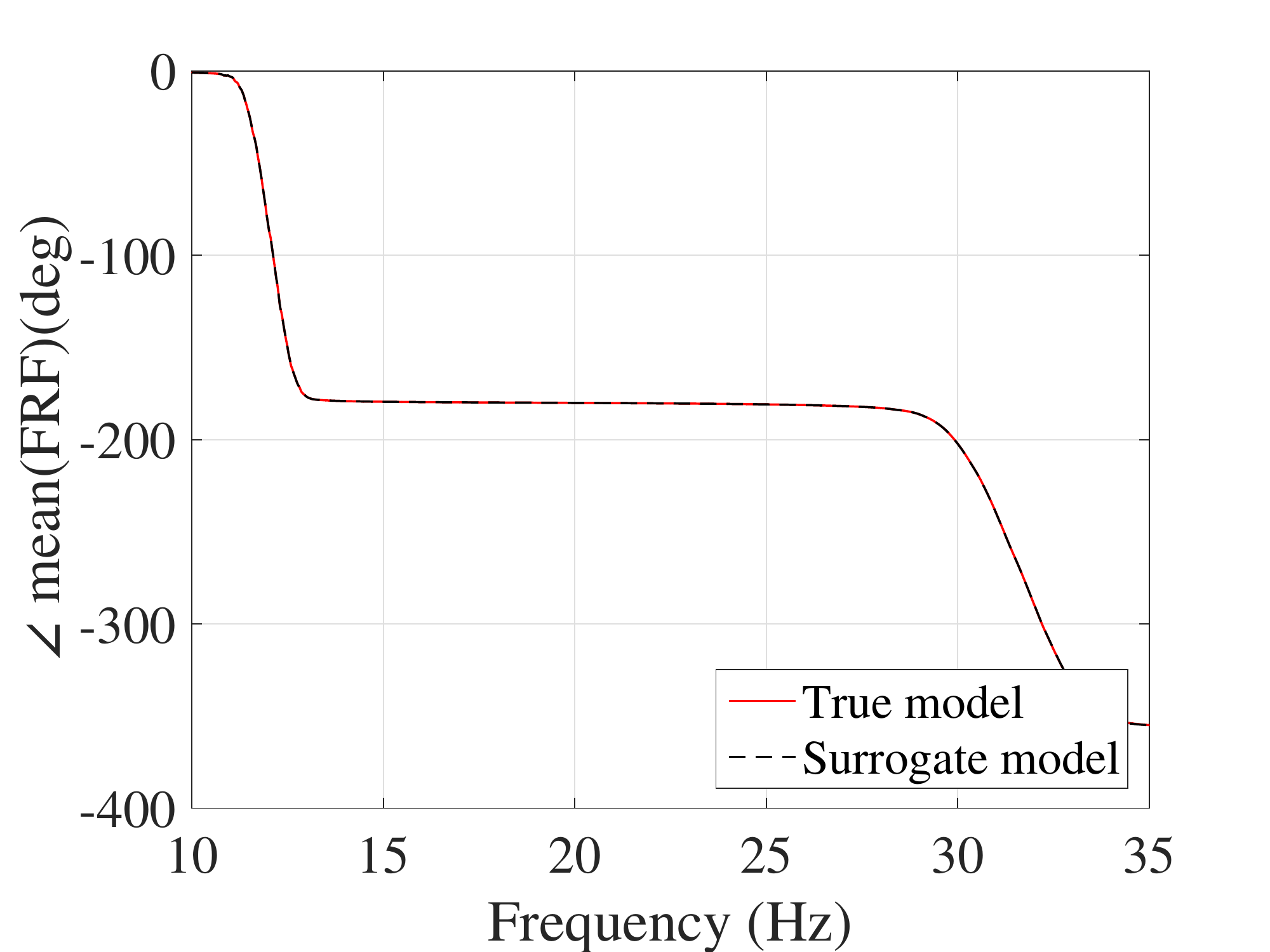}
		\caption{phase of the mean}
		\label{fig:2DOF:FRF:mean2:phase}
	\end{subfigure}%
	\\
	\begin{subfigure}[b]{.5\columnwidth}
		\centering
		\includegraphics[width=1\columnwidth]{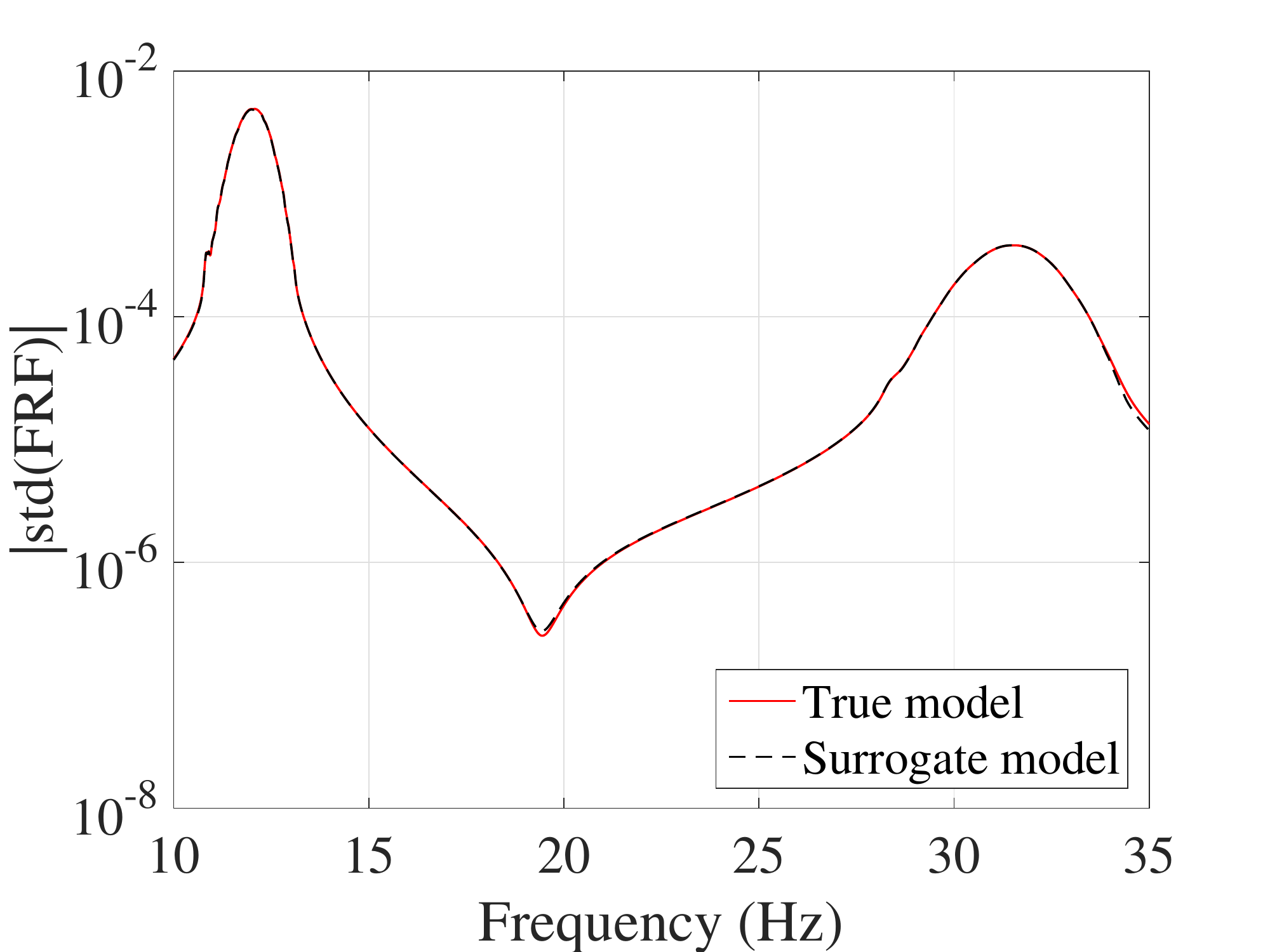}
		\caption{Magnitude of the standard deviation}
		\label{fig:2DOF:FRF:std2}
	\end{subfigure}
	\begin{subfigure}[b]{.5\columnwidth}
		\centering
		\includegraphics[width=1\columnwidth]{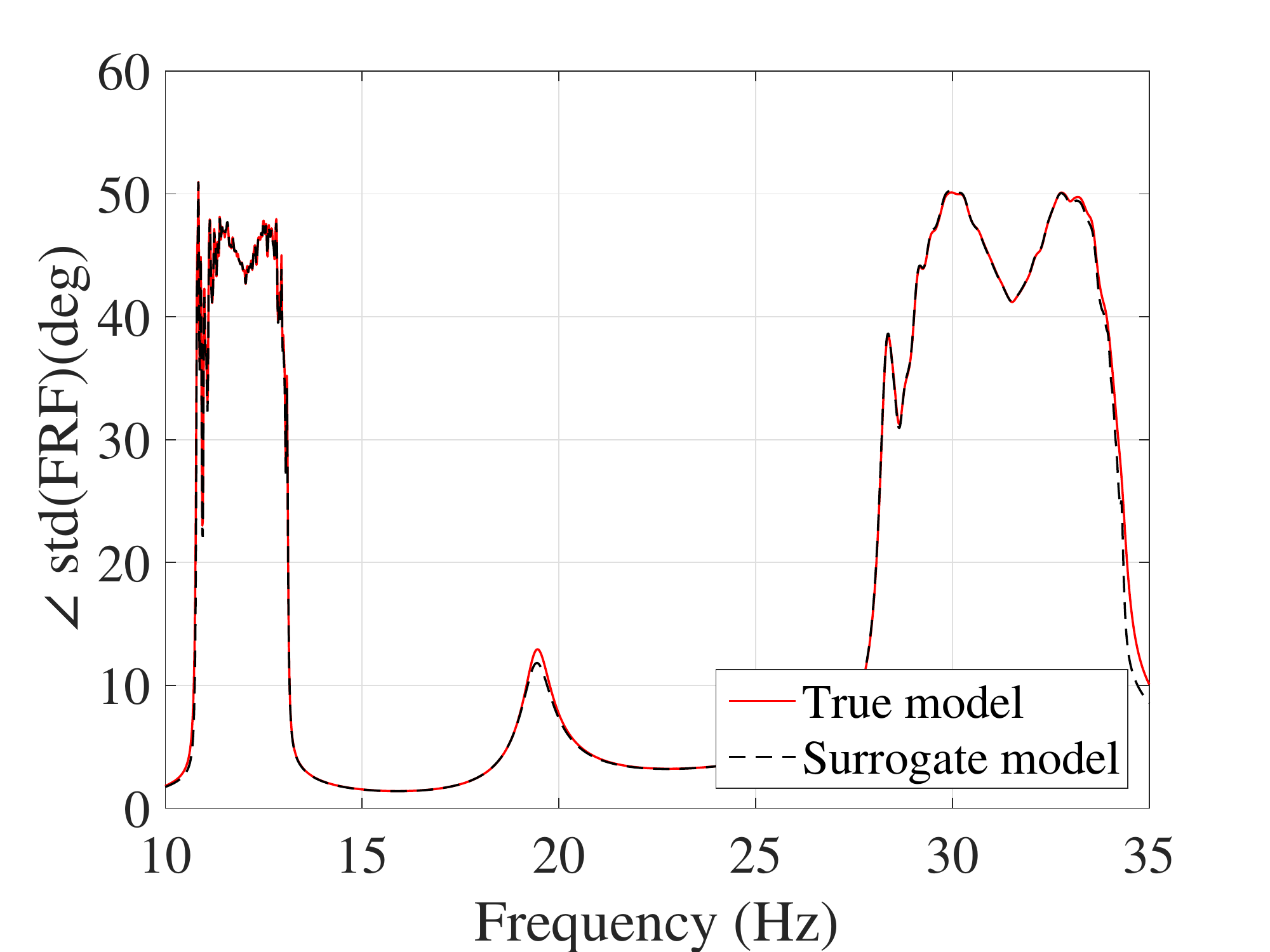}
		\caption{Phase of the standard deviation}
		\label{fig:2DOF:FRF:std2:phase}
	\end{subfigure}%
	\caption{Mean and standard deviation of the FRFs at $2^{nd}$ output evaluated over 10,000 sample points, by the true model (red) and by the surrogate model (black).} 
	\label{fig:FRF:2DOF:out2}
\end{figure}

\newpage
\section{6-DOF system}
\label{appendix:6DOF}

\subsection{Convergence analysis}
\label{app:6DOF:conv}

In this appendix, the results of the convergence analysis of the statistics of the FRFs of the 6-DOF system is presented. The reference results are obtained by running the true model at 10,000 points sampled randomly from the parameter space. 

\subsubsection{Convergence of the mean of the FRFs}
\label{app:6DOF:conv:mean}

\begin{figure}[H]
	\centering
	\begin{subfigure}[b]{.5\columnwidth}
		\centering
		\includegraphics[width=1\columnwidth]{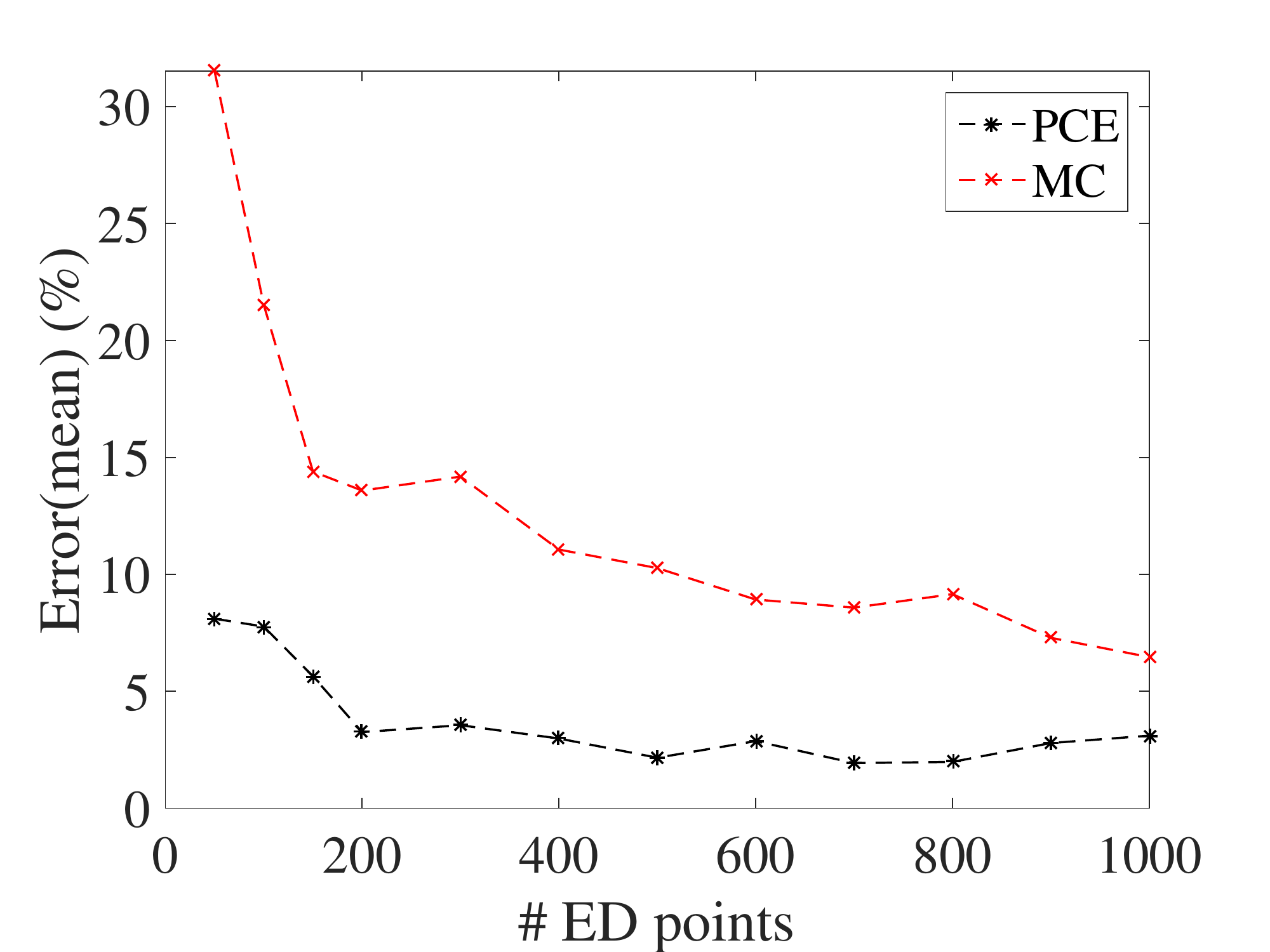}
		\caption{First output}
		\label{fig:6DOF:FRF:mean:conv:out2}
	\end{subfigure}
	\begin{subfigure}[b]{.5\columnwidth}
		\centering
		\includegraphics[width=1\columnwidth]{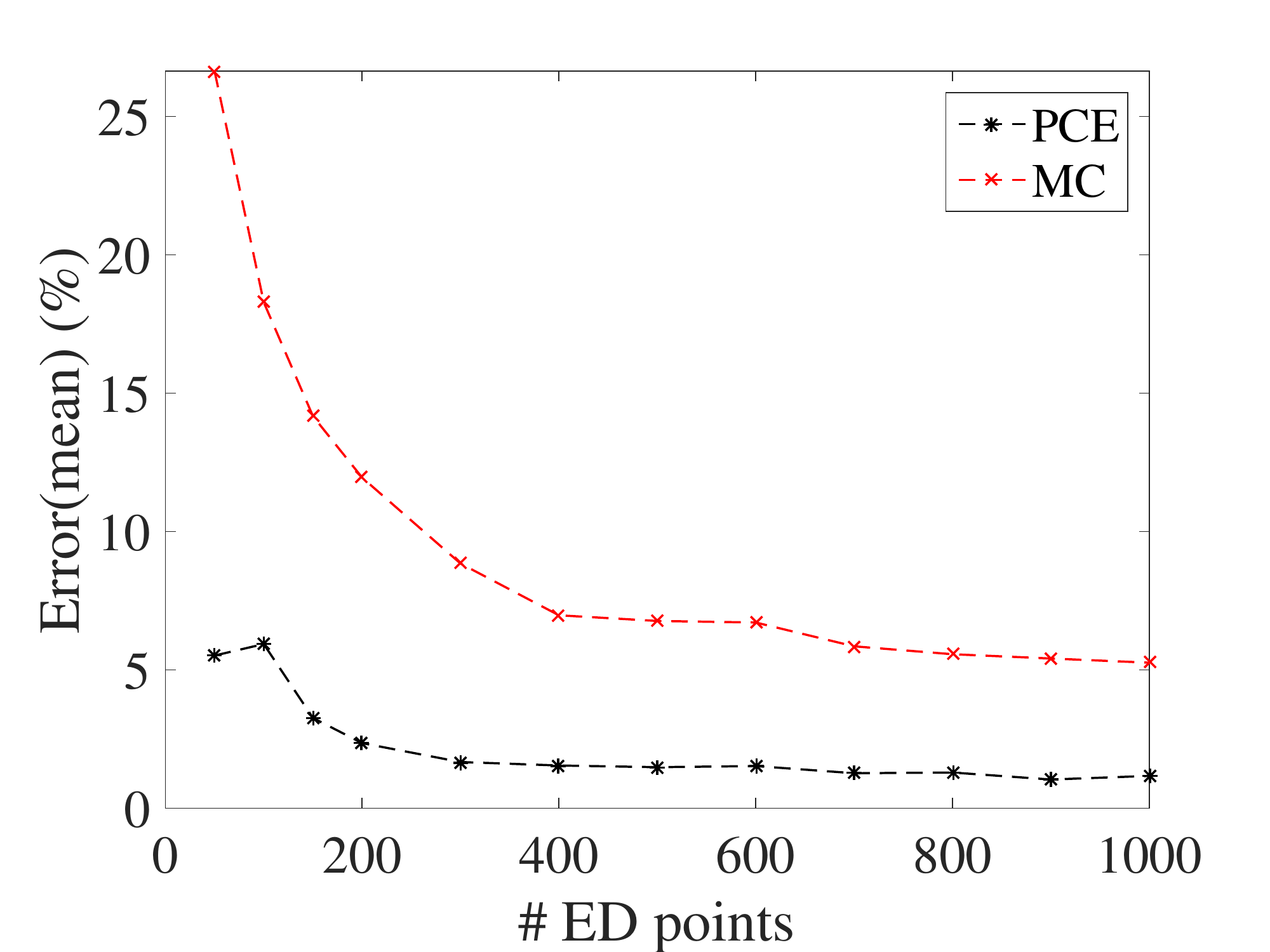}
		\caption{Second output}
		\label{fig:6DOF:FRF:mean:conv:out3}
	\end{subfigure}
	\\
	\begin{subfigure}[b]{.5\columnwidth}
		\centering
		\includegraphics[width=1\columnwidth]{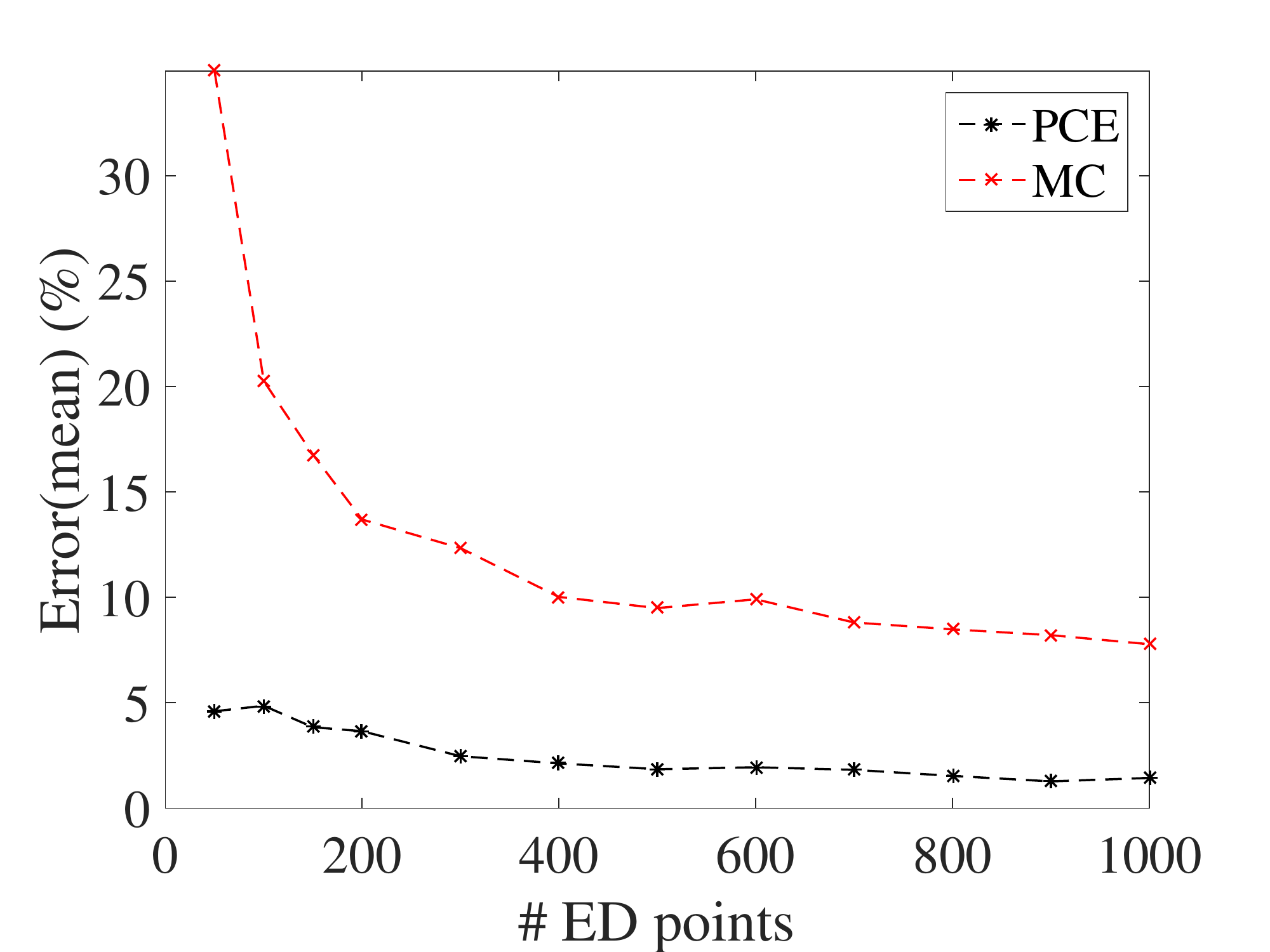}
		\caption{Third output}
		\label{fig:6DOF:FRF:mean:conv:out4}
	\end{subfigure}
	\begin{subfigure}[b]{.5\columnwidth}
		\centering
		\includegraphics[width=1\columnwidth]{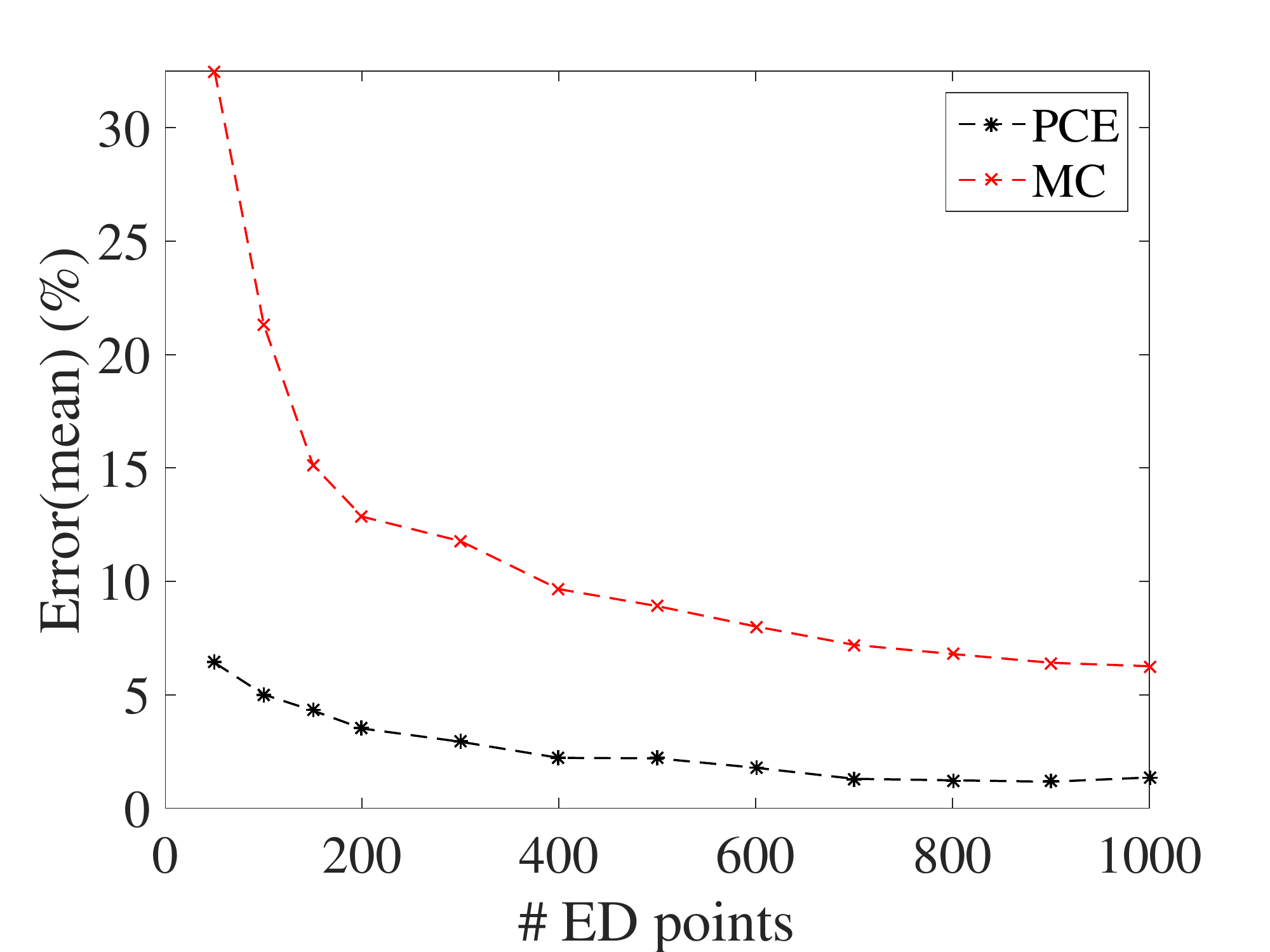}
		\caption{Fourth output}
		\label{fig:6DOF:FRF:mean:conv:out5}
	\end{subfigure}
	\caption{Convergence plot of the mean of the FRFs at 4 outputs obtained by the PCE (black $\ast$) and the true model (red $\times$) by enlarging the experimental design. The reference results were obtained by 10,000 Monte-Carlo simulation of the true model}
	\label{fig:6DOF:frf:mean:conv:4output}
\end{figure}

\subsubsection{Convergence of the standard deviation of the FRFs}
\label{app:6DOF:conv:std}

\begin{figure}[H]
	\centering
	\begin{subfigure}[b]{.5\columnwidth}
		\centering
		\includegraphics[width=1\columnwidth]{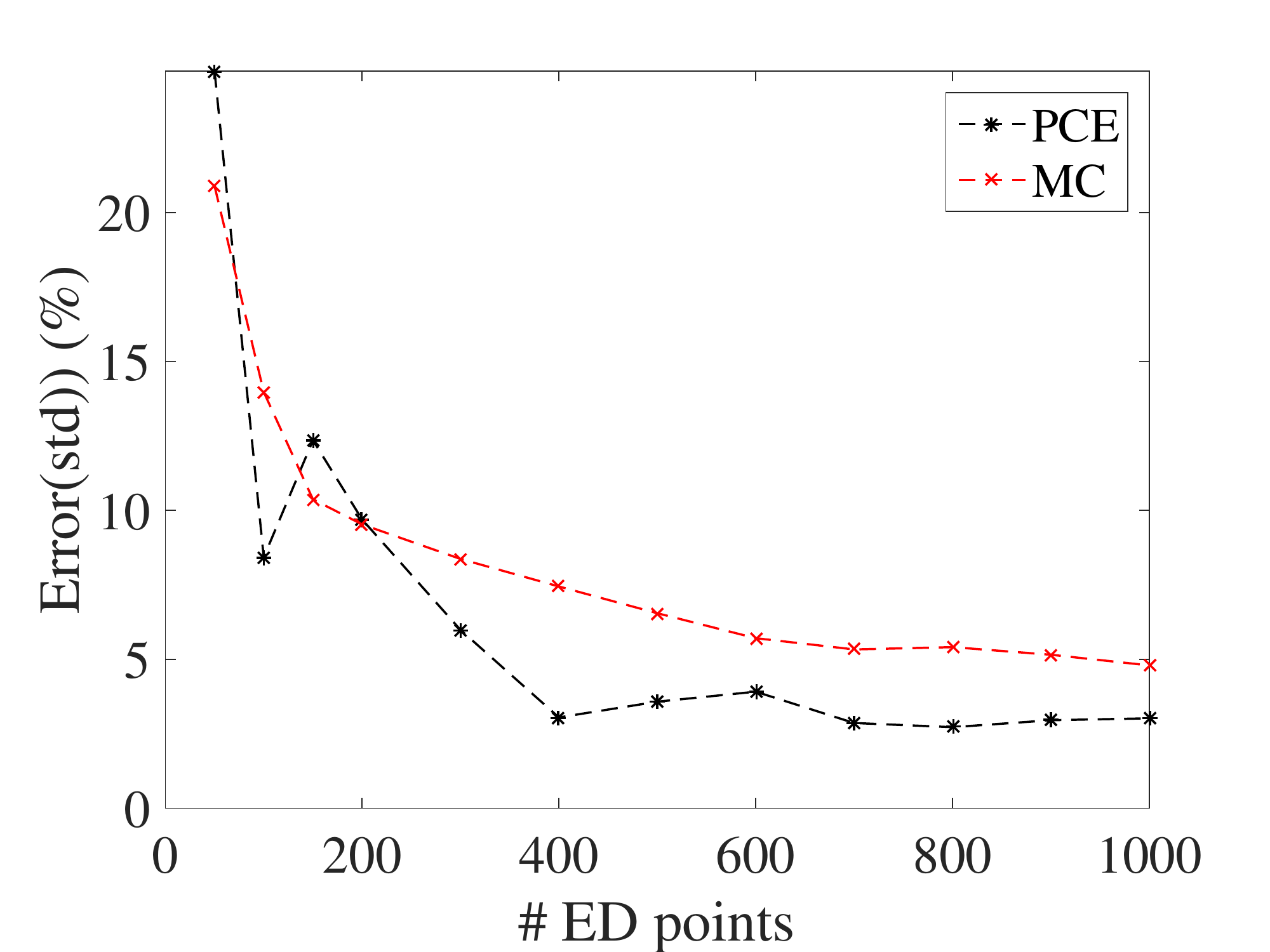}
		\caption{First output}
		\label{fig:6DOF:FRF:std:conv:out2}
	\end{subfigure}
	\begin{subfigure}[b]{.5\columnwidth}
		\centering
		\includegraphics[width=1\columnwidth]{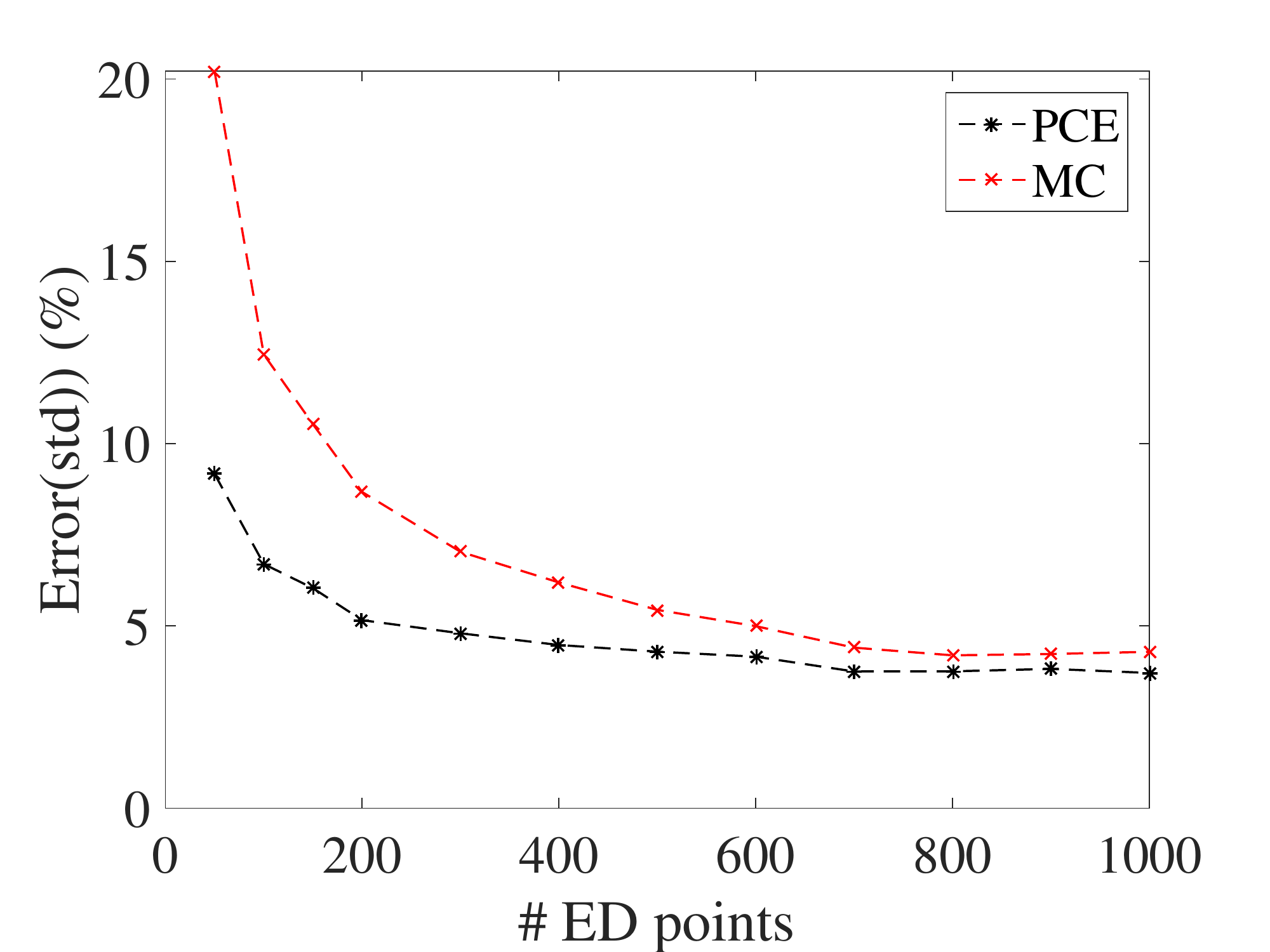}
		\caption{Second output}
		\label{fig:6DOF:FRF:std:conv:out3}
	\end{subfigure}
	\\
	\begin{subfigure}[b]{.5\columnwidth}
		\centering
		\includegraphics[width=1\columnwidth]{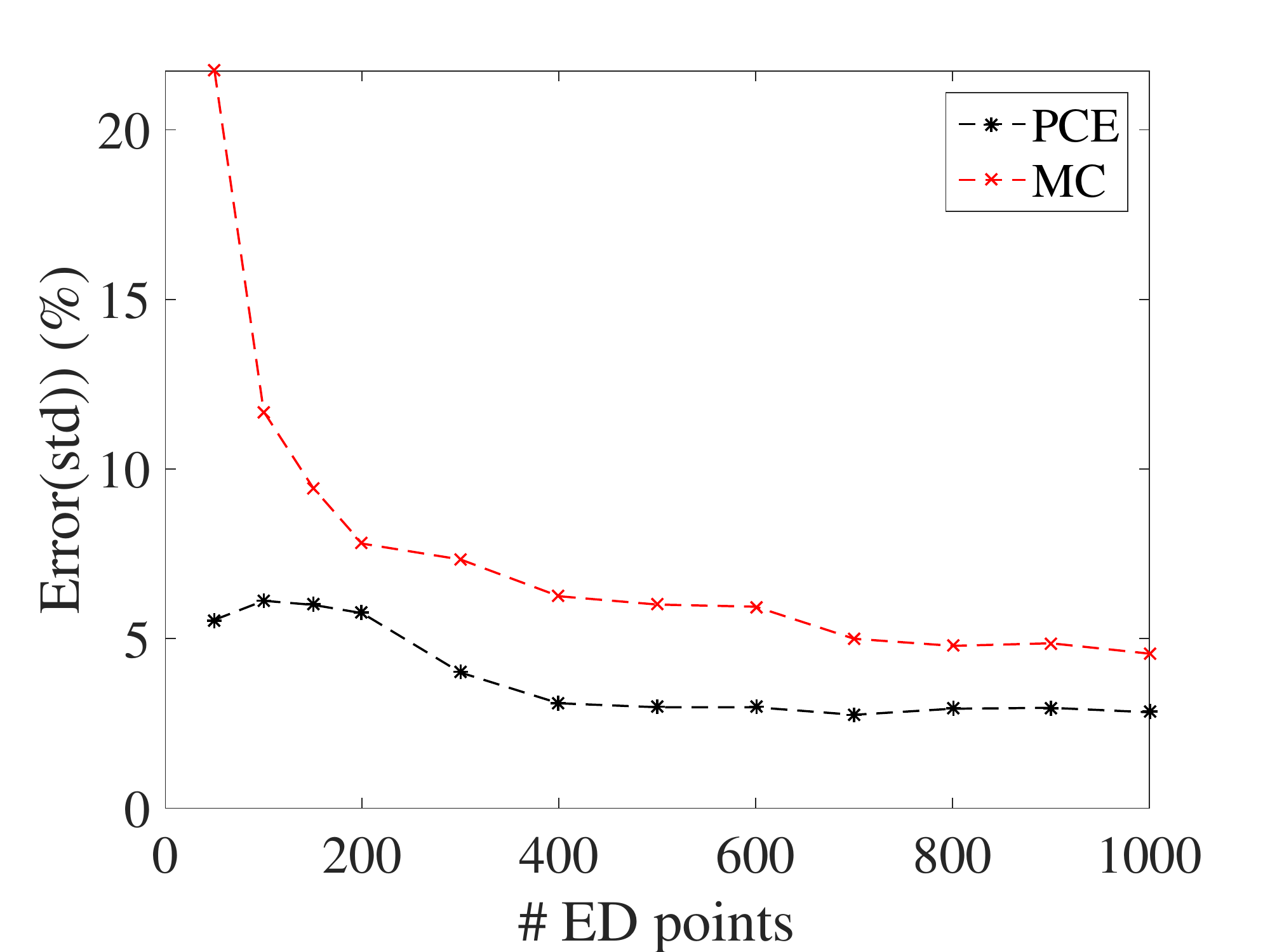}
		\caption{Third output}
		\label{fig:6DOF:FRF:std:conv:out4}
	\end{subfigure}
	\begin{subfigure}[b]{.5\columnwidth}
		\centering
		\includegraphics[width=1\columnwidth]{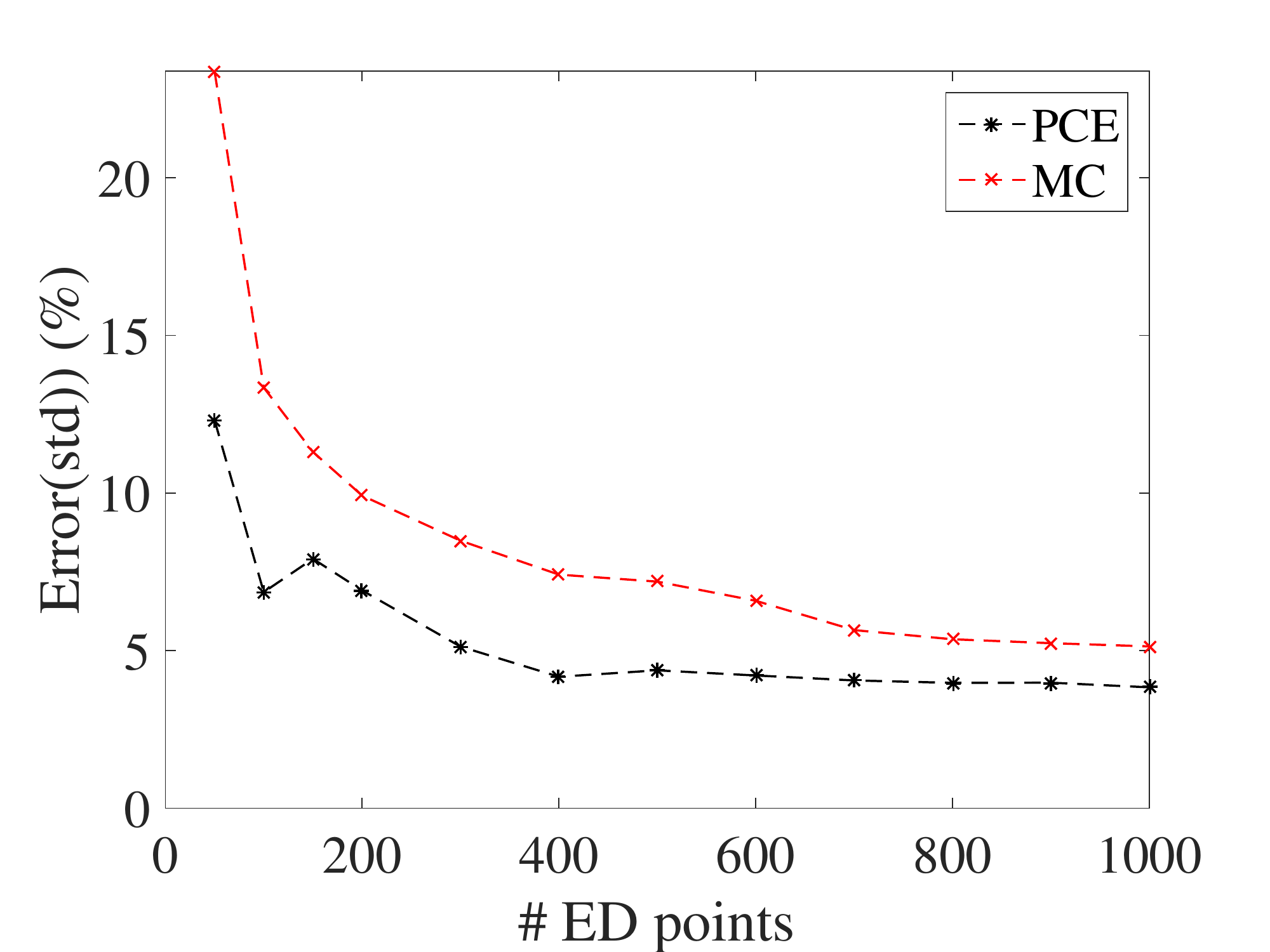}
		\caption{Fourth output}
		\label{fig:6DOF:FRF:std:conv:out5}
	\end{subfigure}
	\caption{Convergence plot of the standard deviation of the FRFs at 4 outputs obtained by the PCE (black $\ast$) and the true model (red $\times$) by enlarging the experimental design. The reference results were obtained by 10,000 Monte-Carlo simulation of the true model}
	\label{fig:6DOF:frf:std:conv:4output}
\end{figure}

\subsection{Moments of the FRFs of the 6-DOF system}
\label{app:6DOF:moments:FRF}
In this appendix, the statistics of the FRFs obtained by evaluating the surrogate model and the true model are compared.

\subsubsection{Mean of the FRFs}
\label{app:6DOF:moments:mean}

\begin{figure}[H]
	\centering
	\begin{subfigure}[b]{.5\columnwidth}
		\centering
		\includegraphics[width=1\columnwidth]{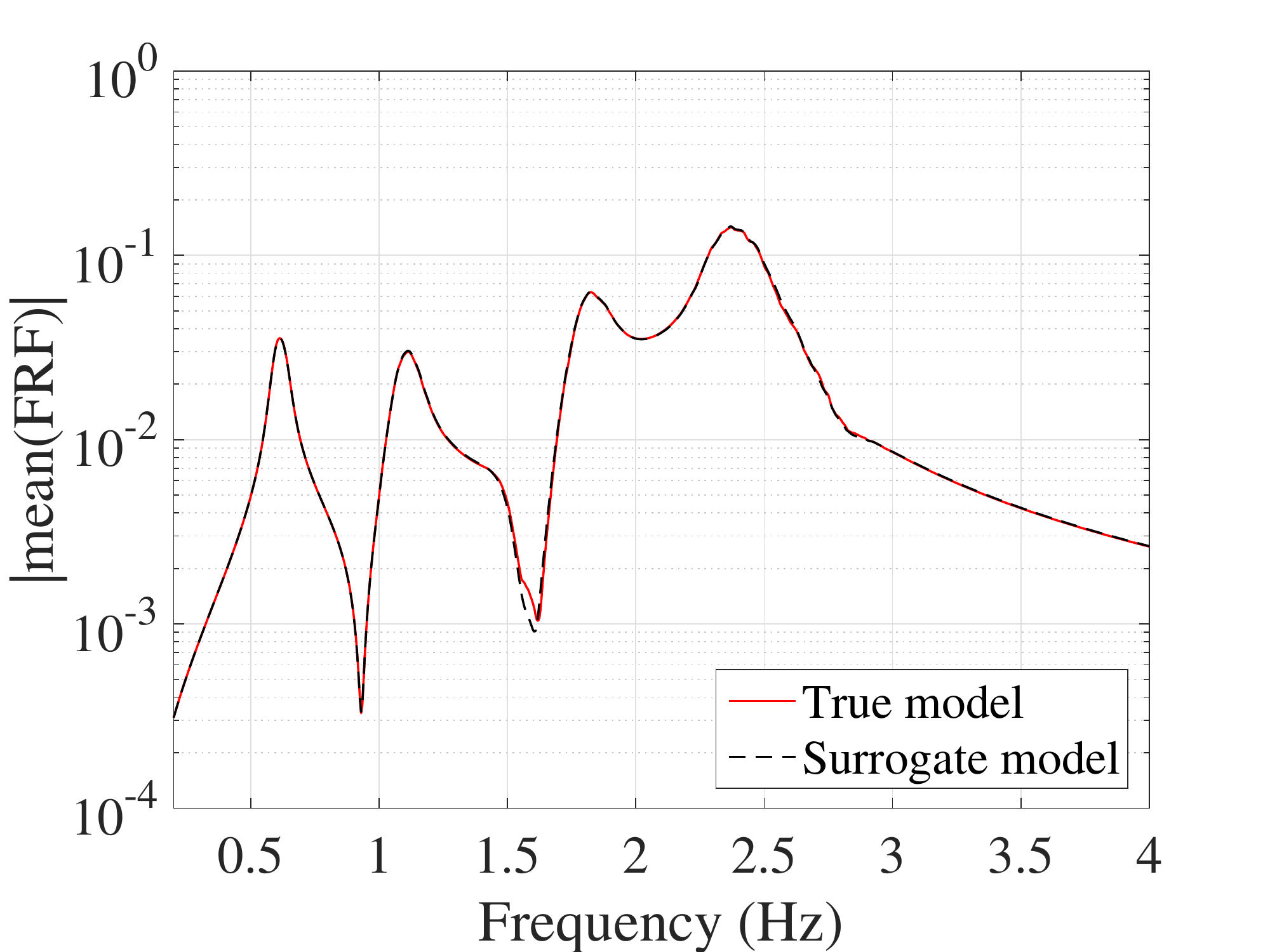}
		\caption{First output}
		\label{fig:6DOF:FRF:mean:out2}
	\end{subfigure}
	\begin{subfigure}[b]{.5\columnwidth}
		\centering
		\includegraphics[width=1\columnwidth]{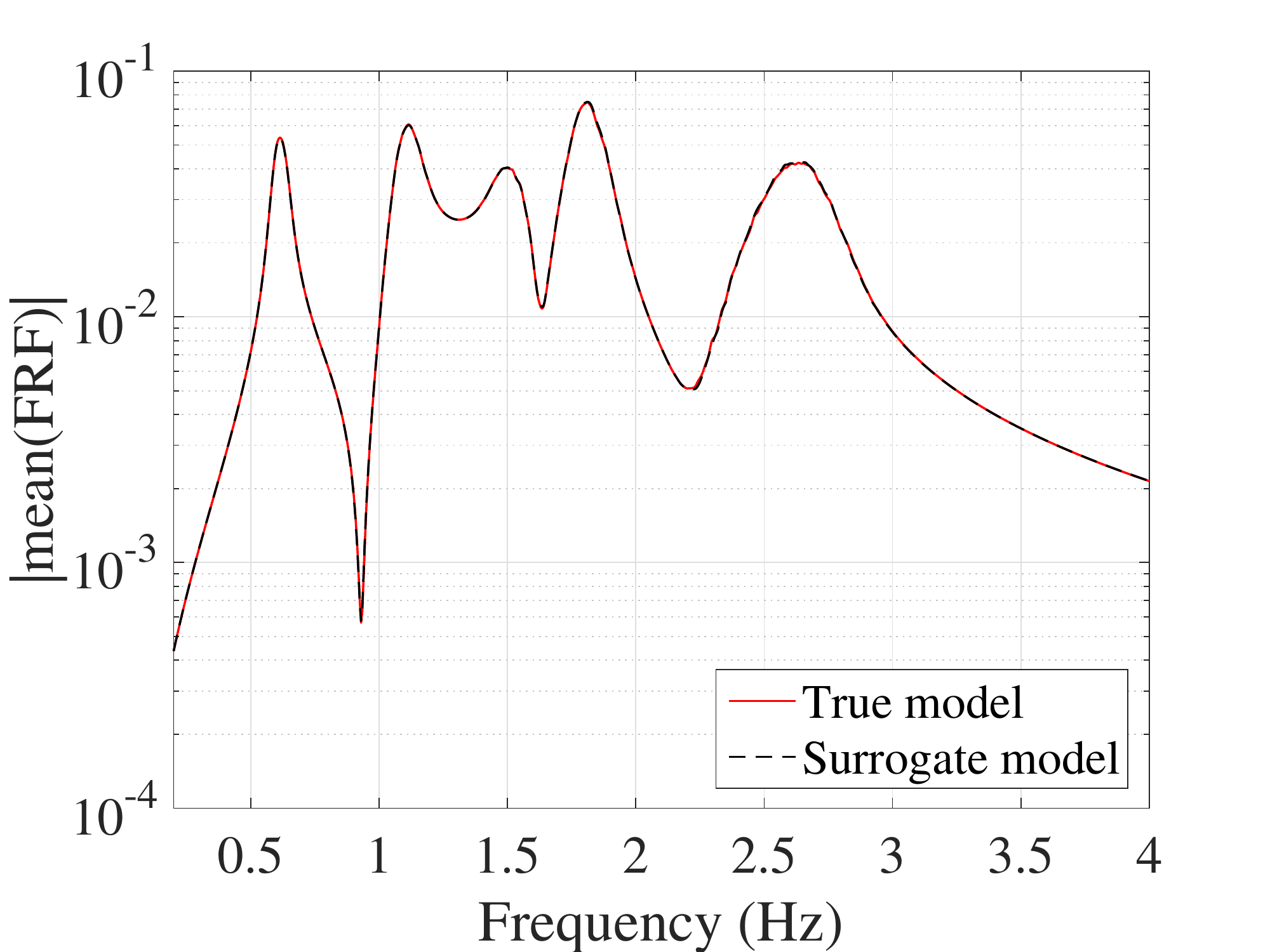}
		\caption{Second output}
		\label{fig:6DOF:FRF:mean:out3}
	\end{subfigure}
	\\
	\begin{subfigure}[b]{.5\columnwidth}
		\centering
		\includegraphics[width=1\columnwidth]{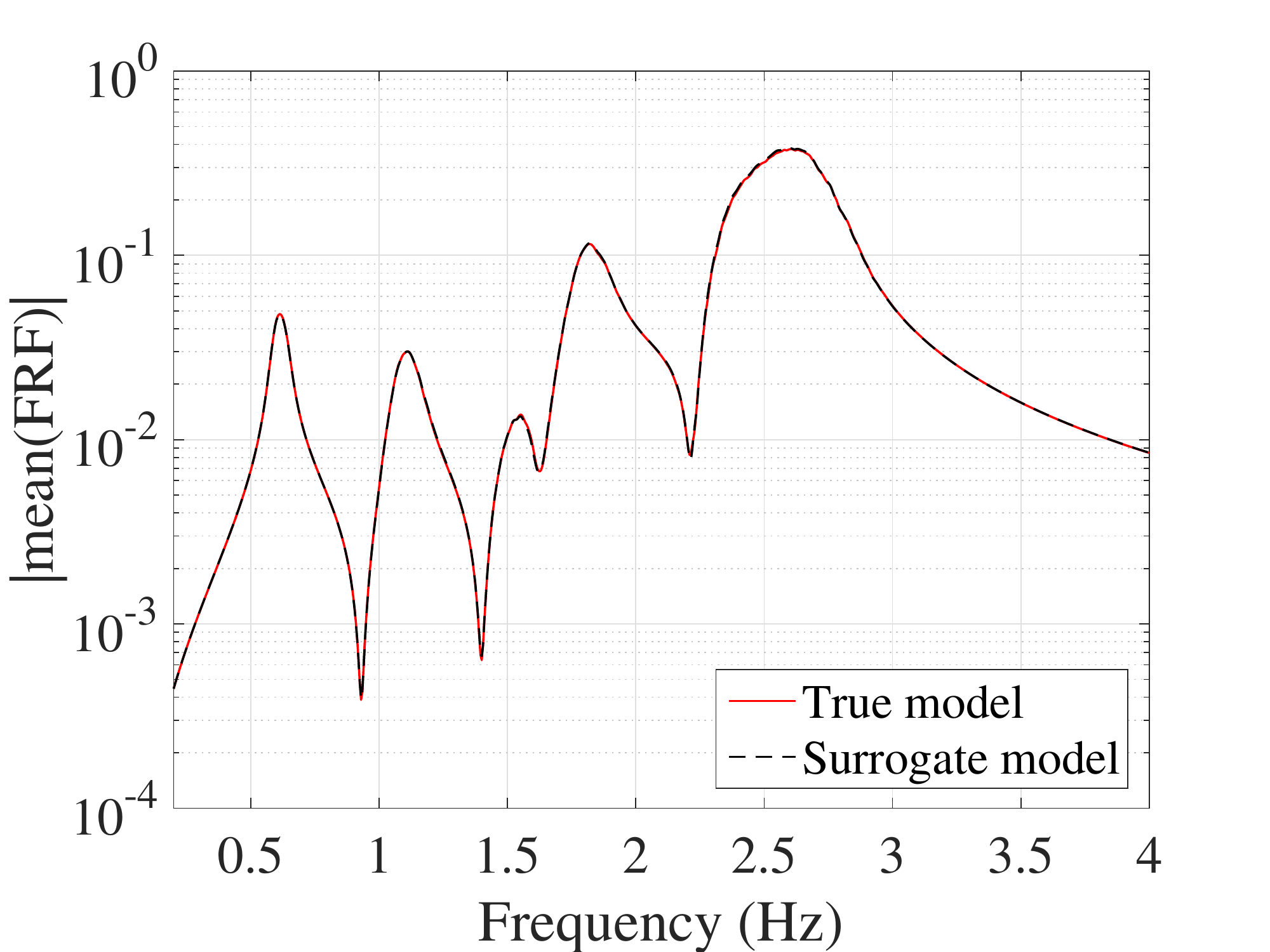}
		\caption{Third output}
		\label{fig:6DOF:FRF:mean:out4}
	\end{subfigure}
	\begin{subfigure}[b]{.5\columnwidth}
		\centering
		\includegraphics[width=1\columnwidth]{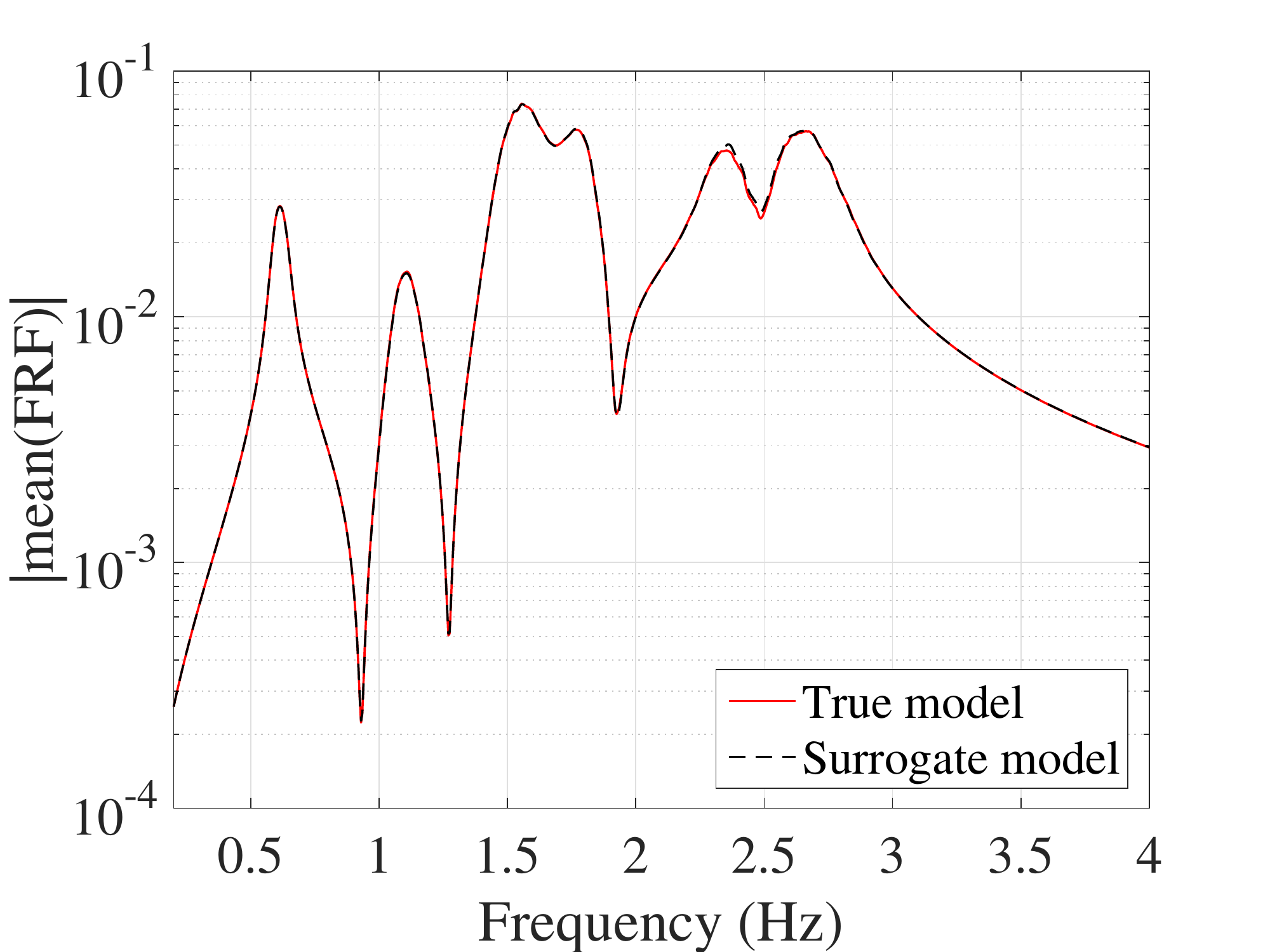}
		\caption{Fourth output}
		\label{fig:6DOF:FRF:mean:out5}
	\end{subfigure}
	\caption{Mean of the FRF of the 6-DOF system at 4 outputs, evaluated at 10,000 MC sample points by the true model (red line) and the surrogate model (black line).}
	\label{fig:6DOF:frf:mean:4output}
\end{figure}

\subsubsection{Standard deviation of the FRFs}
\label{app:6DOF:moments:std}

\begin{figure}[H]
	\centering
	\begin{subfigure}[b]{.5\columnwidth}
		\centering
		\includegraphics[width=1\columnwidth]{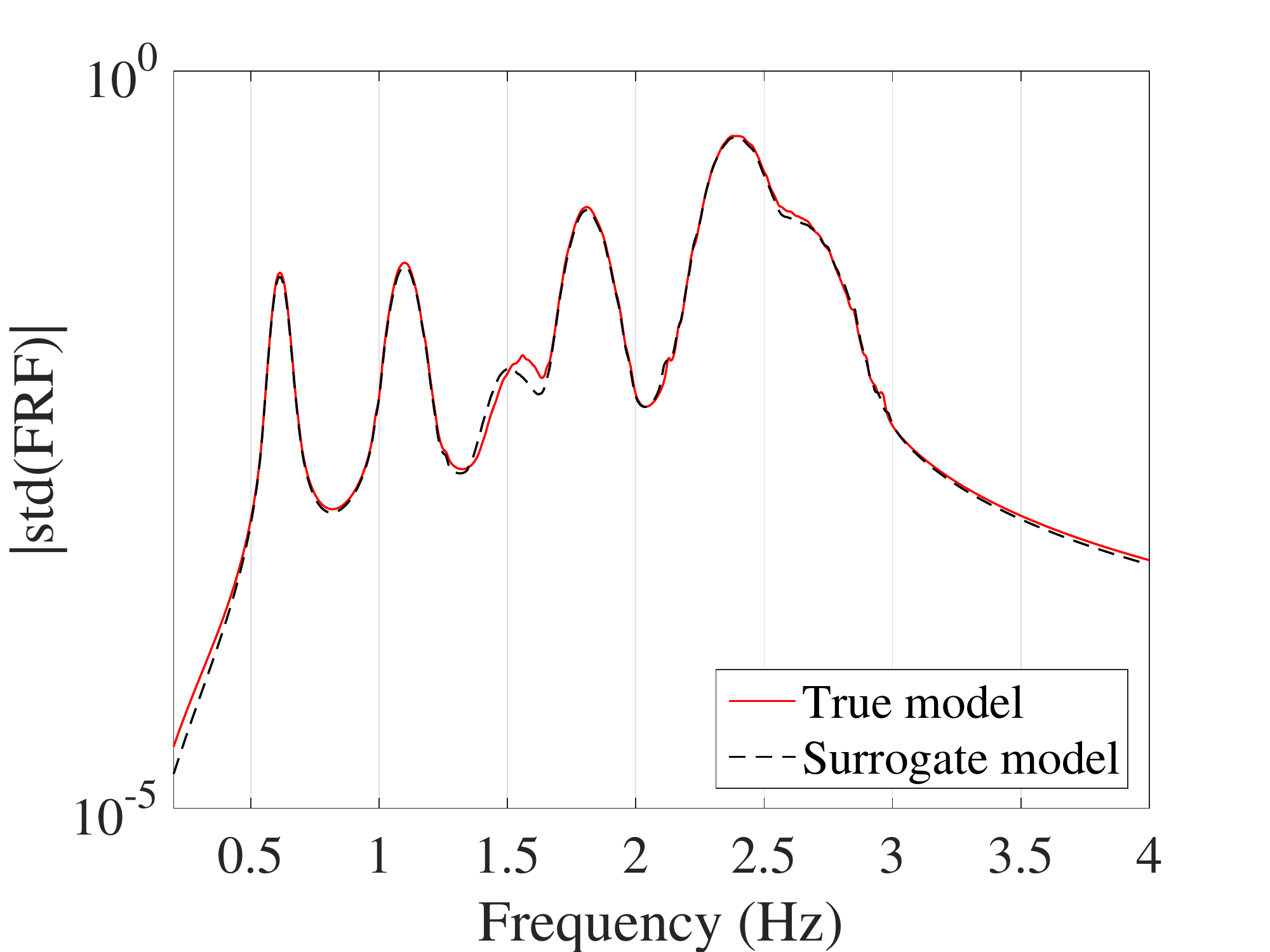}
		\caption{First output}
		\label{fig:6DOF:FRF:std:out2}
	\end{subfigure}
	\begin{subfigure}[b]{.5\columnwidth}
		\centering
		\includegraphics[width=1\columnwidth]{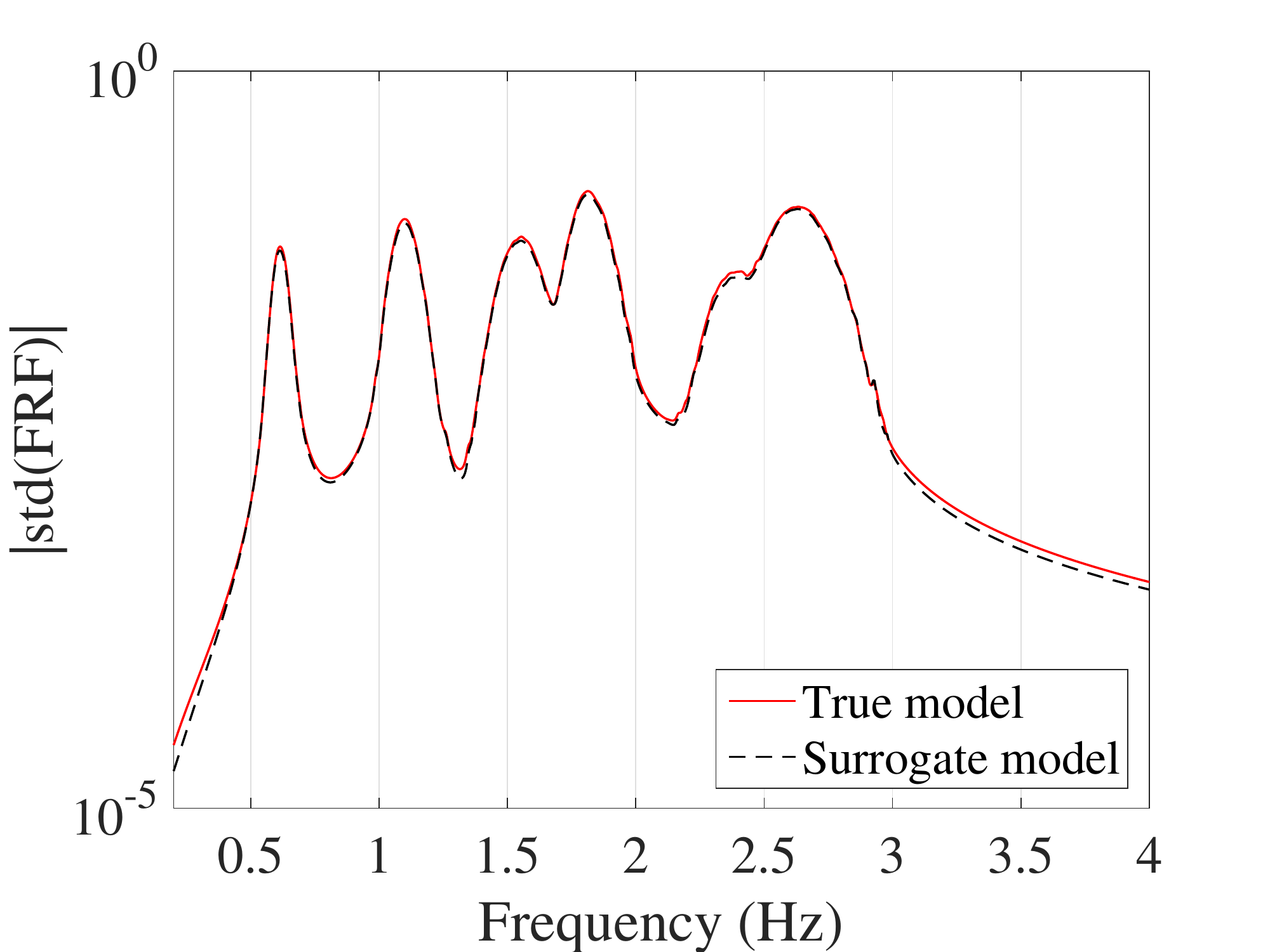}
		\caption{Second output}
		\label{fig:6DOF:FRF:std:out3}
	\end{subfigure}
	\\
	\begin{subfigure}[b]{.5\columnwidth}
		\centering
		\includegraphics[width=1\columnwidth]{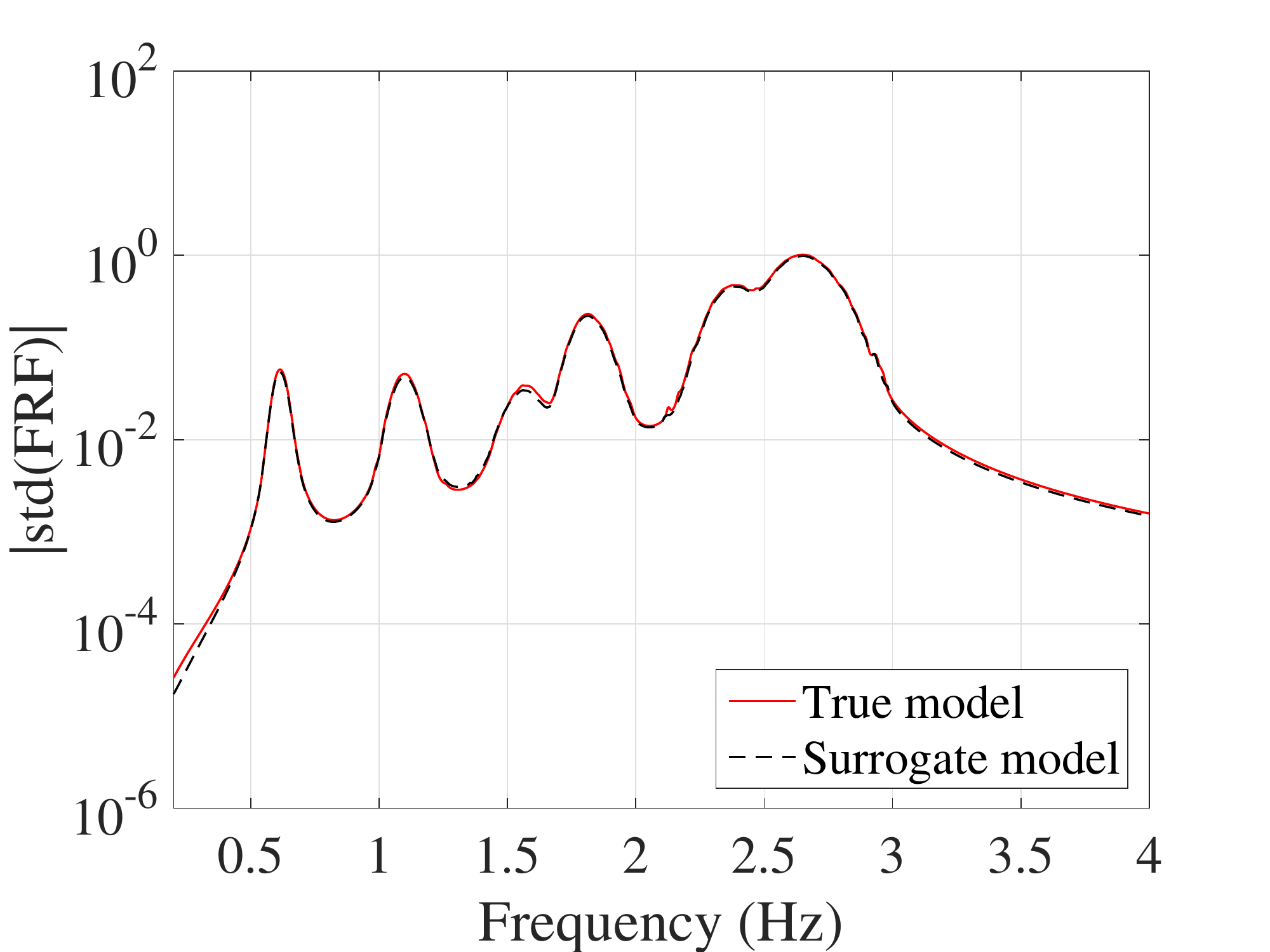}
		\caption{Third output}
		\label{fig:6DOF:FRF:std:out4}
	\end{subfigure}
	\begin{subfigure}[b]{.5\columnwidth}
		\centering
		\includegraphics[width=1\columnwidth]{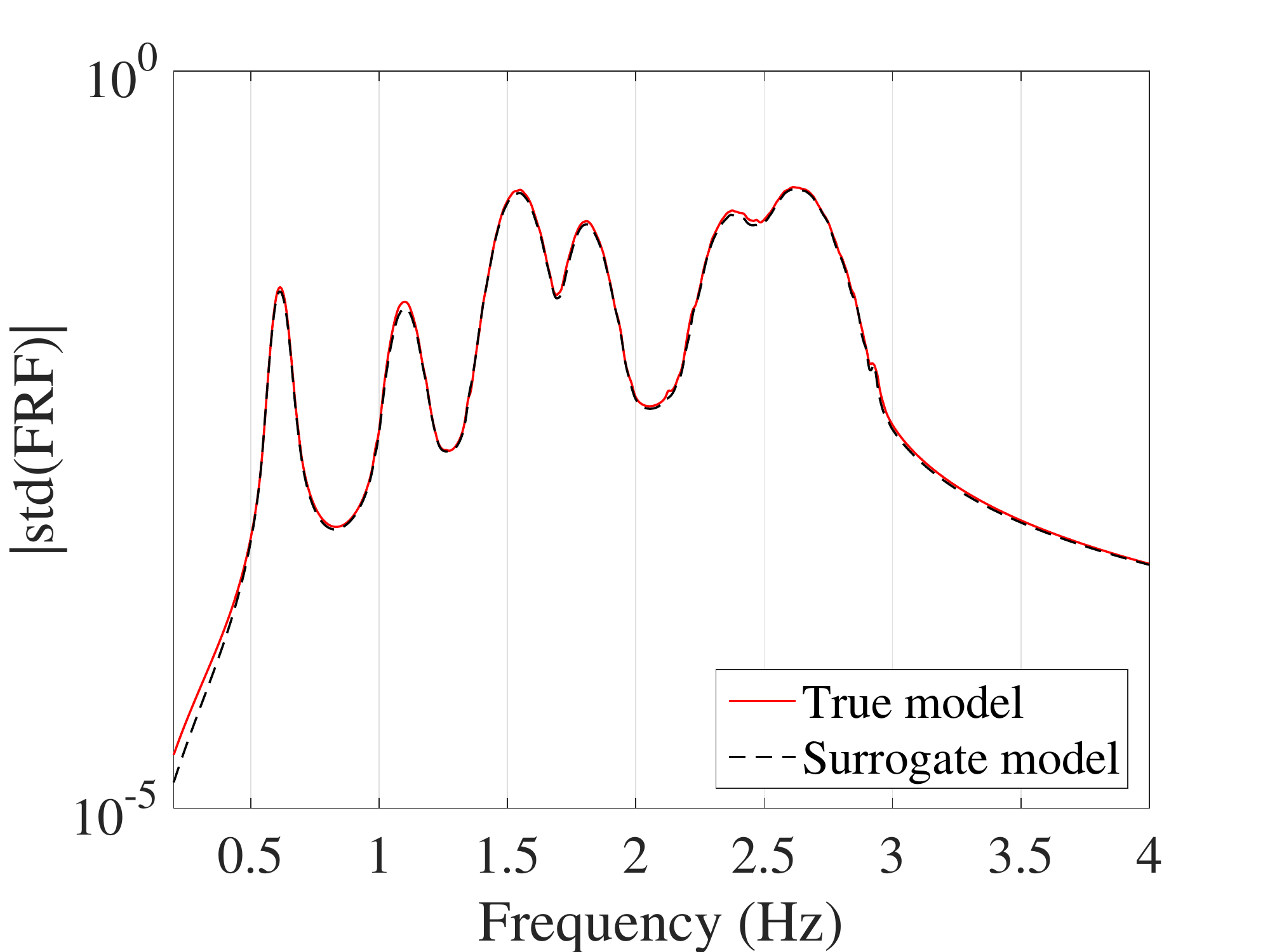}
		\caption{Fourth output}
		\label{fig:6DOF:FRF:std:out5}
	\end{subfigure}
	\caption{Standard deviation of the FRF of the 6-DOF system at 4 outputs, evaluated at 10,000 MC sample points by the true model (red line) and the surrogate model (black line).}
	\label{fig:6DOF:frf:std:4output}
\end{figure}

\subsection{Individual FRFs comparison of the 6-DOF system}
\label{app:6DOF:individual:FRF}

In this appendix, individual FRFs are compared for two particular cases. One FRF that has average error comparing to the true model and one FRF with the maximum error among 10,000 realizations.

\subsubsection{Typical FRFs}
\label{app:6DOF:typic}
\begin{figure}[H]
	\centering
	\begin{subfigure}[b]{.5\columnwidth}
		\centering
		\includegraphics[width=1\columnwidth]{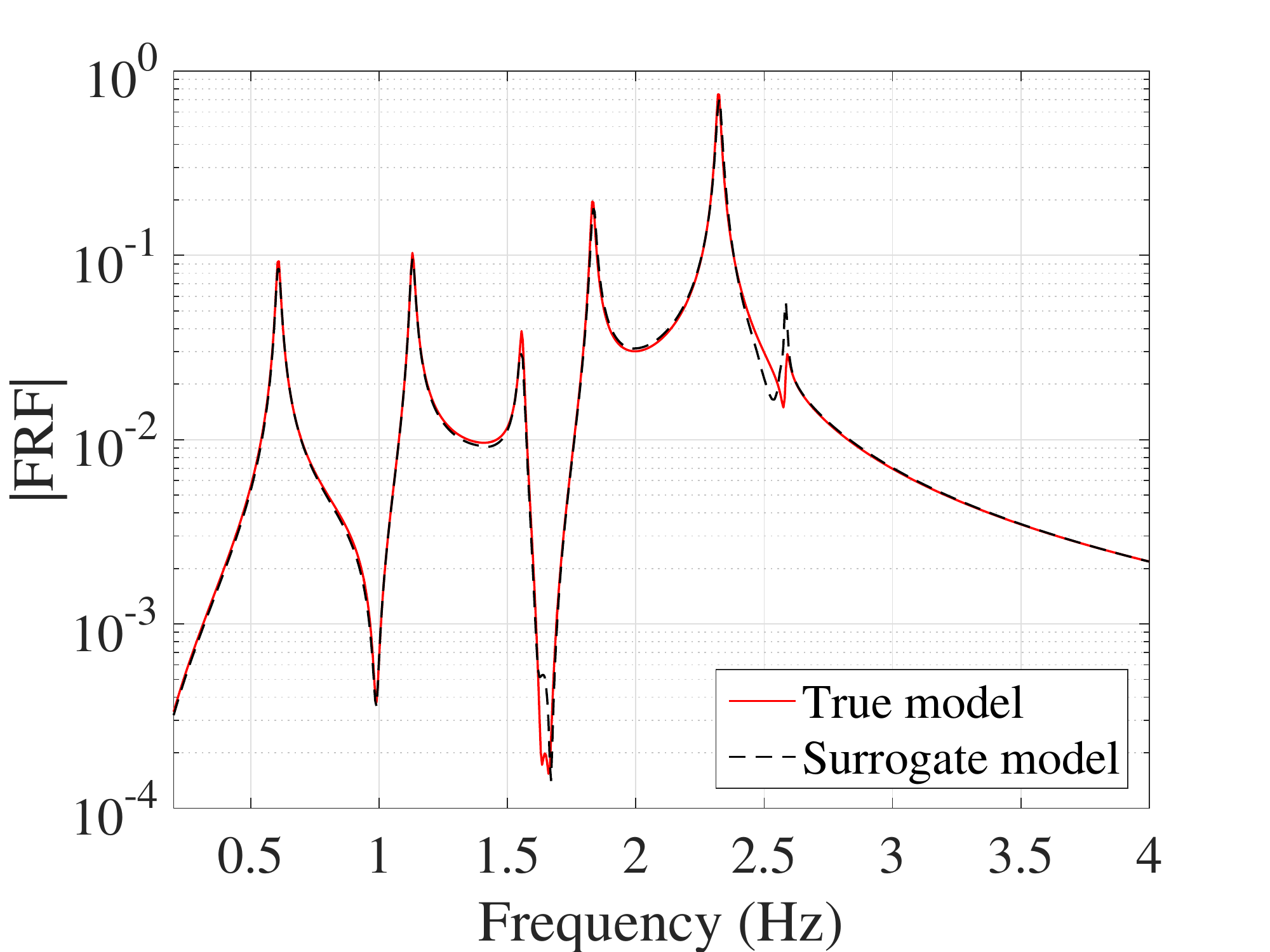}
		\caption{First output}
		\label{fig:6DOF:FRF:typ:out2}
	\end{subfigure}
	\begin{subfigure}[b]{.5\columnwidth}
		\centering
		\includegraphics[width=1\columnwidth]{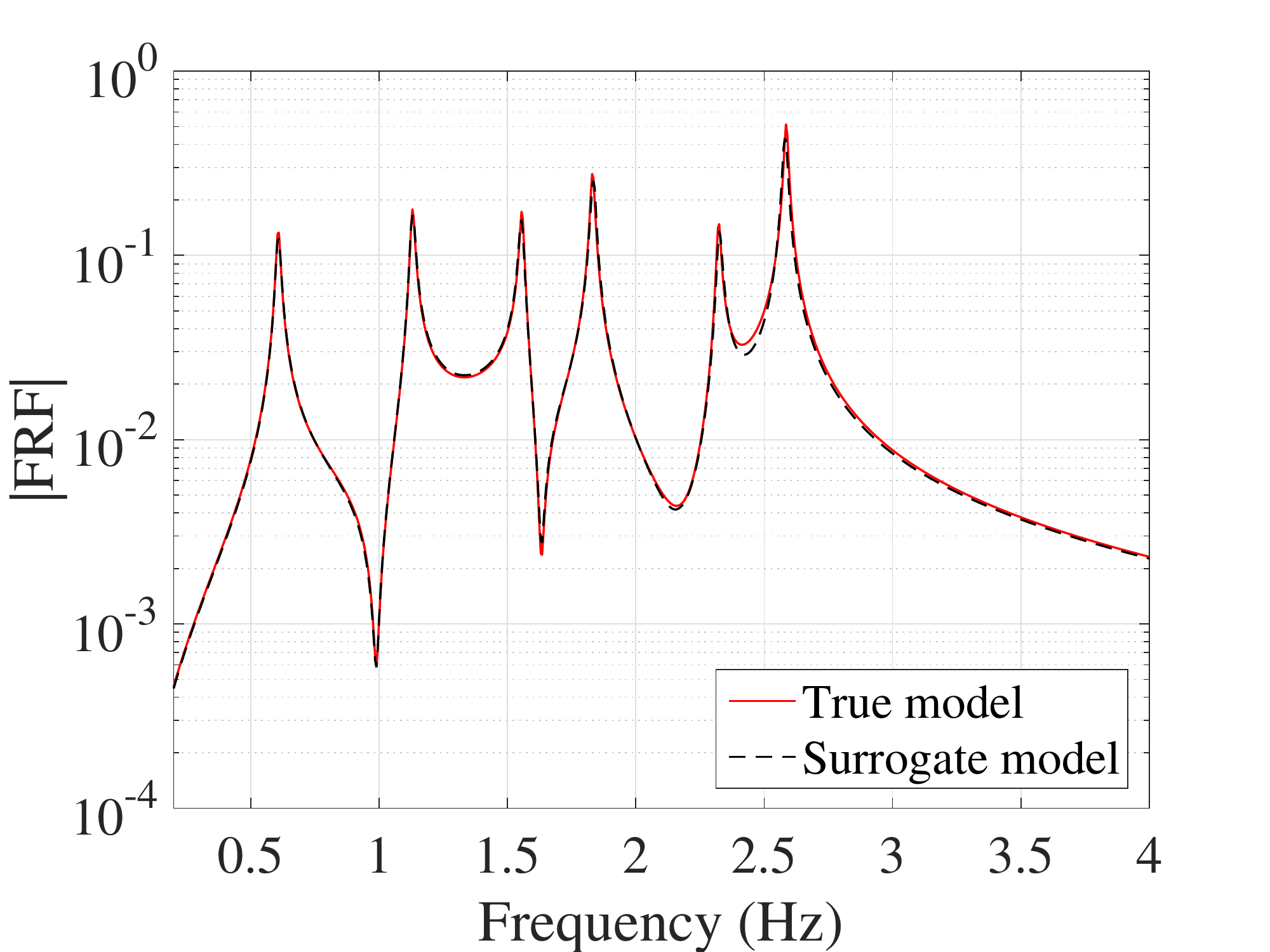}
		\caption{Second output}
		\label{fig:6DOF:FRF:typ:out3}
	\end{subfigure}
	\\
	\begin{subfigure}[b]{.5\columnwidth}
		\centering
		\includegraphics[width=1\columnwidth]{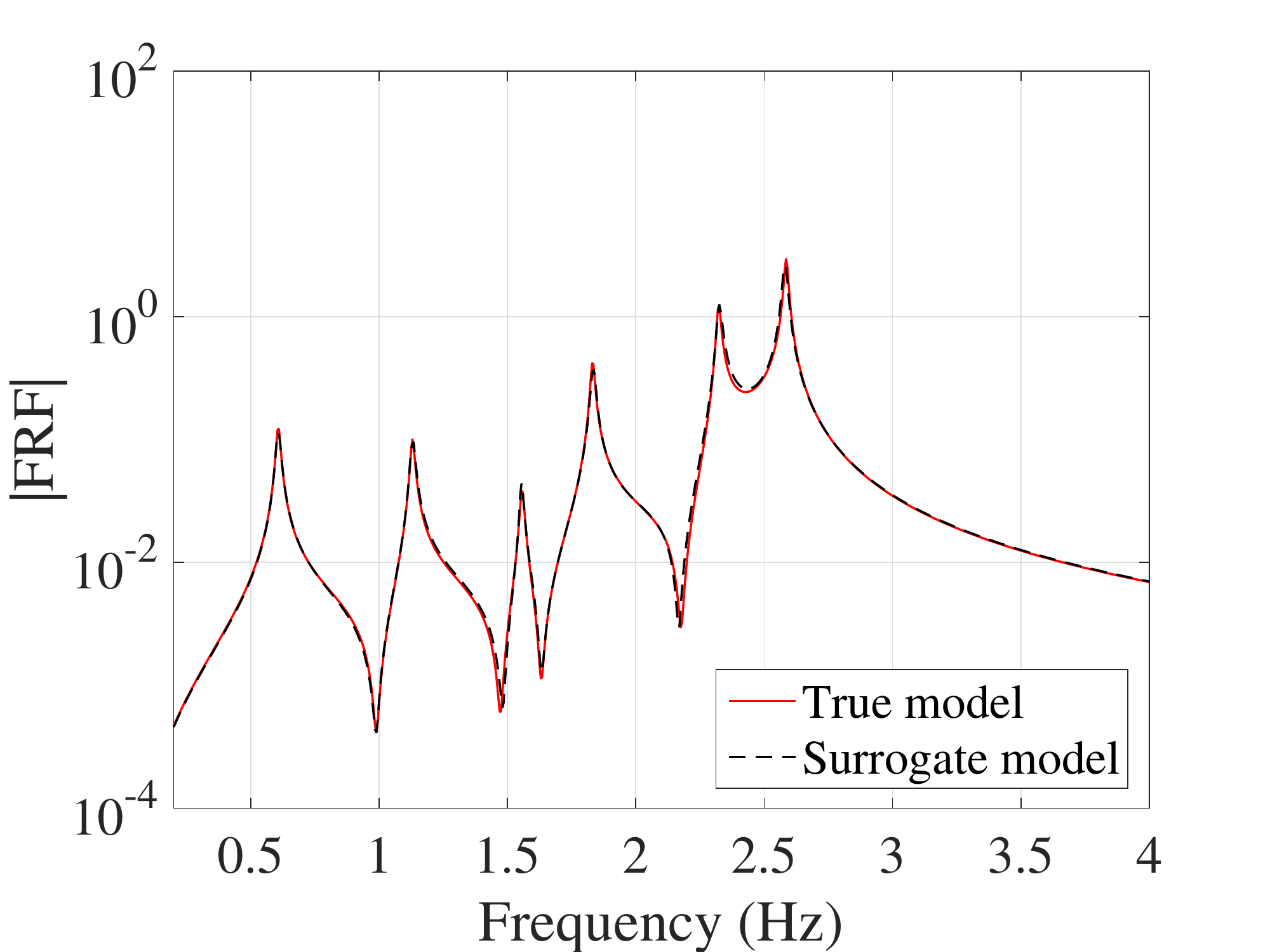}
		\caption{Third output}
		\label{fig:6DOF:FRF:typ:out4}
	\end{subfigure}
	\begin{subfigure}[b]{.5\columnwidth}
		\centering
		\includegraphics[width=1\columnwidth]{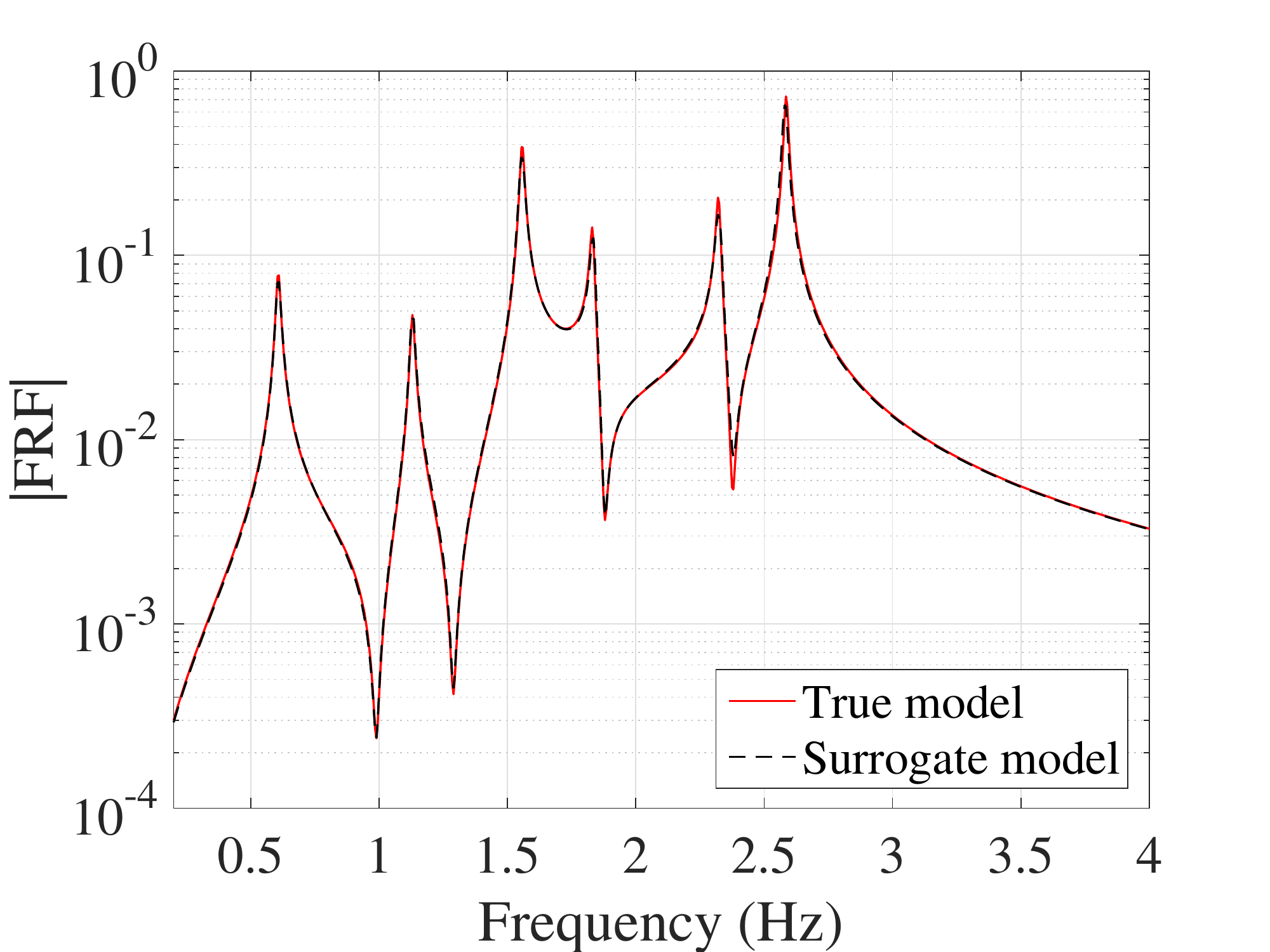}
		\caption{Fourth output}
		\label{fig:6DOF:FRF:typ:out5}
	\end{subfigure}
	\caption{Typical FRFs predicted by the surrogate model at 4 outputs, evaluated by the true model (red line) and the surrogate model (black line).}
	\label{fig:FRF:6DOF:predict:individual:typic:4output}
\end{figure}

\subsection{The worst FRFs}
\label{app:6DOF:worst}

\begin{figure}[H]
	\centering
	\begin{subfigure}[b]{.5\columnwidth}
		\centering
		\includegraphics[width=1\columnwidth]{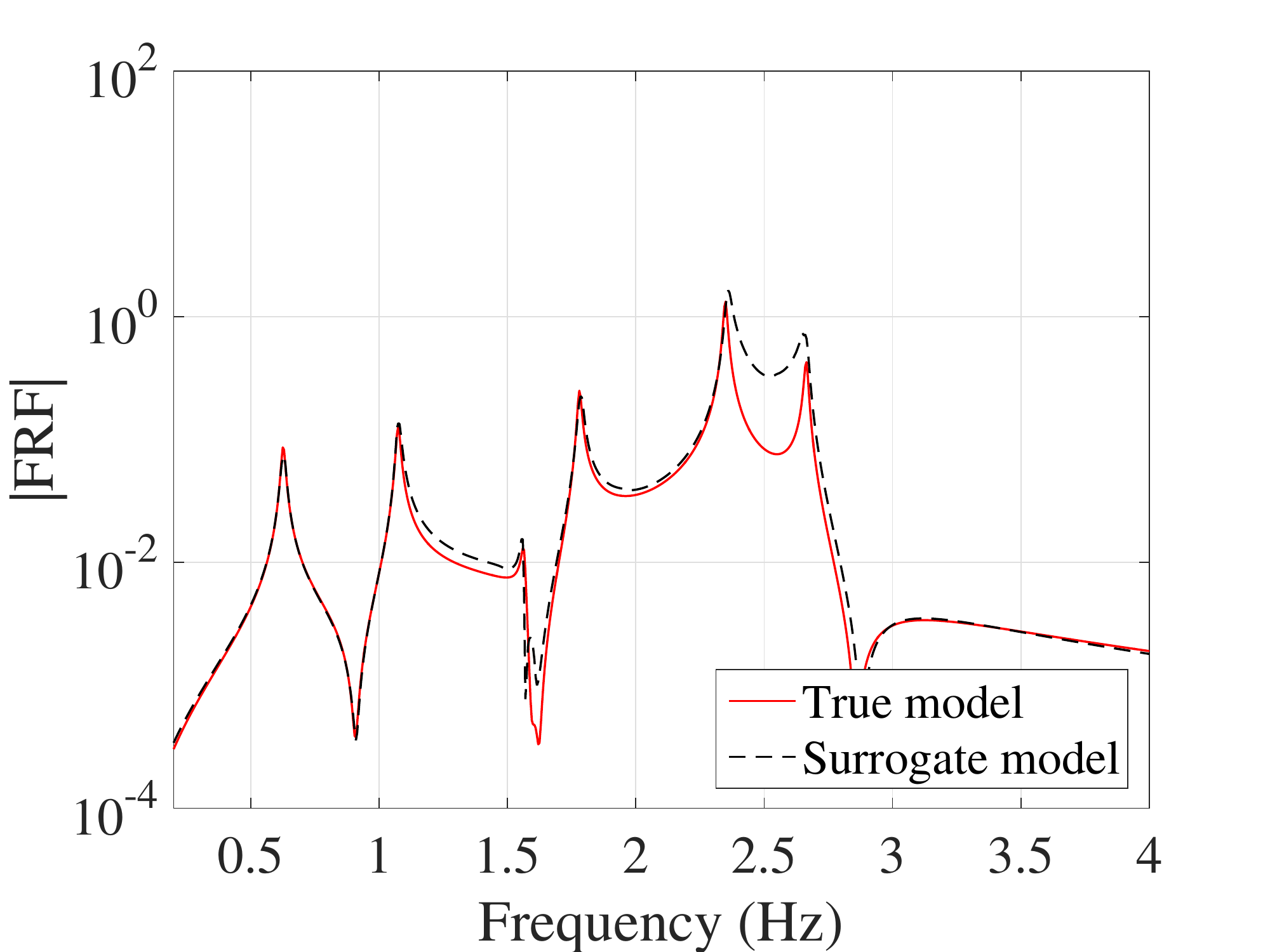}
		\caption{First output}
		\label{fig:6DOF:FRF:worst:out2}
	\end{subfigure}
	\begin{subfigure}[b]{.5\columnwidth}
		\centering
		\includegraphics[width=1\columnwidth]{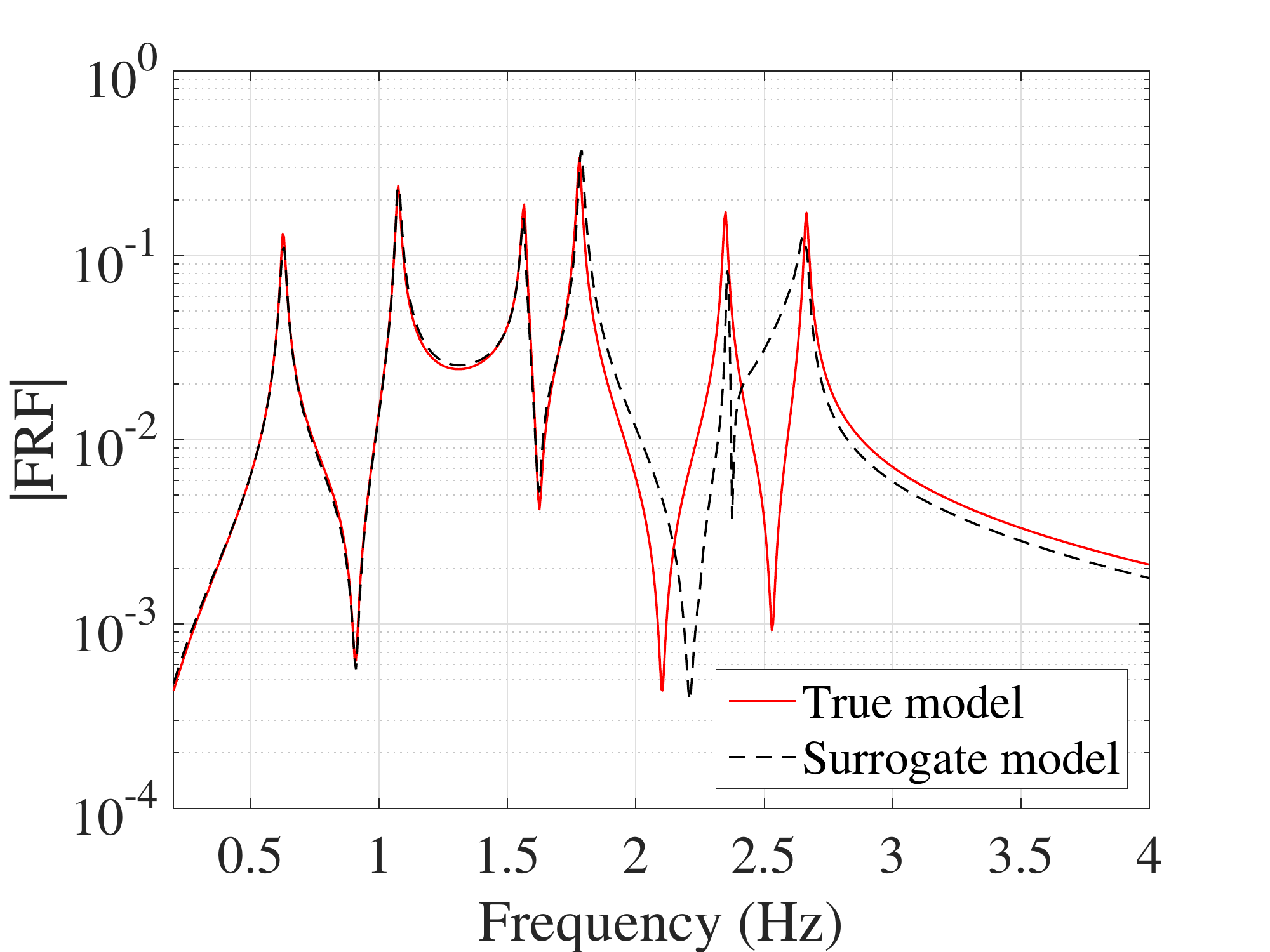}
		\caption{Second output}
		\label{fig:6DOF:FRF:worst:out3}
	\end{subfigure}
	\\
	\begin{subfigure}[b]{.5\columnwidth}
		\centering
		\includegraphics[width=1\columnwidth]{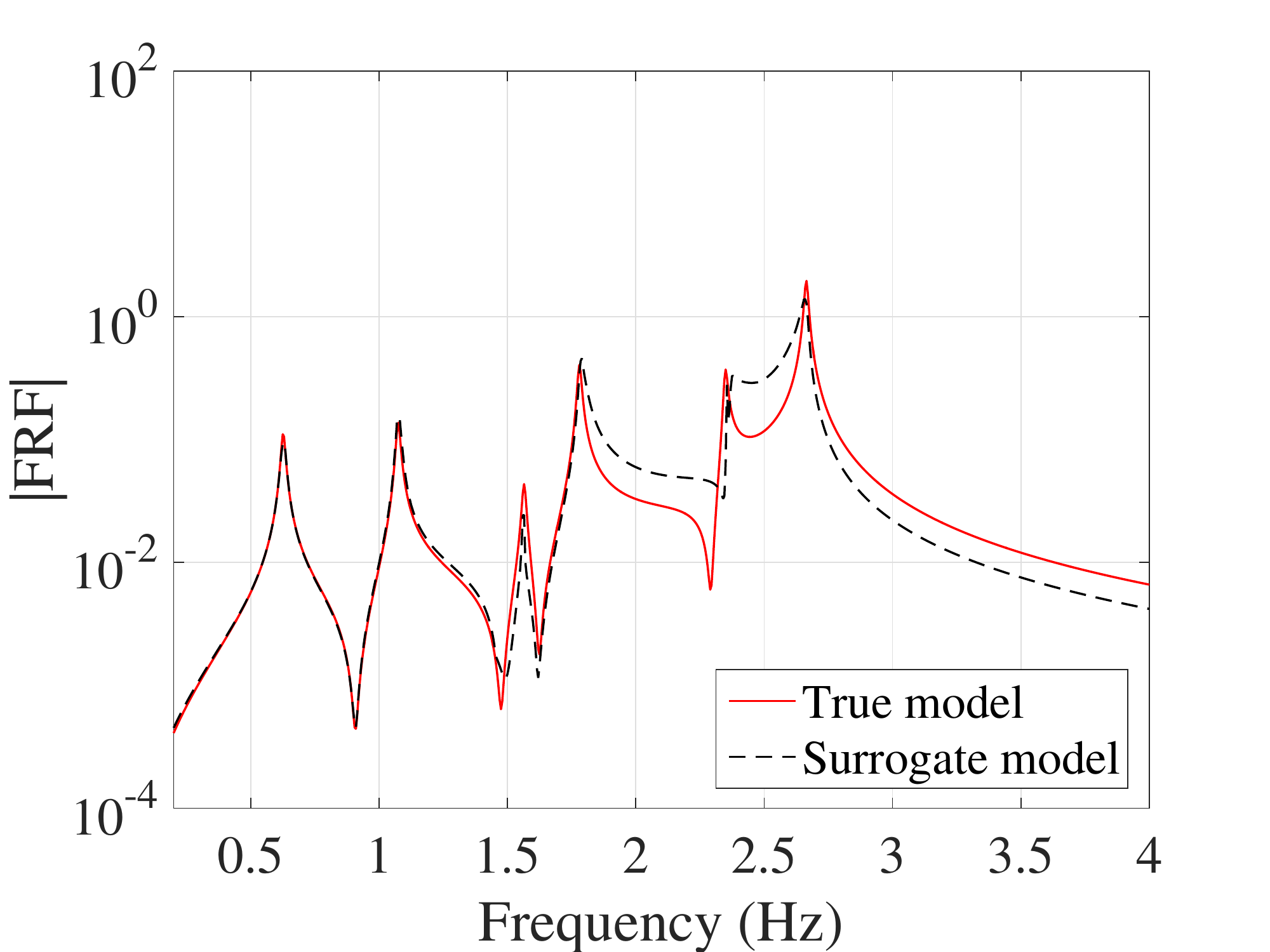}
		\caption{Third output}
		\label{fig:6DOF:FRF:worst:out4}
	\end{subfigure}
	\begin{subfigure}[b]{.5\columnwidth}
		\centering
		\includegraphics[width=1\columnwidth]{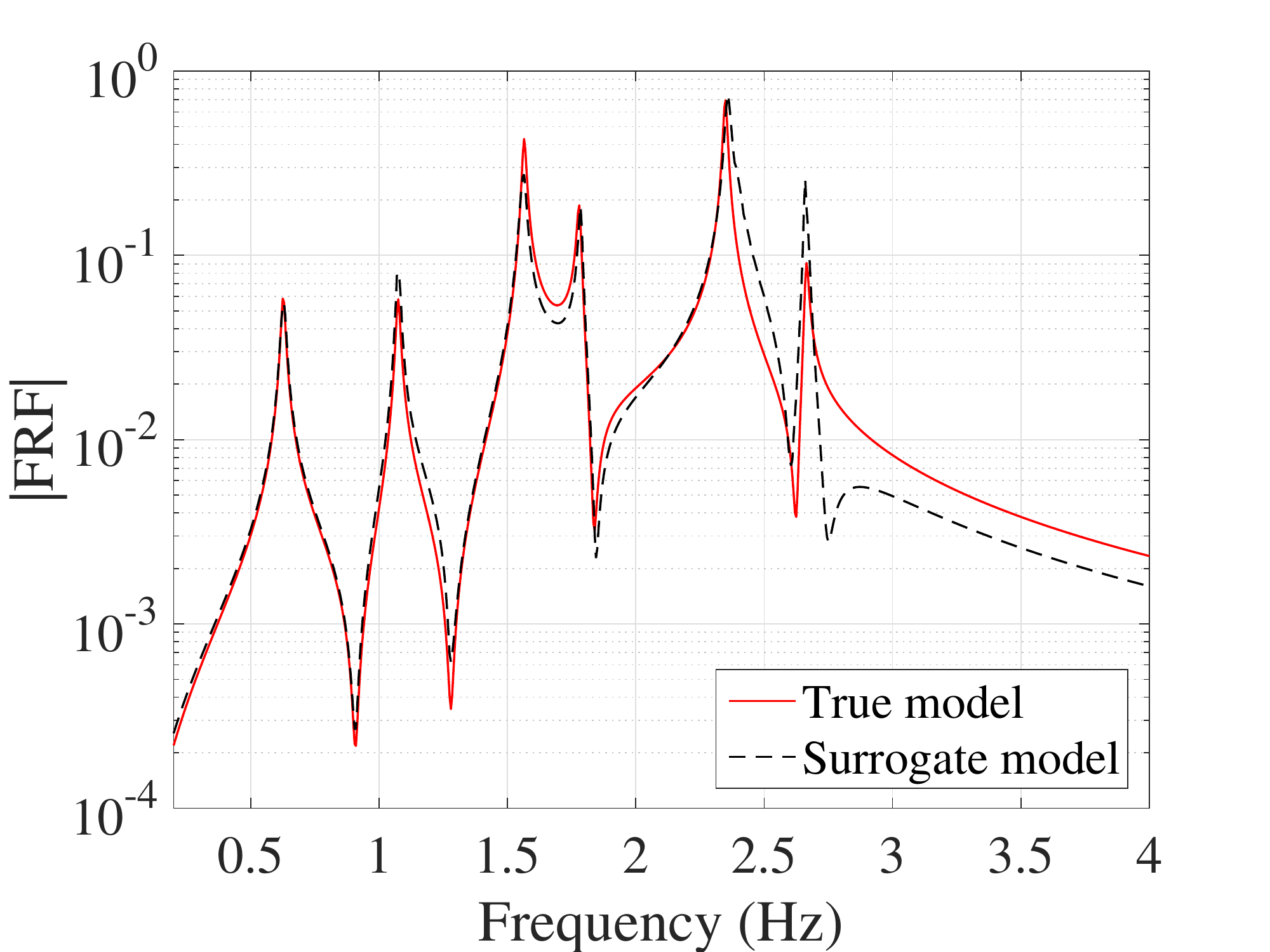}
		\caption{Fourth output}
		\label{fig:6DOF:FRF:worst:out5}
	\end{subfigure}
	\caption{The worst FRFs (out of 10,000) predicted by the surrogate model at 4 outputs, evaluated by the true model (red line) and the surrogate model (black line).}
	\label{fig:FRF:6DOF:predict:individual:worst:4output}
\end{figure}

\section*{Acknowledgment}

The first author would like to express his gratitude to the ETH Zurich for its kind host during this work.


\bibliographystyle{chicago}
\bibliography{bibliography}

\begin{thebibliography}{}

\bibitem[\protect\citeauthoryear{Adhikari}{Adhikari}{2011}]{adhikari2011doubly}
Adhikari, S. (2011).
\newblock Doubly spectral stochastic finite-element method for linear
  structural dynamics.
\newblock {\em J. Aerospace Eng.\/}~{\em 24}, 264--276.

\bibitem[\protect\citeauthoryear{Adhikari and Pascual}{Adhikari and
  Pascual}{2016}]{adhikari2016damping}
Adhikari, S. and B.~Pascual (2016).
\newblock The ‘damping effect’in the dynamic response of stochastic
  oscillators.
\newblock {\em Probabilistic Engineering Mechanics\/}~{\em 44}, 2--17.

\bibitem[\protect\citeauthoryear{Amsallem and Farhat}{Amsallem and
  Farhat}{2011}]{amsallem2011online}
Amsallem, D. and C.~Farhat (2011).
\newblock An online method for interpolating linear parametric reduced-order
  models.
\newblock {\em SIAM J. Sci. Comput.\/}~{\em 33}, 2169--2198.

\bibitem[\protect\citeauthoryear{Avitabile and O’Callahan}{Avitabile and
  O’Callahan}{2009}]{avitabile2009efficient}
Avitabile, P. and J.~O’Callahan (2009).
\newblock Efficient techniques for forced response involving linear modal
  components interconnected by discrete nonlinear connection elements.
\newblock {\em Mech. Syst. Signal Pr.\/}~{\em 23}, 45--67.

\bibitem[\protect\citeauthoryear{Berveiller, Sudret, and Lemaire}{Berveiller
  et~al.}{2006}]{berveiller2006stochastic}
Berveiller, M., B.~Sudret, and M.~Lemaire (2006).
\newblock Stochastic finite element: a non intrusive approach by regression.
\newblock {\em Eur. J. Comput. Mech.\/}~{\em 15}, 81--92.

\bibitem[\protect\citeauthoryear{Blatman and Sudret}{Blatman and
  Sudret}{2008}]{blatman2008sparse}
Blatman, G. and B.~Sudret (2008).
\newblock Sparse polynomial chaos expansions and adaptive stochastic finite
  elements using a regression approach.
\newblock {\em Comptes Rendus M{\'e}canique\/}~{\em 336}, 518--523.

\bibitem[\protect\citeauthoryear{Blatman and Sudret}{Blatman and
  Sudret}{2010}]{blatman2010adaptive}
Blatman, G. and B.~Sudret (2010).
\newblock An adaptive algorithm to build up sparse polynomial chaos expansions
  for stochastic finite element analysis.
\newblock {\em Probabilist. Eng. Mech.\/}~{\em 25}, 183--197.

\bibitem[\protect\citeauthoryear{Blatman and Sudret}{Blatman and
  Sudret}{2011a}]{Blatman2011a}
Blatman, G. and B.~Sudret (2011a).
\newblock Adaptive sparse polynomial chaos expansion based on {L}east {A}ngle
  {R}egression.
\newblock {\em J. Comput. Phys.\/}~{\em 230}, 2345--2367.

\bibitem[\protect\citeauthoryear{Blatman and Sudret}{Blatman and
  Sudret}{2011b}]{blatman2011adaptive}
Blatman, G. and B.~Sudret (2011b).
\newblock Adaptive sparse polynomial chaos expansion based on least angle
  regression.
\newblock {\em J. Comput. Phys.\/}~{\em 230}, 2345--2367.

\bibitem[\protect\citeauthoryear{Blatman and Sudret}{Blatman and
  Sudret}{2013}]{Blatman2013}
Blatman, G. and B.~Sudret (2013).
\newblock Sparse polynomial chaos expansions of vector-valued response
  quantities.
\newblock In G.~Deodatis (Ed.), {\em {Proc. 11th~Int. Conf. Struct. Safety and
  Reliability (ICOSSAR'2013)}, New York, USA}.

\bibitem[\protect\citeauthoryear{Chatterjee, Chakraborty, and
  Chowdhury}{Chatterjee et~al.}{2016}]{Chatterjee2015}
Chatterjee, T., S.~Chakraborty, and R.~Chowdhury (2016).
\newblock A bi-level approximation tool for the computation of {FRFs} in
  stochastic dynamic systems.
\newblock {\em Mech. Syst. Signal Pr.\/}~{\em 70}, 484 -- 505.

\bibitem[\protect\citeauthoryear{Craig and Kurdila}{Craig and
  Kurdila}{2006}]{craig2006fundamentals}
Craig, R.~R. and A.~J. Kurdila (2006).
\newblock {\em Fundamentals of structural dynamics}.
\newblock John Wiley \& Sons.

\bibitem[\protect\citeauthoryear{Efron, Hastie, Johnstone, Tibshirani,
  et~al.}{Efron et~al.}{2004}]{efron2004least}
Efron, B., T.~Hastie, I.~Johnstone, R.~Tibshirani, et~al. (2004).
\newblock Least angle regression.
\newblock {\em Ann. Stat.\/}~{\em 32}, 407--499.

\bibitem[\protect\citeauthoryear{Frangos, Marzouk, Willcox, and van
  Bloemen~Waanders}{Frangos et~al.}{}]{frangos2010surrogate}
Frangos, M., Y.~Marzouk, K.~Willcox, and B.~van Bloemen~Waanders.
\newblock Surrogate and reduced-order modeling: A comparison of approaches for
  large-scale statistical inverse problems.

\bibitem[\protect\citeauthoryear{Fricker, Oakley, Sims, and Worden}{Fricker
  et~al.}{2011}]{fricker2011probabilistic}
Fricker, T.~E., J.~E. Oakley, N.~D. Sims, and K.~Worden (2011).
\newblock Probabilistic uncertainty analysis of an {FRF} of a structure using a
  {G}aussian process emulator.
\newblock {\em Mech. Syst. Signal Pr.\/}~{\em 25}, 2962--2975.

\bibitem[\protect\citeauthoryear{Ghanem and Ghiocel}{Ghanem and
  Ghiocel}{1998}]{Ghiocel98}
Ghanem, R. and D.~Ghiocel (1998).
\newblock Stochastic seismic soil-structure interaction using the homogeneous
  chaos expansion.
\newblock In {\em Proc. 12th~ASCE Engineering Mechanics Division Conference, La
  Jolla, California, USA}.

\bibitem[\protect\citeauthoryear{Ghanem and Spanos}{Ghanem and
  Spanos}{2003}]{ghanem2003stochastic}
Ghanem, R.~G. and P.~D. Spanos (2003).
\newblock {\em Stochastic finite elements: a spectral approach}.
\newblock Courier Corporation.

\bibitem[\protect\citeauthoryear{Ghiocel and Ghanem}{Ghiocel and
  Ghanem}{2002}]{Ghiocel2002}
Ghiocel, D. and R.~Ghanem (2002).
\newblock Stochastic finite element analysis of seismic soil-structure
  interaction.
\newblock {\em J. Eng. Mech.\/}~{\em 128}, 66--77.

\bibitem[\protect\citeauthoryear{Gilli, Lathouwers, Kloosterman, van~der Hagen,
  Koning, and Rochman}{Gilli et~al.}{2013}]{gilli2013uncertainty}
Gilli, L., D.~Lathouwers, J.~Kloosterman, T.~van~der Hagen, A.~Koning, and
  D.~Rochman (2013).
\newblock Uncertainty quantification for criticality problems using
  non-intrusive and adaptive polynomial chaos techniques.
\newblock {\em Ann. Nucl. Energy\/}~{\em 56}, 71--80.

\bibitem[\protect\citeauthoryear{Goller, Pradlwarter, and Schu{\"e}ller}{Goller
  et~al.}{2011}]{goller2011interpolation}
Goller, B., H.~Pradlwarter, and G.~Schu{\"e}ller (2011).
\newblock An interpolation scheme for the approximation of dynamical systems.
\newblock {\em Comput. Methods Appl. Mech. Engrg.\/}~{\em 200}, 414--423.

\bibitem[\protect\citeauthoryear{Hastie, Taylor, Tibshirani, Walther,
  et~al.}{Hastie et~al.}{2007}]{hastie2007forward}
Hastie, T., J.~Taylor, R.~Tibshirani, G.~Walther, et~al. (2007).
\newblock Forward stagewise regression and the monotone lasso.
\newblock {\em Electron. J. Stat.\/}~{\em 1}, 1--29.

\bibitem[\protect\citeauthoryear{Jacquelin, Adhikari, Sinou, and
  Friswell}{Jacquelin et~al.}{2015a}]{Jacquelin2015144}
Jacquelin, E., S.~Adhikari, J.-J. Sinou, and M.~Friswell (2015a).
\newblock Polynomial chaos expansion in structural dynamics: {A}ccelerating the
  convergence of the first two statistical moment sequences.
\newblock {\em J. Sound. Vib.\/}~{\em 356}, 144 -- 154.

\bibitem[\protect\citeauthoryear{Jacquelin, Adhikari, Sinou, and
  Friswell}{Jacquelin et~al.}{2015b}]{jacquelinpolynomial2015}
Jacquelin, E., S.~Adhikari, J.~J. Sinou, and M.~I. Friswell (2015b).
\newblock Polynomial chaos expansion and {S}teady-{S}tate response of a class
  of random dynamical systems.
\newblock {\em J. Eng. Mech.\/}~{\em 141}, 04014145.

\bibitem[\protect\citeauthoryear{Jones, Schonlau, and Welch}{Jones
  et~al.}{1998}]{jones1998efficient}
Jones, D.~R., M.~Schonlau, and W.~J. Welch (1998).
\newblock Efficient global optimization of expensive black-box functions.
\newblock {\em J. Global Optim.\/}~{\em 13}, 455--492.

\bibitem[\protect\citeauthoryear{Kersaudy, Sudret, Varsier, Picon, and
  Wiart}{Kersaudy et~al.}{2015}]{kersaudy2015new}
Kersaudy, P., B.~Sudret, N.~Varsier, O.~Picon, and J.~Wiart (2015).
\newblock A new surrogate modeling technique combining {K}riging and polynomial
  chaos expansions--application to uncertainty analysis in computational
  dosimetry.
\newblock {\em J. Comput. Phys.\/}~{\em 286}, 103--117.

\bibitem[\protect\citeauthoryear{Khorsand~Vakilzadeh, Rahrovani, and
  Abrahamsson}{Khorsand~Vakilzadeh et~al.}{2012}]{khorsand2012improved}
Khorsand~Vakilzadeh, M., S.~Rahrovani, and T.~Abrahamsson (2012).
\newblock An improved modal approach for model reduction based on input-output
  relation.
\newblock In {\em Int. Conf. on Noise and Vibration Engineering (ISMA)/Int.
  Conf. on Uncertainty in Struct. Dynamics (USD). Leuven, Belgium, 2012}, pp.\
  3451--3459. Katholieke Univ Leuven, Dept Werktuigkunde.

\bibitem[\protect\citeauthoryear{Knio, Najm, Ghanem, et~al.}{Knio
  et~al.}{2001}]{knio2001stochastic}
Knio, O.~M., H.~N. Najm, R.~G. Ghanem, et~al. (2001).
\newblock A stochastic projection method for fluid flow: I. basic formulation.
\newblock {\em J. Comput. Phys.\/}~{\em 173}, 481--511.

\bibitem[\protect\citeauthoryear{Kundu and Adhikari}{Kundu and
  Adhikari}{2015}]{kundu2015dynamic}
Kundu, A. and S.~Adhikari (2015).
\newblock Dynamic analysis of stochastic structural systems using frequency
  adaptive spectral functions.
\newblock {\em Probabilit. Eng. Mech.\/}~{\em 39}, 23--38.

\bibitem[\protect\citeauthoryear{Kundu, DiazDelaO, Adhikari, and
  Friswell}{Kundu et~al.}{2014}]{kundu2014hybrid}
Kundu, A., F.~DiazDelaO, S.~Adhikari, and M.~Friswell (2014).
\newblock A hybrid spectral and metamodeling approach for the stochastic finite
  element analysis of structural dynamic systems.
\newblock {\em Comput. Methods Appl. Mech. Engrg.\/}~{\em 270}, 201--219.

\bibitem[\protect\citeauthoryear{Laub}{Laub}{2004}]{laub2005matrix}
Laub, A.~J. (2004).
\newblock {\em Matrix Analysis For Scientists And Engineers}.
\newblock Philadelphia, PA, USA: Society for Industrial and Applied
  Mathematics.

\bibitem[\protect\citeauthoryear{Liu, Zhao, Li, and Zhang}{Liu
  et~al.}{2012}]{liu2012efficient}
Liu, T., C.~Zhao, Q.~Li, and L.~Zhang (2012).
\newblock An efficient backward {E}uler time-integration method for nonlinear
  dynamic analysis of structures.
\newblock {\em Comput. Struct.\/}~{\em 106}, 20--28.

\bibitem[\protect\citeauthoryear{Mai and Sudret}{Mai and
  Sudret}{2015}]{mai2015polynomial}
Mai, C.~V. and B.~Sudret (2015).
\newblock Polynomial chaos expansions for damped oscillators.
\newblock In {\em Proc. 12th Int. Conf. on Applications of Stat. and Prob. in
  Civil Engineering (ICASP12), Vancouver, Canada}.

\bibitem[\protect\citeauthoryear{Manan and Cooper}{Manan and
  Cooper}{2010}]{manan2010prediction}
Manan, A. and J.~Cooper (2010).
\newblock Prediction of uncertain frequency response function bounds using
  polynomial chaos expansion.
\newblock {\em J. Sound Vib.\/}~{\em 329}, 3348--3358.

\bibitem[\protect\citeauthoryear{Pagnacco, Sarrouy, Sampaio, and
  De~Cursis}{Pagnacco et~al.}{2013}]{pagnacco2013polynomial}
Pagnacco, E., E.~Sarrouy, R.~Sampaio, and E.~S. De~Cursis (2013).
\newblock Polynomial chaos for modeling multimodal dynamical
  systems-investigations on a single degree of freedom system.
\newblock In {\em Mec{\'a}nica Computacional, Mendoza, Argentina}.

\bibitem[\protect\citeauthoryear{Pichler, Gallina, Uhl, and Bergman}{Pichler
  et~al.}{2012}]{PichleraMetamodelNaturalfrequency2012}
Pichler, L., A.~Gallina, T.~Uhl, and L.~A. Bergman (2012).
\newblock A meta-modeling technique for the natural frequencies based on the
  approximation of the characteristic polynomial.
\newblock {\em Comput. Struct.\/}~{\em 102-103}, 108116.

\bibitem[\protect\citeauthoryear{Pichler, Pradlwarter, and
  Schu{\"e}ller}{Pichler et~al.}{2009}]{pichler2009mode}
Pichler, L., H.~Pradlwarter, and G.~Schu{\"e}ller (2009).
\newblock A mode-based meta-model for the frequency response functions of
  uncertain structural systems.
\newblock {\em Comput. Struct.\/}~{\em 87}, 332--341.

\bibitem[\protect\citeauthoryear{Rahrovani, Vakilzadeh, and
  Abrahamsson}{Rahrovani et~al.}{2014}]{rahrovani2014modal}
Rahrovani, S., M.~K. Vakilzadeh, and T.~Abrahamsson (2014).
\newblock Modal dominancy analysis based on modal contribution to frequency
  response function ℋ2-norm.
\newblock {\em Mech. Syst. Signal Pr.\/}~{\em 48}, 218--231.

\bibitem[\protect\citeauthoryear{Sch\"obi, Sudret, and Wiart}{Sch\"obi
  et~al.}{2015}]{SchoebiIJUQ2015}
Sch\"obi, R., B.~Sudret, and J.~Wiart (2015).
\newblock Polynomial-chaos-based {K}riging.
\newblock {\em Int. J. Uncertainty Quantification\/}~{\em 5}, 171--193.

\bibitem[\protect\citeauthoryear{Schu{\"e}ller and Pradlwarter}{Schu{\"e}ller
  and Pradlwarter}{2009}]{schueller2009uncertain}
Schu{\"e}ller, G. and H.~Pradlwarter (2009).
\newblock Uncertain linear systems in dynamics: Retrospective and recent
  developments by stochastic approaches.
\newblock {\em Eng. Struct.\/}~{\em 31}, 2507--2517.

\bibitem[\protect\citeauthoryear{Soize and Ghanem}{Soize and
  Ghanem}{2004}]{soize2004physical}
Soize, C. and R.~Ghanem (2004).
\newblock Physical systems with random uncertainties: chaos representations
  with arbitrary probability measure.
\newblock {\em SIAM J. Sci. Comput.\/}~{\em 26}, 395--410.

\bibitem[\protect\citeauthoryear{Sudret}{Sudret}{2007}]{Sudret2007}
Sudret, B. (2007).
\newblock Uncertainty propagation and sensitivity analysis in mechanical models
  -- contributions to structural reliability and stochastic spectral methods.
\newblock Technical report.
\newblock Habilitation \`a diriger des recherches, Universit\'e Blaise Pascal,
  Clermont-Ferrand, France (229 pages).

\bibitem[\protect\citeauthoryear{Tak and Park}{Tak and
  Park}{2013}]{tak2013high}
Tak, M. and T.~Park (2013).
\newblock High scalable non-overlapping domain decomposition method using a
  direct method for finite element analysis.
\newblock {\em Comput. Methods Appl. Mech. Engrg.\/}~{\em 264}, 108--128.

\bibitem[\protect\citeauthoryear{Wiener}{Wiener}{1938}]{wiener1938homogeneous}
Wiener, N. (1938).
\newblock The homogeneous chaos.
\newblock {\em Amer. J. Math.\/}, 897--936.

\bibitem[\protect\citeauthoryear{Xiu and Karniadakis}{Xiu and
  Karniadakis}{2002}]{xiu2002wiener}
Xiu, D. and G.~E. Karniadakis (2002).
\newblock The {W}iener--{A}skey polynomial chaos for stochastic differential
  equations.
\newblock {\em SIAM J. Sci. Comput.\/}~{\em 24}, 619--644.

\bibitem[\protect\citeauthoryear{Yaghoubi, Abrahamsson, and Johnson}{Yaghoubi
  et~al.}{2016}]{yaghoubi2016efficient}
Yaghoubi, V., T.~Abrahamsson, and E.~A. Johnson (2016).
\newblock An efficient exponential predictor-corrector time integration method
  for structures with local nonlinearity.
\newblock {\em Eng. Struct.\/}~{\em 128}, 344 -- 361.

\bibitem[\protect\citeauthoryear{Yaghoubi, Vakilzadeh, and
  Abrahamsson}{Yaghoubi et~al.}{2015}]{yaghoubiparallel}
Yaghoubi, V., M.~K. Vakilzadeh, and T.~Abrahamsson (2015).
\newblock A parallel solution method for structural dynamic response analysis.
\newblock In {\em Dynamics of Coupled Structures, Volume 4}, pp.\  149--161.
  Springer.

\bibitem[\protect\citeauthoryear{Yu, Gillot, and Ichchou}{Yu
  et~al.}{2011}]{yuhermite}
Yu, H., F.~Gillot, and M.~Ichchou (2011).
\newblock Hermite polynomial chaos expansion method for stochastic frequency
  response estimation considering modal intermixing.
\newblock In {\em ECCOMAS Thematic Conf. on Computational Methods in Structural
  Dynamics and Earthquake Engineering, Corfu, Greece, 2011}.

\end{thebibliography}
\end{document}